\documentclass[twocolumn]{aastex631}

\usepackage{booktabs} 
\usepackage{multirow}
\usepackage{amsmath, cleveref}

\begin{document}

\title{A luminous and hot infrared through X-ray transient at a 5\,kpc offset from a dwarf galaxy}

\author[0000-0001-8426-5732]{Jean J. Somalwar}\email{jsomalwa@caltech.edu}
\affil{Cahill Center for Astronomy and Astrophysics, MC\,249-17 California Institute of Technology, Pasadena CA 91125, USA.}

\author[0000-0002-7252-5485]{Vikram Ravi}
\affil{Cahill Center for Astronomy and Astrophysics, MC\,249-17 California Institute of Technology, Pasadena CA 91125, USA.}

\author[0000-0003-4768-7586]{Raffaella Margutti}
\affiliation{Department of Astronomy, University of California, Berkeley, CA 94720-3411, USA}
\affiliation{Department of Physics, University of California, 366 Physics North MC 7300, Berkeley, CA 94720, USA}

\author[0000-0002-7706-5668]{Ryan Chornock}
\affiliation{Department of Astronomy, University of California, Berkeley, CA 94720-3411, USA}

\author[0000-0002-5554-8896]{Priyamvada Natarajan}
\affiliation{Department of Astronomy, Yale University, P.O. Box 208101, New Haven, CT 06520, USA}
\affiliation{Black Hole Initiative, Harvard University, Cambridge, MA 02138, USA}
\affiliation{Department of Physics, Yale University, P.O. Box 208121, New Haven, CT 06520, USA}

\author[0000-0002-1568-7461]{Wenbin Lu}
\affiliation{Department of Astronomy, University of California, Berkeley, CA 94720-3411, USA}
\affiliation{Theoretical Astrophysics Center, University of California, Berkeley, CA 94720, USA}

\author[0000-0002-4269-7999]{Charlotte Angus}
\affiliation{Astrophysics Research Centre, School of Mathematics and Physics, Queens University Belfast, Belfast BT7 1NN, UK}

\author[0000-0002-3168-0139]{Matthew J. Graham}
\affil{Cahill Center for Astronomy and Astrophysics, MC\,249-17 California Institute of Technology, Pasadena CA 91125, USA.}

\author[0000-0002-5698-8703]{Erica Hammerstein}
\affiliation{Department of Astronomy, University of California, Berkeley, CA 94720-3411, USA}

\author[0000-0002-9633-9193]{Edward Nathan}
\affil{Cahill Center for Astronomy and Astrophysics, MC\,249-17 California Institute of Technology, Pasadena CA 91125, USA.}

\author[0000-0002-2555-3192]{Matt Nicholl}
\affiliation{Astrophysics Research Centre, School of Mathematics and Physics, Queens University Belfast, Belfast BT7 1NN, UK}

\author[0000-0002-4477-3625]{Kritti Sharma}
\affil{Cahill Center for Astronomy and Astrophysics, MC\,249-17 California Institute of Technology, Pasadena CA 91125, USA.}

\author[0000-0003-2434-0387]{Robert Stein}
\affiliation{Department of Astronomy, University of Maryland, College Park, MD 20742, USA}
\affiliation{Joint Space-Science Institute, University of Maryland, College Park, MD 20742, USA} 
\affiliation{Astrophysics Science Division, NASA Goddard Space Flight Center, Mail Code 661, Greenbelt, MD 20771, USA} 

\author[0009-0005-3565-4164]{Frank Verdi}
\affil{Cahill Center for Astronomy and Astrophysics, MC\,249-17 California Institute of Technology, Pasadena CA 91125, USA.}

\author[0000-0001-6747-8509]{Yuhan Yao}
\affiliation{Miller Institute for Basic Research in Science, 468 Donner Lab, Berkeley, CA 94720, USA}
\affiliation{Department of Astronomy, University of California, Berkeley, CA 94720-3411, USA}

\author[0000-0001-8018-5348]{Eric C. Bellm}
\affiliation{DIRAC Institute, Department of Astronomy, University of Washington, 3910 15th Avenue NE, Seattle, WA 98195, USA}

\author[0000-0001-9152-6224]{Tracy X. Chen}
\affiliation{IPAC, California Institute of Technology, 1200 E. California
             Blvd, Pasadena, CA 91125, USA}

\author[0000-0002-8262-2924]{Michael W. Coughlin}
\affiliation{School of Physics and Astronomy, University of Minnesota, Minneapolis, MN 55455, USA} 

\author{David Hale}
\affiliation{Caltech Optical Observatories, California Institute of Technology, Pasadena, CA  91125}

\author[0000-0002-5619-4938]{Mansi M. Kasliwal}
\affil{Cahill Center for Astronomy and Astrophysics, MC\,249-17 California Institute of Technology, Pasadena CA 91125, USA.}

\author[0000-0003-2451-5482]{Russ R. Laher}
\affiliation{IPAC, California Institute of Technology, 1200 E. California
             Blvd, Pasadena, CA 91125, USA}

\author{Reed Riddle}
\affiliation{Caltech Optical Observatories, California Institute of Technology, Pasadena, CA  91125}

\author[0000-0003-1546-6615]{Jesper Sollerman}
\affiliation{Department of Astronomy, The Oskar Klein Center, Stockholm University, AlbaNova University Center, SE 106 91 Stockholm, Sweden}

\begin{abstract}
We are searching for hot, constant-color, offset optical flares in the Zwicky Transient Facility (ZTF) data stream that are ${>}10''$ from any galaxy in public imaging data from the PanSTARRS survey. Here, we present the first discovery from this search: AT\,2024puz, a luminous multiwavelength transient offset by $5\,$kpc from a ${\sim}10^8\,M_\odot$ galaxy at $z=0.356$ with a low-moderate star formation rate. It produced luminous $10^{44.79 \pm 0.04}\,{\rm erg\,s}^{-1}$ optical/UV emission that evolved on a ${\sim}20$\,day timescale, as well as $10^{44.12\pm0.03}\,{\rm erg\,s}^{-1}$ X-ray emission with a photon-index $\Gamma=1.7$. No associated radio or millimeter emission was detected. We show that the early-time optical emission is likely powered by reprocessing of high-energy, accretion-powered radiation, with a possible contribution from a shock in a dense circum-transient medium. If the shock is dominant at early-times, the circum-transient medium has a mass ${\sim}0.1-1\,M_\odot$, radius $10^{15}\,$cm, and a density profile shallower than ${\sim}r^{-1}$. A near-infrared excess appears at late-times and is suggestive of reprocessing within a wind or other circum-transient medium. The X-rays are most consistent with a central engine. We suggest that AT\,2024puz may be associated with an accretion event onto a $50-10^5\,M_\odot$ BH, where the lower masses are preferred based on the large projected offset from the host galaxy. AT2024puz exhibits properties similar to both luminous fast blue optical transients (LFBOTs) and tidal disruption events (TDEs), but is intermediate between them in its energetics and evolution timescale. This highlights the need for broader exploration of the landscape of hot optical transients to trace their origins.
\end{abstract}


\section{Introduction} \label{sec:intro}


Many open questions surround the formation and demographics of black holes (BHs) with masses $10^{2-5}\,M_\odot$. The mass function of stellar mass BHs informs models of stellar evolution. The ``upper mass gap'' of stellar mass black holes, near ${\sim}50\,M_\odot$, is thought to be the result of implosion of more massive stars in pulsational pair instability supernovae (SNe), but gravitational wave searches have begun detecting objects above this limit \citep[][]{Abbott2019BinaryVirgo, Woosley2017PulsationalSupernovae}, possibly due to mergers of massive stellar BHs in dense environments \citep[][]{Kremer2020PopulatingClusters}.

These massive stellar-mass BHs may be the local universe analogues of the seeds of SMBHs (BH mass $M_{\rm BH} > 10^{6}\,M_\odot$). SMBH seeding and growth is one of the biggest open problems in astrophysics \citep[][]{Inayoshi2020TheHoles,PNseeds2014}. If BHs with masses ${\sim}100\,M_\odot$, formed from the stellar remnants of the first stars at high redshift, can grow rapidly through super-Eddington accretion, they may be able to form SMBHs detected at both high and low redshifts as proposed by the light seed models of SMBH growth \citep[][]{Madau2001MassiveRemnants,PNseeds2014}. Additionally, amplified growth of light seeds might be incubated in nuclear star clusters that might operate throughout cosmic time leading to the formation of IMBHs even in the nearby universe \citep[][]{PN-IMBH2021}.  Heavy seed models invoke ${\sim}10^4\,M_\odot$ gas clouds that directly collapse into BHs in the early universe \citep[][]{Loeb1994CollapseHoles,DCBHLN2006} or stellar mass black holes that rapidly accrete through runaway mergers and/or rapid accretion in dense environments \citep[][]{Miller2002FourBodyCoalescence,DevecchiMV2009,TalPN2014}. The resulting massive BHs may subsequently grow through Eddington-limited accretion to SMBH masses over time. These models produce distinct, observable predictions for the present-day demographics, occupation fraction, and locations of intermediate mass BHs (IMBHs; $M_{\rm BH} \sim 10^{2-5}\,M_\odot$) \citep[][]{greene_imbh}.

To solve these challenging open problems, we must discover more massive stellar and supermassive BHs. Transient emission is proving to be the most promising method of doing so as they uncover both active and quiescent BHs. In particular, transient optical flaring is enabling us to discover {\it populations} of time-domain varying events associated with compact objects and pushing to new regimes of their physics. Accretion flares from low mass SMBHs and IMBHs form a clear example of this, which we briefly outline here.

Tidal disruption events (TDEs) occur when a massive black hole (MBH, $M_{\rm BH} \sim 10^{2-9}\,M_\odot$) tidally shreds a nearby star \citep[][]{phinney_tde, rees_tde, evans_tde}. TDEs produce luminous electromagnetic flares and illuminate otherwise non-accreting and unobservable MBHs. To date ${\sim}100$ TDEs have been discovered, largely thanks to wide-field, cadenced, optical surveys, like the Zwicky Transient Facility (ZTF; \citealp[][]{Bellm2019TheResults, Graham2019TheObjectives, Dekany2020TheSystem, Masci2019TheArchive}), All Sky Automated Survey for SuperNovae
 (ASASSN; \citealp[][]{Shappee2014THENGC2617, Kochanek2017TheV1.0}), Panoramic Survey Telescope and Rapid Response System (Pan-STARRS1; PS1; \citealp{Chambers2016TheSurveys}), and Asteroid Terrestrial-impact Last Alert System (ATLAS). These have contributed ${\gtrsim}100$ events in the last decade \citep[e.g.][]{vanvelzen_firstztftde, hammerstein_finalseason, vanvelzen_ztf, YaoTDESamp, Hoogendam2024DiscoveryDate, Holoien2020TheTimes, Hinkle2020DiscoveryGalaxy, Holoien2019DiscoveryTESS, Hinkle2020DiscoveryGalaxy, Hinkle2023SCATGalaxy}. TDEs produce blue, hot ($T\sim 10^{4-5}\,$K) optical flares that show minimal cooling and evolve over weeks-months.

TDE discoveries are beginning to push down to the lowest mass SMBHs known (${\sim}10^5\,M_\odot$; \citealp[e.g.][]{angus_imbhtde, Yao2024SubrelativisticAT2022lri}), providing unique access and potential for discovering the elusive population of bona-fide IMBHs. Selection effects, however, may be preventing discoveries of a population of IMBH TDEs. In ZTF, TDEs are typically identified as blue ($g-r < 0.2$\,mag), constant color ($d(g-r)/dt < 0.02$\,mag/day) coincident with the location of a galactic nucleus, defined by pre-TDE images from public survey data (typically Pan-STARRS; \citealp[][]{ps1, vanvelzen_firstztftde, YaoTDESamp}). The last host galaxy cut is particularly significant for ZTF IMBH TDE sensitivity. First, many IMBHs are expected to be hosted by dwarf galaxies, which are not detectable to large comoving volumes in typical survey data (e.g., a $10^8\,M_\odot$ galaxy can be detected to $z\approx 0.1$ in Pan-STARRS). Requiring a detectable host galaxy typically rules out these sources. Likewise, dwarf galaxies may not have well-defined nuclei, or gravitational potential minima \citep[][]{Weller2022DynamicsSimulation}, but TDEs are required to reside in their hosts' nuclei. Off-center TDEs are expected to be more numerous in more massive galaxies from wandering IMBHs in the hierarchical build up of structure in the standard $\Lambda$ cold dark matter paradigm via mergers of galaxies and the BHs hosted by them \citep[][]{ricarte2021a}. These will likewise be excluded from current searches \citep[but see][]{Yao2025AAT2024tvd}. 

Searches of optical transient surveys that focus on non-TDE phenomena do not suffer from the same selection effects, and so may be more sensitive to IMBHs. One candidate for an IMBH-triggered transient is the luminous fast blue optical transient (LFBOT; \citealp[][]{Metzger_LFBOT, prentice_18cow, Kuin2019AnTransients, coppejans_css161010, gutierrez_css161010}). LFBOTs have optical evolution similar to TDEs (blue, constant color), but vary on much faster timescales $\lesssim 1$ week and always offset from their host galaxy center \citep[e.g.][]{ho_18cow, perley_18cow, ho_fbots, ho_koala, chrimes_finch, coppejans_css161010, gutierrez_css161010, bright_xnd, ho_2020xnd, chrimes_fhnlate, inkenhaag_cow, chrimes_finch, migliori_cow}. They are produced by a compact object of an as yet unknown nature: stellar mass BHs, IMBHs, and neutron stars (NSs) have all been proposed as potential sources \citep[][]{margutti_18cow, li_lfbot_magnetar, Metzger_LFBOT, 2025arXiv250103316T, Pasham2022EvidenceAT2018cow}. Like TDE searches, LFBOT searches may be limited by other selection effects: events are typically required to rise and fade on a $\lesssim$week timescale \citep[][]{ho_fbots}, which is significantly faster than almost all SMBH TDEs \citep[][]{YaoTDESamp}, leaving a gap between these populations. If LFBOTs are not TDEs, they are likely triggered by accretion onto a NS or stellar mass BH, or spindown of a young magnetar \citep[see][and references therein]{margutti_18cow}. 

If we want to identify the nature of the compact object that triggers LFBOTs, or find definitive accreting IMBHs, we must extend optical transient searches to span the parameter space intermediate to TDEs and LFBOTs. In other words, we need to map out the full range of these extreme, energetic, hot optical transients, regardless of timescale or host galaxy. This will allow us to conclusively connect these populations, or show that they are distinct and that LFBOTs are more likely associated with stellar mass compact objects, as well as potentially identify new types of transients.

This paper presents the first discovery of our ongoing effort to explore the full landscape of energetic, hot optical transients with ZTF. Here, we present AT\,2024puz (hereafter 24puz), a luminous, multiwavelength transient in a dwarf galaxy that may be associated with an extreme compact object accretion event. We favor models with an accreting BH ranging in mass $50-10^5\,M_\odot$. In Section~\ref{sec:discovery}, we present the parameters of the search in which we discovered 24puz. In Section~\ref{sec:obs}, we detail our extensive multiwavelength follow-up effort and data reduction. In Section~\ref{sec:ana}, we analyze our observations to constrain the basic physical parameters of 24puz. In Section~\ref{sec:res}, we constrain the emission mechanisms and sizes of the emitting regions. In Section~\ref{sec:disc}, we compare 24puz to similar transient classes, constrain the rate of 24puz-like events, and comment on the most likely kind of compact object powering the event. Finally, we present our conclusions and future prospects in Section~\ref{sec:conc}.

We adopt the Plank2015 \citep{Ade2015PlanckParameters} cosmology with $H_0 = 67.7$\,km\,s$^{-1}$ and $\Omega_{M} = 0.3075$. We correct for Milky Way extinction using the \citep[][]{Fitzpatrick1999CorrectingExtinction} extinction law with $A_V=0.1\,$mag and $R_V=3.1$ \citep[][]{Schlegel1998MapsForegrounds}.

\section{Discovery} \label{sec:discovery}

\begin{figure*}
\centering
\includegraphics[width=\textwidth]{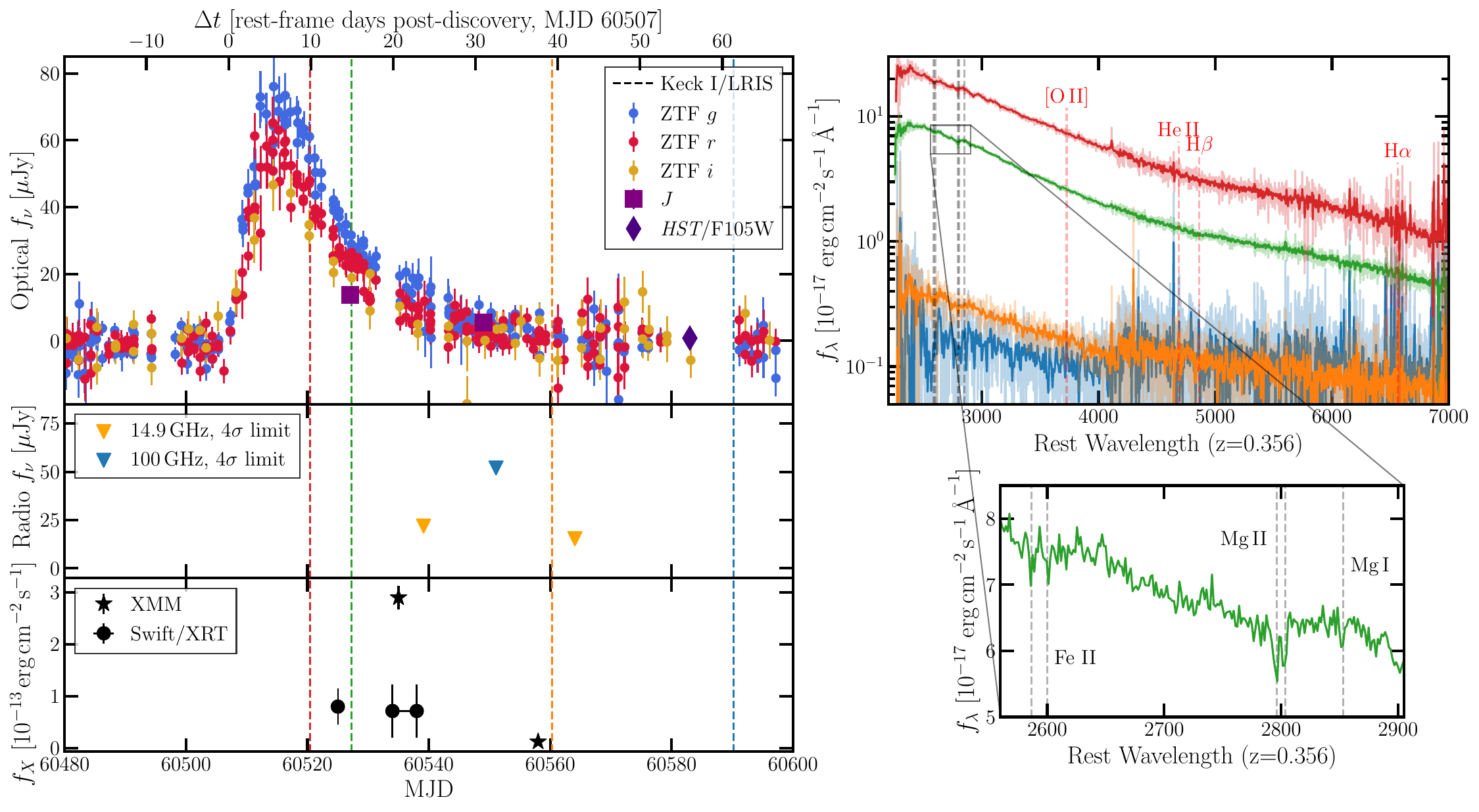}
\caption{Summary of emission from 24puz. The {\it left} panel shows the ZTF lightcurve on the {\it top} panel, the radio/millimeter upper limits in the {\it middle} panel, and the $0.3-10$\,keV Swift/XRT and XMM-Newton lightcurves in the {\it bottom} panel. The X-ray fluxes are computed as the unabsorbed flux assuming a $\Gamma = 1.77$ power-law. The two Swift/XRT observations that are joined together by a line were individually non-detections, so we show the flux measured by stacking the two observations, which is significant. Dates of Keck I/LRIS spectroscopy are shown as dashed lines. The {\it top right} panels shows the optical spectral sequence, which no significant features detected. The solid lines are smoothed by a Gaussian with width of 5\,pix. The faded lines show the unsmoothed spectra. Commonly detected transient lines are shown as dashed red lines, none are detected. The apparent line in the red, MJD 60520 spectrum near $4000\,\AA$ is a poorly subtracted sky line. The {\it bottom right} panel shows a zoom-in on the detected absorption lines, with the lines labeled in grey. }\label{fig:ztf_lc}
\end{figure*}

\begin{figure*}
\centering
\includegraphics[width=\textwidth]{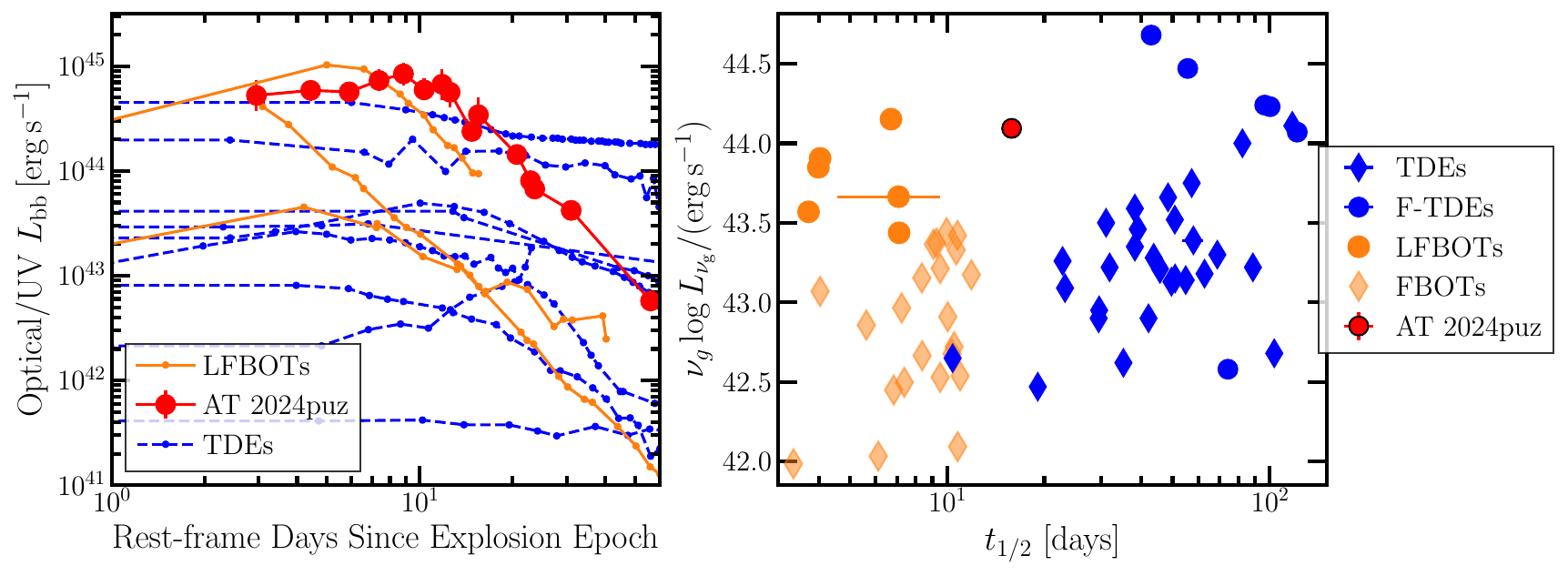}
\caption{Comparison between the optical/UV emission from 24puz and LFBOTs/TDEs from \cite{YaoTDESamp, ho_fbots, Pursiainen_wpp, chrimes_finch}. In all panels, LFBOTs are shown in orange, TDEs are shown in blue, and 24puz in red. The {\it left} panel shows the black body luminosity evolution. LFBOTs show rapid fading with a similar late-time power-law decay as 24puz, but on overall slower timescales and typically with fainter luminosities. TDEs show a range of lightcurve shapes, which generally evolve slower than that of 24puz. Some TDEs show a rapid fade like 24puz, but these are all at least one order of magnitude fainter than 24puz. These trends are highlighted in the {\it right} panel, which shows the peak $g$-band luminosity versus time above half-peak luminosity. We also include fast blue optical transients (FBOTs) as faint diamonds. TDEs are divided into featureless TDEs (F-TDEs, circles), which show featureless spectra like 24puz, and those that have transient spectral features (diamonds). 24puz is intermediate to LFBOTs and TDEs, and is notably more luminous than all events at a similar timescale and is much faster evolving than F-TDEs. 
}\label{fig:optcomp}
\end{figure*}

AT 2024puz was the first discovery from an ongoing, real-time search for hostless or highly offset, consistently blue, optical transients with ZTF (\citealp[][]{Bellm2019TheResults, Graham2019TheObjectives, Dekany2020TheSystem, Masci2019TheArchive}). ZTF conducts several public and private surveys using the Palomar 48-inch Schmidt telescope \citep{Bellm2019TheScheduler}. The public survey covers the northern sky once every 2 days in the g and r bands, while private surveys include i-band as well as high cadence observations of smaller areas.  We identify candidates in ZTF alert photometry from the \texttt{kowalski} broker \citep[][]{Walt2019SkyPortal:Platform, Coughlin2023AAstronomy}. Our cuts are identical to those for the TDE sample in \cite{YaoTDESamp} (see Section 2.2), except that we require a significant offset from the nearest PanSTARRS source (${>}10''$). The goal of this search is to identify TDE-like events that are either at higher redshift or in low mass galaxies. 

We manually scan through the resulting candidates on a ${\sim}$weekly basis using the \texttt{fritz.science} instance of the SkyPortal platform \citep[][]{Walt2019SkyPortal:Platform, Coughlin2023AAstronomy}. We identify sources that have lightcurves that are inconsistent with a Cataclysmic Variable, which form our main background. We do this by rejecting sources that have no observations during the flare rise and a perfectly linear decay (by eye) in magnitude space, with no color evolution. We obtain rapid follow-up with the Spectral Energy Distribution Machine (SEDM) on the Palomar 60-inch telescope \citep[][]{Blagorodnova2018TheClassification, Rigault2019FullyPysedm} and/or the Low Resolution Imaging Spectrometer (LRIS) on the Keck I telescope \citep[][]{lris}, with some dependence on our telescope allocations. We rule out any objects with spectra that have features that are well-modelled as a supernova or that are at $z\approx 0$. We continue to follow-up any remaining sources.

24puz was first reported on the Transient Name Survey by \cite{Sollerman2024ZTF2024-07-20} on 2024-07-20 or MJD 60511 and was the first source to pass all these cuts, a few months after we began our search. The ZTF lightcurve for this source (described in detail below) is shown in the top left panel of Figure~\ref{fig:ztf_lc}. We initiated an expansive multiwavelength follow-up for 24puz upon obtaining a Keck I/LRIS spectrum that showed a featureless spectrum with galaxy ISM absorption features at $z=0.356$. Figure~\ref{fig:optcomp} compares the optical properties of 24puz to other classes of featureless, blue transients (LFBOTs and TDEs), we will discuss this figure in detail in Section~\ref{sec:disc}, but include it here for context.

The first ZTF forced-photometry detection (${>}3\sigma$) of 24puz ocurred on MJD 60507, or 2024-07-16. We adopt this MJD as the discovery date $t_0$ throughout this work.

\section{Multiwavelength Observations and Data Reduction} \label{sec:obs}

In this section, we summarize archival and follow-up observations and our data reduction procedures. 

\subsection{Zwicky Transient Facility}

While we discovered 24puz in alert photometry from ZTF, we perform all analysis using forced photometry from the IPAC ZTF forced photometry server \citep[][]{ipacztffp} in the $gri$ bands. This server performs point spread function photometry on ZTF difference images. We processed the resulting lightcurve following recommended procedures\footnote{\url{https://irsa.ipac.caltech.edu/data/ZTF/docs/ztf_forced_photometry.pdf}} and it is shown in Figure~\ref{fig:ztf_lc}.

\subsection{Palomar 60in{.} Rainbow Camera}

We obtained optical $ugri$ photometry of 24puz with the Spectral Energy Distribution Machine rainbow imager on the Palomar 60in{.} telescope (PI: R{.} Stein; \citealp[][]{Blagorodnova2018TheClassification}). The observations are summarized in Table~\ref{tab:sedm}. We reduce the data using the automated pipeline\footnote{\url{https://sites.astro.caltech.edu/sedm/Pipeline.html}}.

\subsection{Palomar 200in{.} Wide Field Infrared Camera}

We observed 24puz with the Wide Field Infrared Camera (WIRC) on the Palomar 200 inch (P200) telescope on MJDs 60526 (2024-08-04; PI S{.} Ocker) and 60546 (2024-08-24; PI V{.} Ravi). We used the $J$-band and exposure times of 1620 and 6525 sec, respectively. We reduced both epochs using the \texttt{irImagePipe} code \citep[][]{De2020PalomarResults} using default settings, including flux calibration against sources from the Two Micron All Sky Survey (2MASS) catalog \citep[][]{Cutri20032MASSSources.}. We measured the source flux in an aperture of radius $7$\,pix{.} and subtracted the median background in an annulus of $[10,50]$ pix. The source was detected at $J$-band flux densities of $14.2\pm1.6$ and $5.9\pm 1.2\,{\rm \mu Jy}$ in each epoch. 


\subsection{The Neil Gehrels Swift Observatory} 

We obtained observations of 24puz with the Neil Gehrels Swift Observatory (Swift; \citealp[][]{Gehrels2004TheMission}) through ToOs 20912 and 20962 (PI Somalwar, object ID 16746). The observations are summarize in Table~\ref{tab:xrt} and Table~\ref{tab:uvot}. The X-Ray Telescope (XRT) was used in photon counting mode and the Ultra-violet Optical Telescope (UVOT) observations used the u, uvm2, uvw1, and uvw2 filters. 

We first describe the UVOT data reduction and then the XRT. For UVOT, we used the default data reduction provided by the observatory. We measured the source flux in all bands using an aperture centered on the position of 24puz with radius $5''$ and a background region offset from the source region with radius $20''$. We used the \texttt{uvotsource} tool in \texttt{heasoft} (v6.34) to perform the photometry. We used the \texttt{ssstype=high} flag to reject observations where the transient was located in a low sensitivity region. The measured fluxes for all good observations are detailed in Table~\ref{tab:uvot}.

We processed the XRT data using the online Swift/XRT data products tool\footnote{\url{https://www.swift.ac.uk/user_objects/}}. We used the lightcurve tool to measure the count rate in each observation, adopting the ZTF coordinates of 24puz and a $3\sigma$ detection threshold. We used default settings and binned by observation ID. We converted counts to fluxes assuming a power-law spectrum with photon-index $\Gamma=1.7$, motivated by our X-ray spectral modeling in Section~\ref{sec:xrayem}, and Milky-Way absorption with $n_H = 2.88\times10^{20}$ cm$^{-2}$ \citep[][]{BenBekhti2016HI4PI:GASS}. The resulting count rates and upper limits are shown in Table~\ref{tab:xrt}.

\subsection{The XMM-Newton Telescope}

We observed 24puz with the XMM-Newton telescope on the dates and with the exposure times listed in Table~\ref{tab:xmm}. We used the thin filters for all observations. We retrieved the PPS data for each observation from the XMM-Newton archive. For each epoch, we extracted a spectrum of 24puz following recommended procedures \footnote{\url{https://www.cosmos.esa.int/web/xmm-newton/sas-threads}}, including flagging bad time intervals using the \texttt{tabtigen} command with a rate upper limit found by manual inspection of the lightcurve. We process the data from all three EPIC cameras and perform all fitting to the data from the three detectors together. We centered the source apertures on the ZTF position of 24puz and used radii of $20''$ to avoid contamination from a nearby source. We used $30''$ background regions close to the source but uncontaminated by any nearby sources. Our first EPIC-PN exposure had the strongest contamination and so the effective exposure time quoted in Table~\ref{tab:xmm} is lower than the on-source time as observed by the telescope.

\subsection{The \textit{Nuclear Spectroscopic Telescope Array}}

The \textit{Nuclear Spectroscopic Telescope Array} (\textit{NuSTAR}) is a hard X-ray telescope, with two independent detectors known as Focal Plane Module (FPM) A and B, observing between 3--79 keV \citep[][]{nustar_Harrison2013}.  We obtained two observations approximately a day apart, details in Table~\ref{tab:nustar}. We reduced the observations with the data reduction software \textsc{NuSTARDAS v2.1.2}, with \textsc{CALDB v20240520}. For each observation, we used the \textsc{nupipeline} task to produce a cleaned event list from each detector.  From these event lists, we extract events from a $60"$ circular region around the expected source location, and also from a $60"$ circular background region located close to the source region on the detector.  We report the observed count rates from each region on each detector in Table~\ref{tab:nustar}.
As none of the detectors showed a clear detection, we use the method of \cite{Kraft1991} to estimate the $90\%$ upper confidence limit for the number of counts seen by each \textit{NuSTAR} detector.

\subsection{The Low Resolution Imaging Spectrometer}

\begin{figure}
\centering
\includegraphics[width=0.5\textwidth]{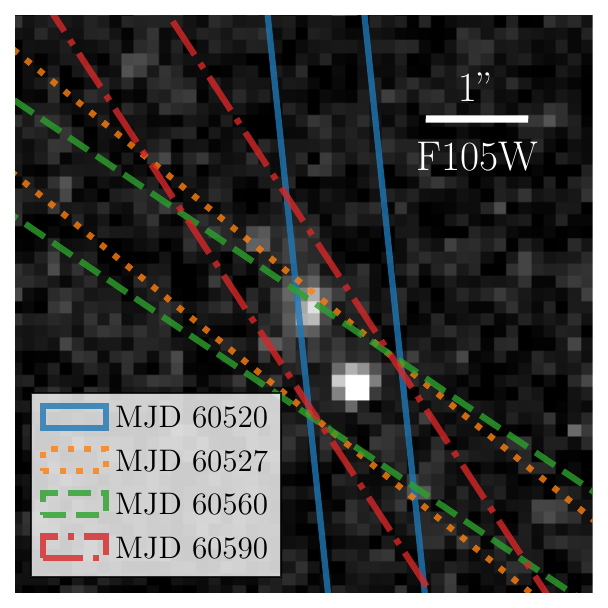}
\caption{Position angle of the Keck I/LRIS longslit for each observation, overlaid on the {\it HST}/WFC3 F105W observation. 24puz is towards the bottom right corner of the image whereas the galaxy is in the center. The MJD 60590 observation longslit was positioned to include the nearby galaxy and 24puz. }\label{fig:spec_loc}
\end{figure}

We obtained both optical spectra and photometry for 24puz with the Low Resolution Imaging Spectrometer (LRIS) on the Keck I telescope \citep[][]{lris} (PI: V{.} Ravi). The observations are summarized in Table~\ref{tab:lris}.

We centered all observations on the transient location as measured by ZTF. The slit angle was set to parallactic for all observations except MJD\,60590, for which we set the angle to $337.18^\circ$ to include the closest galaxy to 24puz that is detected in our {\it Hubble Space Telescope} ({\it HST}) observations, described later in this section. The slit positions are shown in Figure~\ref{fig:spec_loc}. We used the 400/3400 grism, the 400/8500 grating with central wavelength 7830, and the 560 dichroic. We reduced the observations using the \texttt{lpipe} data reduction pipeline \citep[][]{lpipe} with default settings.

The resulting wavelength range from ${\sim}3100-10000\,{\rm \AA}$ and the resolution $R\approx 700$. Our spectroscopic observations are summarized in Table~\ref{tab:lris}. The slit positions are overlaid on a {\it Hubble Space Telescope} image of the transient in Figure~\ref{fig:spec_loc}.


\subsection{The Lowell Discovery Telescope}

We obtained observations of 24puz with the Large Monolithic Imager (LMI) mounted on the 4.3-m Lowell Discovery Telescope (LDT; PI E{.} Hammerstein) on 2024-08-27. We reduced the LDT data using standard data reduction techniques, including bias subtraction, flat-fielding, and cosmic ray rejection. We used the Scamp code to align frames and the Swarp code to combine. We flux calibrate using the PanSTARRS DR2 catalog. The data is summarized in Table~\ref{tab:LDT}.

\subsection{The Liverpool Telescope}

We obtained imaging of 24puz in the $u, g, r$ and $i$ bands with the Liverpool Telescope (LT; PI C{.} Angus) between Aug. 5 2024 and Aug. 20 2024. All raw LT images were reduced using standard reduction techniques, and aperture photometry was performed using the \lq~Photometry Sans Frustration\rq~ pipeline \citep[{\tt{PSF}};][]{Nicholl2023ATGalaxies}, using a PSF optimized aperture and archival images from the Sloan Digital Sky Survey \citep[SDSS; ][]{2017ApJS..233...25A} for template subtraction. The observations are summarized in Table~\ref{tab:LT}. 

\subsection{The {\it Hubble Space Telescope}}

We observed 24puz for two orbits with four bands using the Wide Field Camera 3 on the {\it Hubble Space Telescope} on Sept{.} 30, 2024 (PID 17854, PI Somalwar). The data is summarized in Table~\ref{tab:hst}. We retrieved the reduced data from the \texttt{mast} archive. From some basic data quality checks, we concluded that all the default data was sufficient for our analysis, except the IR/F160W image. This image suffered from scattered light, so we followed recommended procedures to manually flag the individual reads that were most affected and then redrizzled the data using recommended parameters \footnote{see \url{https://github.com/spacetelescope/WFC3Library/tree/main/notebooks/ir_scattered_light_calwf3_corrections} for details}. 

The transient was significantly detected ($>5\sigma$) in all images, alongside many other sources. We performed aperture photometry for both the transient and all other sources in the image using the \texttt{sep} package \citep[][]{Barbary2016SEP:Library, Bertin1996SExtractor:Extraction.}. We first background subtract each image using background boxes of size $32\times32$\,pix$^2$ and a smoothing kernel of $3\times3$\,pix$^2$. We also measure the uncertainty in the images using these parameters. We detect sources on the F606W image, because of its combination of resolution and sensitivity. We verify by eye that we are not missing any IR or UV-bright sources because of this choice. We extract sources within $200''$ of the transient ($1\,$Mpc at $z=0.356$) using a $3\times3$ smoothing kernel and requiring at least $5$ pixels above $1.5\sigma$. We measure elliptical kron radii using \texttt{sep}. We then perform aperture photometry using the \texttt{photutils} package. We use elliptical apertures scaled to twice the kron radii, which should include ${\sim}90\%$ of the enclosed flux for typical galaxies. For the IR images, which have poorer spatial resolutions, we increase the aperture radii by the quadrature difference of the IR and F606W PSFs. We measure the background around each source and the local flux uncertainty using the mean and root-mean-squared error in a circular aperture from $1.8''-3''$, centered on each source. We reject any source that is not detected at a $3\sigma$ level in at least $3$ bands.

For most of the sources in the images, our goal is to measure approximate stellar masses (to be described in Section~\ref{sec:host}) and half-light radii. We measure half-light radii using the \texttt{sep} \texttt{flux\_radius} function. For the transient, we modify our photometry slightly. We adopt a circular aperture with radius twice the kron radius. We then correct the photometry in each band for the encircled energy curves provided by {\it MAST}.

\subsection{The Very Large Array} 

We observed 24puz with the Jansky Very Large Array (VLA) on MJDs 60539 (2024-08-17; PI Somalwar, Project 24B-456) and 60564 (2024-09-11; PI J{.} Somalwar, Project 24B-456). The VLA was in B configuration and we observed in the Ku band. We used J1740+5211 as a gain calibrator and 3C286 as a flux calibrator. We followed recommended procedures for high frequency observations. We manually reduced each epoch using the \texttt{casa} package \citep{casa} version 6.5.3.28 following recommended procedures. No source was detected in a 5\arcsec~box around the known transient location in either observation. The root-mean-square (RMS) uncertainties were $5.5, 3.8\,{\rm \mu Jy}$, respectively. We set $4\sigma$ upper limits of $<22, 15.2\,{\rm \mu Jy}$ respectively, where we adopt the rms measured in a 30\arcsec~box. 

\subsection{The Northern Extended Millimetre Array} 

We observed 24puz with the Northern Extended Millimetre Array (NOEMA) on MJD 60551 (2024-08-29) at 100\,GHz with proposal D24AB (PI: V{.} Ravi), using a standard continuum setup. We used MWC349 as the flux calibrator, and 1739+522 as the gain calibrator. The observations were reduced using the standard NOEMA pipeline. No source was detected in a 5\arcsec~box around the known transient location in either observation, with an RMS of $13\,{\rm \mu Jy}$. We set a $3\sigma$ upper limit of $<52,{\rm \mu Jy}$, where we adopt the rms measured in a 20\arcsec~box. 

\section{Data Analysis} \label{sec:ana}

In this section, we describe the physical properties of 24puz and its environment that can be inferred from our observations. We begin with constraints on the redshift and host galaxy of 24puz. Then, we consider the transient emission. We first discuss the ultraviolet-infrared lightcurve, followed by the limits on transient optical spectral features, the X-ray lightcurve and spectral evolution, and finally the implications of the radio limits. 

\subsection{Host galaxy and environment}\label{sec:host}

In this section, we identify possible host galaxies of 24puz. We first constrain the redshift of 24puz using the detected ISM absorption lines. Then we consider whether 24puz is associated with a galaxy (or galaxy overdensity) that is detected in imaging of the field. We finally constrain the possibility that 24puz is associated with an undetected (i.e., faint and/or compact and obscured by the transient) host galaxy. 

\subsubsection{Optical spectroscopy: ISM Absorption and redshift constraints} \label{sec:opt_abs}

\begin{figure*}
\centering
\includegraphics[width=\textwidth]{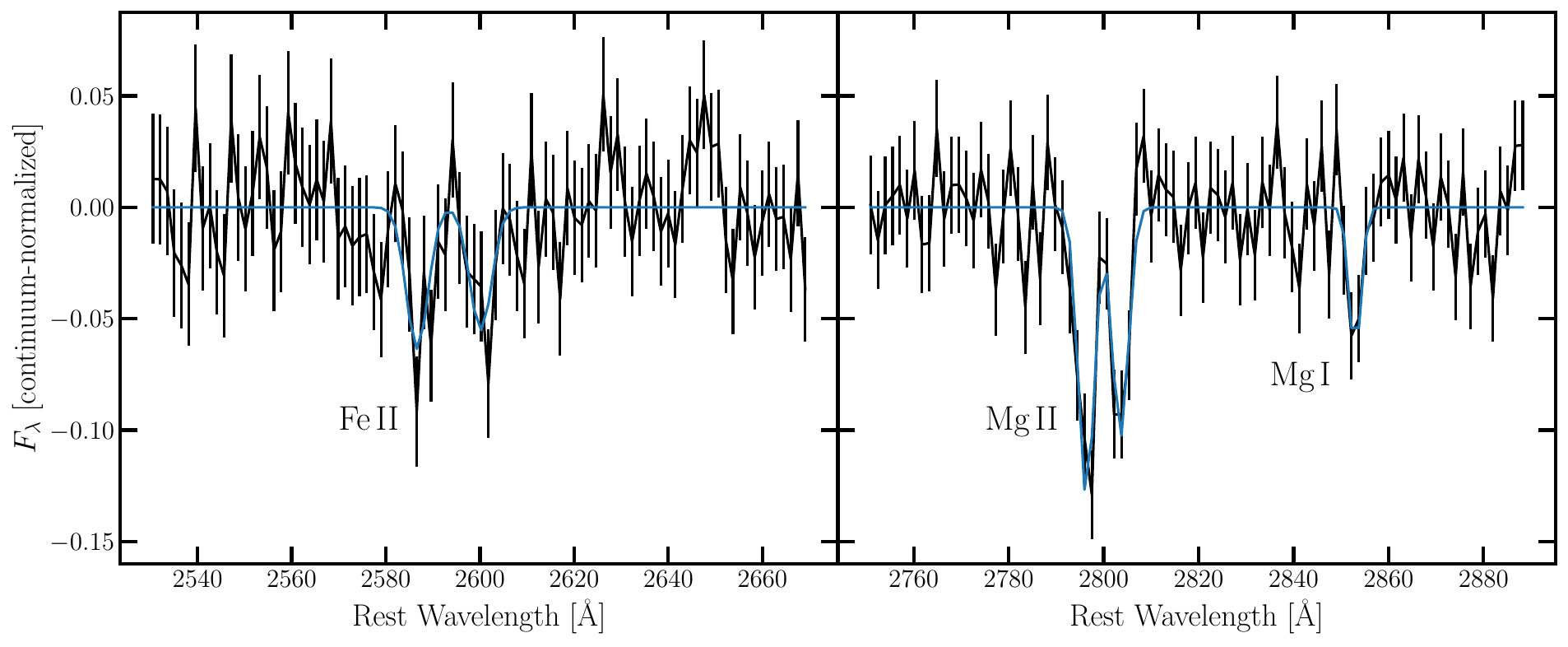}
\caption{Fits to the absorption lines ${\rm Fe\,II}\,\lambda\lambda 2586,2600$, ${\rm Mg\,II}\,\lambda\lambda2796,2803$, and ${\rm Mg\,I}\,\lambda 2852$. The spectra are fit as linear continuum components and Gaussian lines, as described in Section~\ref{sec:opt_abs}. The Gaussian widths and redshifts are fixed to the same value for all lines. The best fit redshift is $z=0.35614\pm0.00009$. The Mg\,II absorption is consistent with a strong-absorber (${\rm EW}_{2796} = 0.753 \pm 0.104 {\rm \AA} > 0.3 {\rm \AA}$), suggesting that this absorption is more likely occurring within the nearest galaxy to 24puz (G1) or a nearby group/cluster.} \label{fig:ism_abs}
\end{figure*}

\begin{deluxetable}{ccc}
\tablecaption{ Absorption line equivalent widths \label{tab:ISM_EW}}
\tablehead{Line & Equivalent Width [${\rm \AA}$] & }
\startdata
Fe\,II$\,\lambda 2586$ & $0.502 \pm 0.148$ \cr
Fe\,II$\,\lambda 2600$ & $0.438 \pm 0.14$  \cr
\hline
Mg\,II$\,\lambda 2796$ & $0.753 \pm 0.104$\cr
Mg\,II$\,\lambda 2803$ & $0.599 \pm 0.099$\cr
\hline
Mg\,I$\,\lambda 2852$ & $0.286 \pm 0.098$
\enddata
\tablecomments{The line redshifts were tied together in fitting. The line widths were tied together for doublets, but were otherwise allowed to float freely. The best-fit redshift was $z=0.35614\pm0.00009$.
}%
\end{deluxetable}

In this section, we measure the redshift and equivalent widths of the ISM absorption lines detected in the spectra of 24puz, as shown in Figure~\ref{fig:ism_abs}. We consider the ${\rm Fe\,II}\,\lambda\lambda 2586,2600$, ${\rm Mg\,II}\,\lambda\lambda2796,2803$, and the ${\rm Mg\,I}\,\lambda 2852$ lines. We do not detect any Ca absorption. We measure these lines using our spectrum from MJD 60527.3, which is the deepest spectrum we obtained that includes bright transient emission (as required to detect the absorption lines). We fit the spectrum in the regions [$2630\,{\rm \AA}$, $2670\,{\rm \AA}$] and [$2760\,{\rm \AA}$, $2880\,{\rm \AA}$], which include sufficient continuum to perform a local continuum fit around each line. We model the continuum in each region separately as first degree polynomials. We chose these continuum models as the lowest degree polynomials required such that the $\chi^2/{\rm dof}$ was consistent in each region with a p-value $>10\%$. We fit each absorption line as a Gaussian. The Gaussian amplitudes were independent. The widths were tied for lines within doublets but were otherwise free. We perform the fit using the \texttt{scipy} least-squares function and report best-fit equivalent widths and $1\sigma$ uncertainties in Table~\ref{tab:ISM_EW}. The fit is shown in Figure~\ref{fig:ism_abs}. The best-fit redshift was $z=0.35614\pm0.00009$. The lines are largely unresolved. The $\chi^2/{\rm dof}$ of the final fit was $140/167$, for a fully consistent p-value of $0.94$. 

The Mg\,II\,$\lambda 2796$ equivalent width is ${\rm EW}_{2796} = 0.753 \pm 0.104\,{\rm \AA}$, placing the absorber in the class of strong absorbers ($W_{2796} \gtrsim 0.3\,{\rm \AA}$; \citealp[][]{Steidel1992MGGalaxies}). This gas could originate from the host galaxy of 24puz or from the extended circumgalactic medium of a nearby galaxy/galaxy group. We will discuss these possibilities in the following section. 

\subsubsection{{\it HST} imaging: nearby galaxies and probability of association}

\begin{figure*}[]
\centering
\includegraphics[width=\textwidth]{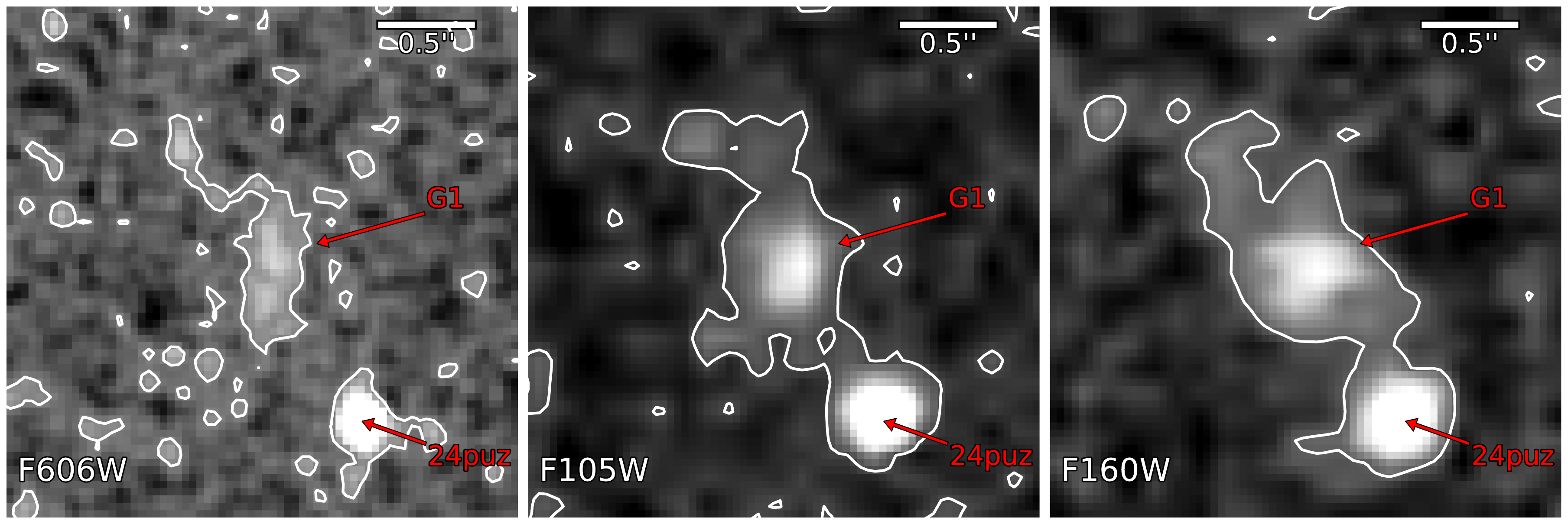}
\caption{Morphology of the nearest galaxy to 24puz, which we name G1. We show {\it HST}/WFC3 imaging of G1 in the F606W ({\it left}), F105W ({\it middle}), and F160W ({\it right}) bands. These images were taken $56$ rest-days post-discovery. We have reprojected all images to the F606W pixel scale for ease of comparison, but this means that the F105W and F160W have been resampled to a smaller pixel scale than the original images. G1 and 24puz are labeled. The white lines show contours, which highlight the morphology of G1. In all bands, there is an extended tidal tail or other irregular component towards the top left of G1. The contour levels are shown for visualization but are not intended to represent sigma levels. }\label{fig:G1_morph}
\end{figure*}

We next consider the galaxies detected in our deep {\it HST} imaging of the field of 24puz (zoom-in in Figure~\ref{fig:G1_morph}). We will show that 24puz is likely associated with a detected galaxy, and in the following sections we will identify and characterize the galaxy that is the most likely host. Henceforth, we will call the nearest galaxy to 24puz in projected distance ``G1''.

In Appendix~\ref{sec:chance_assoc}, we show that the probability of 24puz randomly lying at its location relative the G1 is $p_1=3\times10^{-4}$. We also show that the probability that all the second through tenth-nearest neighbors are closer than those observed for 24puz is small: $p_{2-10} = 1.3\times10^{-3}$. It is thus ${\gtrsim}3\sigma$ unlikely both that 24puz lies close to the observed galaxies by chance.


Given that the probability of 24puz randomly lying near both the nearest and the second-tenth nearest neighbors is low, 24puz is likely part of a galaxy group or cluster. The closest cataloged galaxy group is centered at a $5.89'$ offset and at spectroscopic redshift $z=0.354$, corresponding to a 1.8\,Mpc projected distance \citep[][]{Zou2021GalaxyDetection, Wen2024ASurveys}. This redshift is fully consistent (${<}1\sigma$) with the ISM/CGM absorption lines in the spectrum of 24puz at $z=0.35614\pm0.00009$. The reported mass and radius of this structure are $M_{500} = 8.5\times10^{13}\,M_\odot$ and $r_{500} = 0.612$\,Mpc \citep[][]{Wen2024ASurveys}. The projected offset between 24puz and this group is large: $1.8$\,Mpc corresponds to $2.9r_{500}$. 24puz is well outside the virial radius of the group (note that $r_{200} \approx (1.4-2)r_{500}$, so the offset is ${\sim}(1.5-2)r_{200}$). 
24puz may be associated with a structure that is infalling into this group, or it may be unassociated and instead part of a smaller, nearby group that is undetectable in X-ray or spectroscopic group catalogs. 

Even though 24puz may not be associated with this group, the Mg\,II absorber that is detected in our optical spectroscopy could be associated with the group at 1.8\,Mpc. If the Mg\,II absorber is associated with the group, it is possible that we are simply unlucky and 24puz is a background source at $z\gtrsim 0.356$ that happens to lie along the line-of-sight of this group. At a $1.8$\,Mpc$=2.9r_{500}$ projected offset from a low-mass cluster, the covering fraction of Mg\,II absorbers is low: ${<}1\%$ \citep[][]{Anand2022CoolAbsorbers}. It is unlikely that 24puz would happen to lie along such a line of sight, unless it is associated with a structure (e.g., a galaxy) that hosts Mg\,II absorbers. As we will discuss, 24puz is coincident with a galaxy, so it is feasible that the absorber is within this galaxy \citep[][]{mgii_ew_mass}.

Thus, regardless of the precise large-scale structure that 24puz is associated with, we can come to some reasonable conclusions. First, 24puz is unlikely to be randomly associated with the galaxies in the field. Second, the ISM/CGM absorption lines at $z=0.35614\pm0.00009$ are unlikely to be associated with an absorber in the nearby group given the large projected offset, and instead are probably associated with a galaxy or galaxy group closer to 24puz. We can then reasonably assume that 24puz is located within a more nearby galaxy or group, and is thus also at $z=0.35614\pm0.00009$. We adopt the assumption that 24puz is at this redshift for the rest of this work.



\subsubsection{Host Galaxy Candidates}

From the previous section, we have concluded that 24puz is likely hosted by a large-scale structure at $z=0.35614\pm0.00009$. Now, we narrow down the association: 24puz is either an object that (1) is gravitationally bound to a galaxy, like a star or stationary BH; or, (2) ejected, like a recoiling BH or NS. In the former case, we expect 24puz to be within, roughly, the virial radius of its host. In the latter, 24puz may be offset from its host galaxy.

We first consider the case where 24puz is bound to a galaxy. We have named the closest galaxy (in projected distance) to 24puz ``G1''.  24puz is at a ${\sim}5\,$kpc offset from G1, assuming $z=0.356$. We model our {\it HST}/F606W image of G1 with a Sersic profile, in addition to a Moffat profile to represent 24puz. 24puz is located at ${\sim}3$ projected half-light radii from G1.

In the previous section, we showed that the random association probability with G1 is low ($p_1 = 3\times 10^{-4}$), assuming a uniform probability of 24puz occuring at any point in the field. The galaxy with the second-lowest chance association probability has $p_2=0.02$, a factor ${\sim}70\times$ higher. If 24puz has a higher probability of occuring within the virial radius of the other galaxies in the field, the probability of chance association with G1 would {\it decrease}. Thus, even though there are many galaxies within a virial radius of 24puz, G1 is the most likely host.

We next consider the case that 24puz is produced by an astrophysical object that has received a kick velocity after a dynamical event (a merger, binary/triple interactions; \citealp[e.g.][]{Peres1962ClassicalRecoil, Bekenstein1973Gravitational-RadiationHoles}). In this case, the offset distribution depends on the type of astrophysical object and the process that caused the kick. 
If 24puz is kicked but it still associated with G1, we require a low velocity kick ${\lesssim}100\,{\rm km\,s}^{-1}$: the observed offset is 5\,kpc, so for a mean kick velocity $v_k$, we have a mean delay time ${\sim}$60\,Myr$\times \frac{v_k}{100\,{\rm km\,s}^{-1}}$. Note that G1 is a dwarf galaxy with low escape velocity $\lesssim 150\,$km/s.
If the kick velocities are $v_k \gg 100\,{\rm km\,s}^{-1}$ or the delay times ${\sim}$giga-years, then 24puz cannot be associated with G1. While we cannot exclude this case, the problem of the low probability of chance association with G1 remains. Thus, we still prefer a physical association with G1, and thus small kick velocity and delay time. 

Based on these arguments, we will cautiously associate 24puz with G1 for the rest of this work.

\subsubsection{Host Galaxy Physical Properties}

In this section, we analyze the photometric properties of G1. For completeness, we also constrain the presence of a stellar overdensity (e.g., a stellar cluster) at the location of 24puz, but the luminous emission from 24puz precludes any strong photometric constraints on the overdensity. 

We first consider the morphology of G1. In Figure~\ref{fig:G1_morph}, we show zoom-ins of G1 in the F606W, F105W, and F160W bands. The F606W image is smoothed with a Gaussian kernel of width 0.9\,pix for visualization. All images have been reprojected onto the frame of the F606W image. Contours are overlaid on each image to guide the eye. While the low signal-to-noise of the G1 detection combined with the insufficient resolution in the redder bands preclude quantitative modelling of the structure of G1, the images suggest a faint source detected near G1, which may be connected to G1 via a low surface brightness tail. We cannot confirm this, however, due to the low signal-to-noise of the image: we require deep, space-based follow-up. G1 may be in an interacting pair, have tidal tails, or simply be near a background galaxy. We favor one of the former two scenarios, particularly given the evidence that 24puz is in a galaxy group. The chance association probability of a background galaxy with G1 is even lower than that of 24puz, given the proximity of the candidate background galaxy. 

Next, we consider the stellar mass and age of G1. We model the {\it HST} photometry of G1, including an upper limit from the F336W band, as a simple stellar population. The lack of wavelength coverage and high signal-to-noise observations prevents more detailed modeling. We use the \texttt{prospector} code \citep[][]{Johnson2021StellarProspector, Leja2017DerivingUniverse, Conroy2010FSPS:Synthesis}. We set a normal redshift prior at $z=0.35614\pm0.00009$. We do not include host galaxy extinction, given the lack of photometric constraints on G1 and the low extinction level implied by the UV/optical SED of 24puz, but we do include Milky Way extinction with $A_V = 0.1\,$mag and $E(B-V)=0.03\,$mag \citep[][]{Schlegel1998MapsForegrounds}. We allow the redshift, galaxy age, galaxy metallicity, and galaxy mass to float. We require that the galaxy age is smaller than the age of the universe at each redshift. We fit the data using the \texttt{dynesty} code using the random walk sampling, $400$ initial live points and $200$ live points per batch \citep[][]{Speagle2020DYNESTY:Evidences, Skilling2004NestedSampling, Skilling2006NestedComputation}.

\begin{figure}[h!]
\centering
\includegraphics[width=0.5\textwidth]{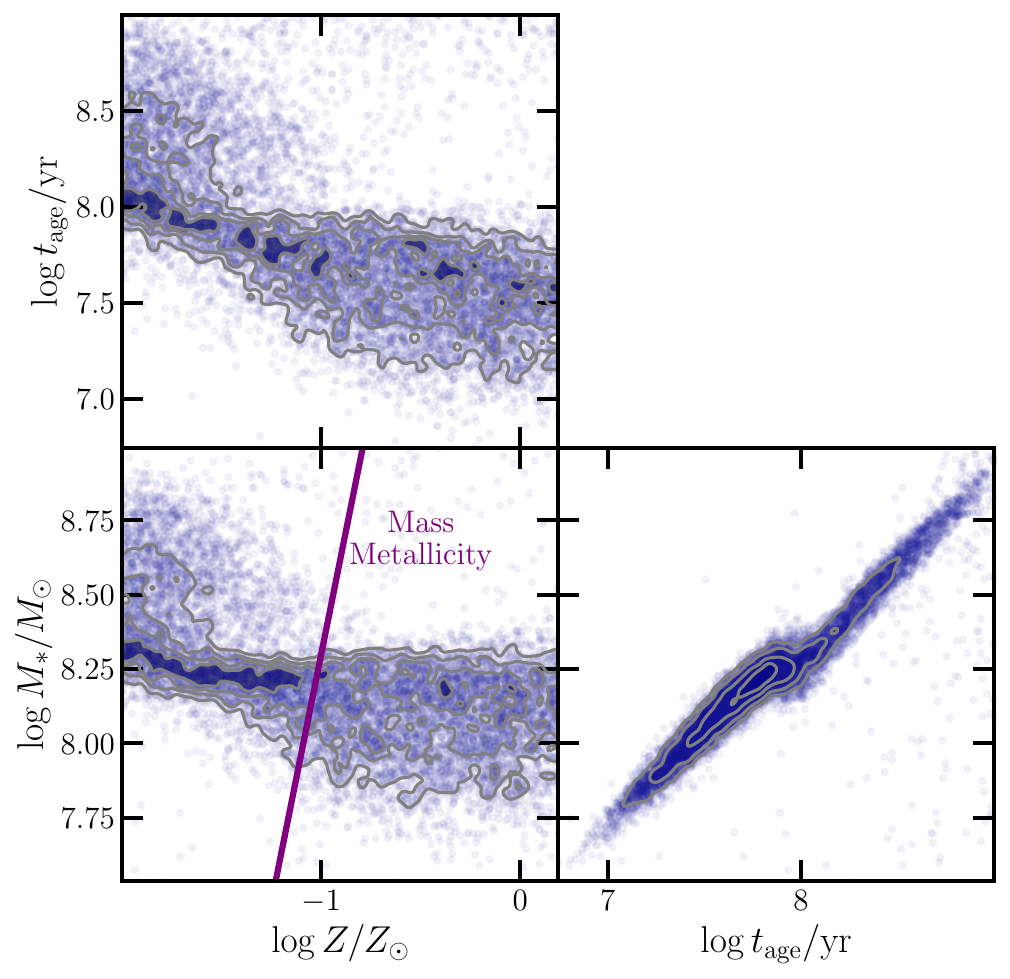}
\caption{Constraints on the stellar population of G1. We assume a simple stellar population, due to a lack of constraints on the galaxy emission. The corner plot shows the distribution of possible stellar ages ($t_{\rm age}$), metallicities $Z$, and stellar masses $M_*$ with grey contours overlaid. The mass-metallicity relation from \cite{Kirby2013THE} is shown in dark purple. We find that G1 is a dwarf galaxy with $M_* \lesssim 10^{8.75}\,M_\odot$. If it lies on the mass-metallicity relation, then the mass is $10^{7.75}\lesssim M_*/M_\odot \lesssim 10^{8.25}$ and the age $t_{\rm age}/{\rm Myr} \sim 100$\,Myr.}\label{fig:G1_ssp}
\end{figure}

The results are shown in Figure~\ref{fig:G1_ssp}. There are strong degeneracies between the stellar mass, age, and metallicity, but we can draw some conclusions. First, G1 is a dwarf galaxy, with a stellar mass $M_* \lesssim 10^{8.75}\,M_\odot$. If G1 lies on the mass-metallicity relation from \cite{Kirby2013THE}, then the stellar mass is $10^{7.75}\lesssim M_*/M_\odot \lesssim 10^{8.25}$. In this case, the stellar population age is $10 \lesssim t_{\rm age}/{\rm Myr} \lesssim 100$\,Myr, although the upper end of this range is preferred.

We briefly consider a stellar population that is hidden beneath the transient emission from 24puz. Because 24puz was still luminous during our {\it HST} observations, our constraints are weak. As we will discuss in Section~\ref{sec:bb_optuv}, however, 24puz shows an evolving red excess. A component, however, could also be produced by a stellar population. We model this by assuming that all the emission in the F105W and F160W bands is produced by stars, with no transient component. This will give us a rough estimate on the maximum mass expected. We use the F336W and F606W constraints on the transient emission as upper limits. We find a 99\% mass upper limit $M_* < 10^{8.9}\,M_\odot$; i.e., a compact dwarf galaxy or stellar cluster could be present. Deep, space-based follow-up once the transient emission has faded will be critical for constraining the stellar population at the location of 24puz.

We conclude that G1 is a low mass dwarf galaxy with a moderately young stellar population. Current constraints on any stars hidden underneath the transient emission allow for a compact dwarf galaxy or stellar cluster, with a mass upper limit $M_* < 10^{8.9}\,M_\odot$.

\subsubsection{Star formation rate constraints} \label{sec:SF}

\begin{deluxetable}{cccccc}
\tablecaption{ Emission line fluxes \label{tab:sfr}}
\tablehead{Date & $\Delta t$ & $L_{\rm H \alpha}$ & ${\rm SFR}_{\rm H \alpha}$ & $L_{\rm [O\,II]}$ & ${\rm SFR}_{\rm [O\,II]}$ 
 }
\startdata
2024-07-29 & 60520.4 & $<3.04$ & $<0.09$ & $<2.62$ & $<0.17$ \cr
2024-08-05 & 60527.3 & $<1.44$ & $<0.04$ & $<1.0$ & $<0.07$ \cr
2024-09-07 & 60560.3 & $<0.65$ & $<0.02$ & $<0.19$ & $<0.01$ \cr
2024-10-07 & 60590.2 & $<1.56$ & $<0.05$ & $0.15^{+0.05}_{-0.05}$ & $0.01^{+0.003}_{-0.003}$ \cr
\enddata
\tablecomments{Line emission and star formation rate constraints. All luminosities are in units $10^{40}\,{\rm erg\,s}^{-1}$ and star formation rates in units ${\rm M_\odot\,yr}^{-1}$. $\Delta t$ gives rest-frame days post-discovery.}%
\end{deluxetable}

\begin{figure*}[!ht]
\centering
\includegraphics[width=\textwidth]{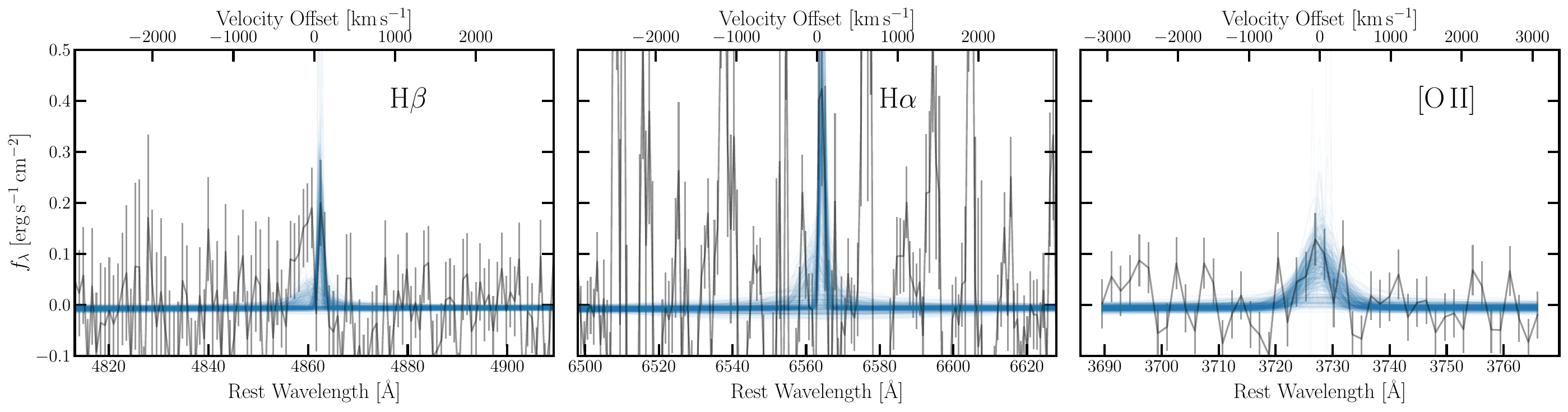}
\caption{Emission line fits to the optical spectrum of 24puz from MJD 60590.2, or $61.4$ rest-days post-discovery. Fits to H${\rm \beta}$, H${\rm \alpha}$ and the $[{\rm O\,II}]\,\lambda\lambda3726,3729$ doublet are shown in the {\it left}, {\it middle}, and {\it right} panels respectively. The data is shown in black. The amplitudes of the Balmer lines are tied to the expected ratio for star formation and the ratio of the [O\,II] doublet amplitudes are likewise tied. The apparent line at the location of H${\rm \beta}$ is a sky subtraction artifact - re-reducing the spectrum with different sky subtraction algorithms removes this feature. The [O\,II] line, on the other hand, is robust to sky subtraction. Blue lines show samples from our Gaussian \texttt{emcee} fits.}\label{fig:lines}
\end{figure*}

\begin{figure*}[!ht]
\centering
\includegraphics[width=0.53\textwidth]{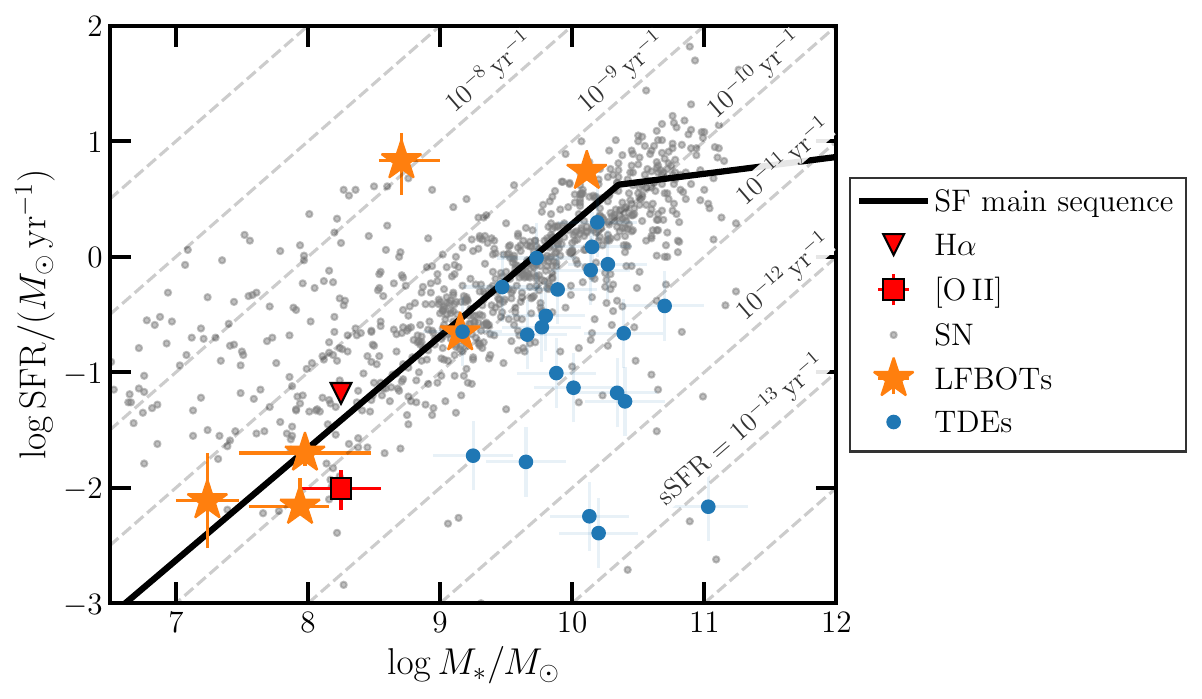}
\includegraphics[width=0.45\textwidth]{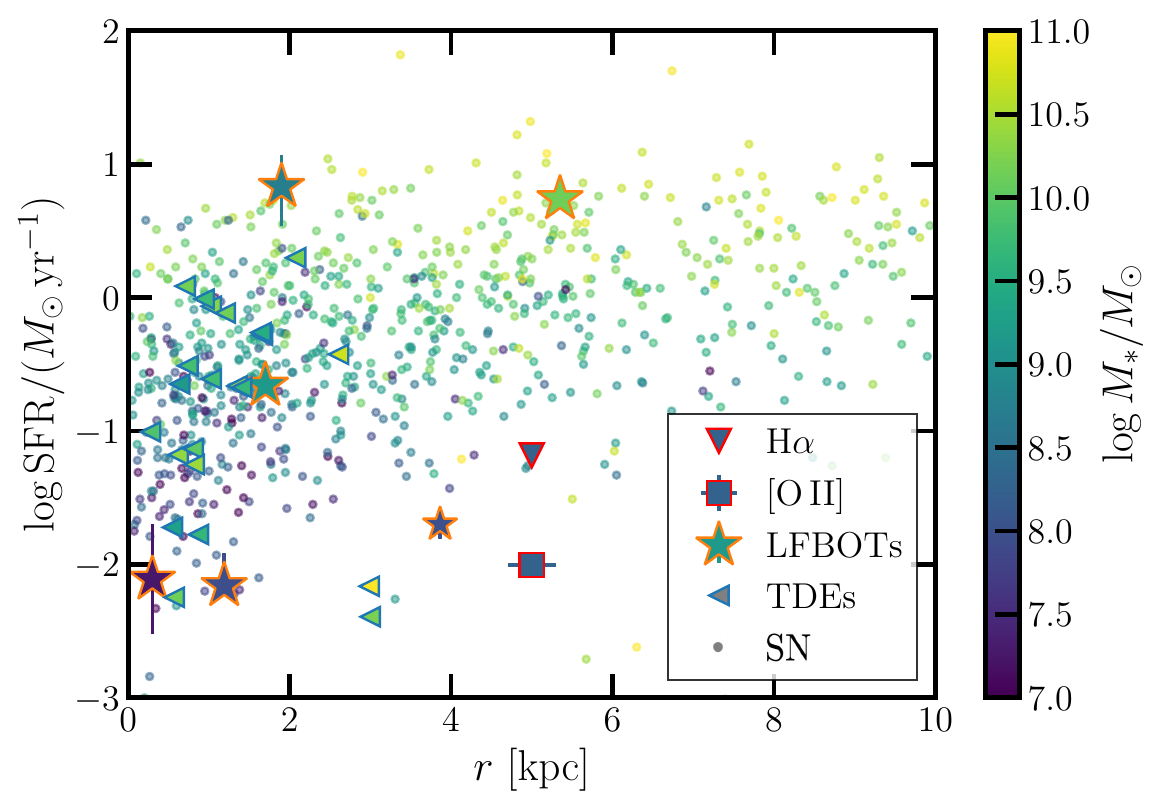}
\caption{{\it Left:} Star-formation vs stellar mass of G1 relative to other galaxy populations. Constraints from the optical spectroscopy of G1 are shown in red. LFBOT hosts from \cite{perley_18cow, perley_xnd, coppejans_css161010, chrimes_fhnlate, yao_mrf, ho_koala} are shown in orange and TDE hosts from \cite{YaoTDESamp} in blue. The core-collapse supernova host galaxy sample from \cite{Schulze2021TheSupernovae} is shown in grey. Lines of constant specific star formation rate (sSFR) are shown in grey dashed. 24puz lies below the star forming main sequence. Its location is consistent with LFBOT host galaxies, but is at a low sSFR and stellar mass relative to core-collapse supernova hosts and a low stellar mass relative to TDE hosts. {\it Right:} Star-formation vs physical offset from host galaxy of 24puz relative to other populations. The format is the same as in the {\it left} panel, except that we have colored the points by their stellar mass but left the marker outline colors the same as in the {\it left} panel. For TDEs, we assume a host galaxy offset $<0\farcs6$. The LFBOT AT\,2020xnd does not have a reported offset so we assume that it is ${\lesssim}1\arcsec$, based on a by-eye approximation from images in \cite{perley_xnd}. 24puz is at a larger offset than expected from its host galaxy mass, if it is associated with a star-forming region. }\label{fig:Ms_SFR}
\end{figure*}

We constrain the star formation rate at the location of 24puz by constraining the H${\rm \alpha}$, H${\rm \beta}$, and $[{\rm O\,II}]\,\lambda\lambda3726,3729$ luminosities from our LRIS spectra. Note that the slits for each spectrum were positioned differently. All the spectra were centered on 24puz and contained fractions of G1. The MJD 60590 (61.4 rest-days) spectrum was positioned to fully cover both 24puz and G1. We will consider each spectrum separately, given the different fractions of G1 included in the slit.


We fit each spectrum after continuum subtraction. We model the continuum by convolving each spectrum with a Gaussian kernel of width 50 pixels and subtracting this from the spectrum. We fit H${\rm \alpha}$ and H${\rm \beta}$ simultaneously because they are expected to have correlated luminosities: we fit for the H${\rm \alpha}$ luminosity, denoted $L_\alpha$ and fix the H${\rm \beta}$ luminosity $L_{\beta} = L_{\alpha}/2.86$, where we have adopted the theoretical Balmer decrement commonly assumed for star-formation (computed assuming photoionized gas at temperature $10^4\,$K; \citealp[][]{Osterbrock2006AstrophysicsNuclei}). This Balmer decrement assumes no host galaxy extinction. We separately fit a Gaussian at the locations of $[{\rm O\,II}]\,\lambda\lambda3726,3729$ and we tie the doublet ratio to be $L_{3726}/L_{3729}=0.35$ \citep[][]{Osterbrock2006AstrophysicsNuclei}. We allow the widths of the lines to vary between $\sigma_v \in [90,600]\,{\rm km\,s}^{-1}$. We assume a redshift $z=0.356$ but allow the line centroids to vary by $\Delta_v \in [-150,150]\,{\rm km\,s}^{-1}$. These velocities are assumed to be the same for the Balmer lines but are allowed to be different for $[{\rm O\,II}]\,\lambda\lambda3726,3729$. We adopt a linear model to absorb any residual continuum. We fit regions of each spectra corresponding to $10^4\,{\rm km\,s}^{-1}$ around each relevant line. We use the \texttt{emcee} sampler with default settings to perform the fit independently for each spectrum \citep[][]{Foreman-Mackey2013EmceeHammer}. We used $200$ walkers and $7000$ burn-in steps, followed by an additional $5000$ steps. We thinned the resulting chains by a factor of thirteen. 

The fit results are summarized in Table~\ref{tab:sfr}. The only $3\sigma$ line detection was $[{\rm O\,II}]\,\lambda\lambda3726,3729$ in the MJD 60590 (61.4 rest-days) spectrum. Assuming that the [O\,II] luminosity correlates with star formation rate as SFR$=6.58 \times 10^{-42} {\rm L([O\,II])/(erg\,s}^{-1})$ \citep[][]{Kewley:2004}, we find that this detection corresponds to a star formation rate $ {\rm SFR} = 0.01^{+0.003}_{-0.003}\,{\rm M_\odot\,yr}^{-1}$, or a specific star formation rate $ {\rm sSFR} \approx 10^{-10}\,{\rm yr}^{-1}$. This detection is consistent with the Balmer upper limit and the upper limits from every other spectrum. Extrapolating to low stellar masses from the star-forming main sequence at $z=0.356$ measured by \cite{leja_sfms}, we find that this emission is below the star forming main sequence (see the right panel of Fig~\ref{fig:Ms_SFR}), although it is within ${\sim}2\sigma$ of the measured spread (extrapolated by eye from Fig. 3 of \cite{leja_sfms}). 

Note that the significance of the $[{\rm O\,II}]\,\lambda\lambda3726,3729$ detection is somewhat sensitive to the prior on the line width. We tested increasing the prior to a maximum of $3000\,$km\,s$^{-1}$ (which is unphysically broad), the significant decreases to ${\sim}2.5\sigma$, so we consider this line a marginal detection. The conclusion that the star formation late is below the star-forming main sequence is still robust.

We conclude that G1 is located below the star forming main sequence, and there is no evidence for strong star formation at the location of 24puz.

\subsection{Broadband UVOIR Lightcurve}\label{sec:bb_optuv}

We modeled the broadband transient emission from ultraviolet through infrared (UVOIR) wavelengths as evolving blackbodies. We find a statistically consistent fit at all epochs without including extinction within the host galaxy. We use the \texttt{dynesty} nested sampler with default settings \citep[][]{Speagle2020DYNESTY:Evidences}. We adopt uninformative Heaviside priors on the temperature $T_{bb}$ and radius $R_{bb}$: $\log T_{bb}/{\rm K} \in [3, 5]$ and $\log R_{bb}/{\rm cm} \in [12, 17]$. We report the best-fit parameters and computed luminosities $L_{bb}$ and absolute $g$-band magnitudes $M_g$ in Table~\ref{tab:OIR_bb}.

The final two epochs of observations showed an infrared excess above a single blackbody fit, so we require a second component. We consider three models: (1) a second blackbody (e.g., warm gas), (2) an absorbed, dusty blackbody, where the absorption is assumed to be caused by graphite grains, and (3) a power-law, defined as $f_\nu \propto \nu^{-\Gamma}$. As we will discuss, these choices are motivated by similar excesses in observations of luminous fast blue optical transients and tidal disruption events \citep[][]{perley_18cow, margutti_18cow, chen_cow_early, metzger_dust}. For the blackbody models, we report infrared luminosities integrated across all wavelengths. For the power-law model, we integrate from $2000-10^7\,{\rm \AA}$ to compute the luminosity, to match previous work \citep[][]{chen_cow_early}. We will discuss the red excess and these models in more detail in Section~\ref{sec:redex}.

\begin{figure}[!ht]
\centering
\includegraphics[width=0.45\textwidth]{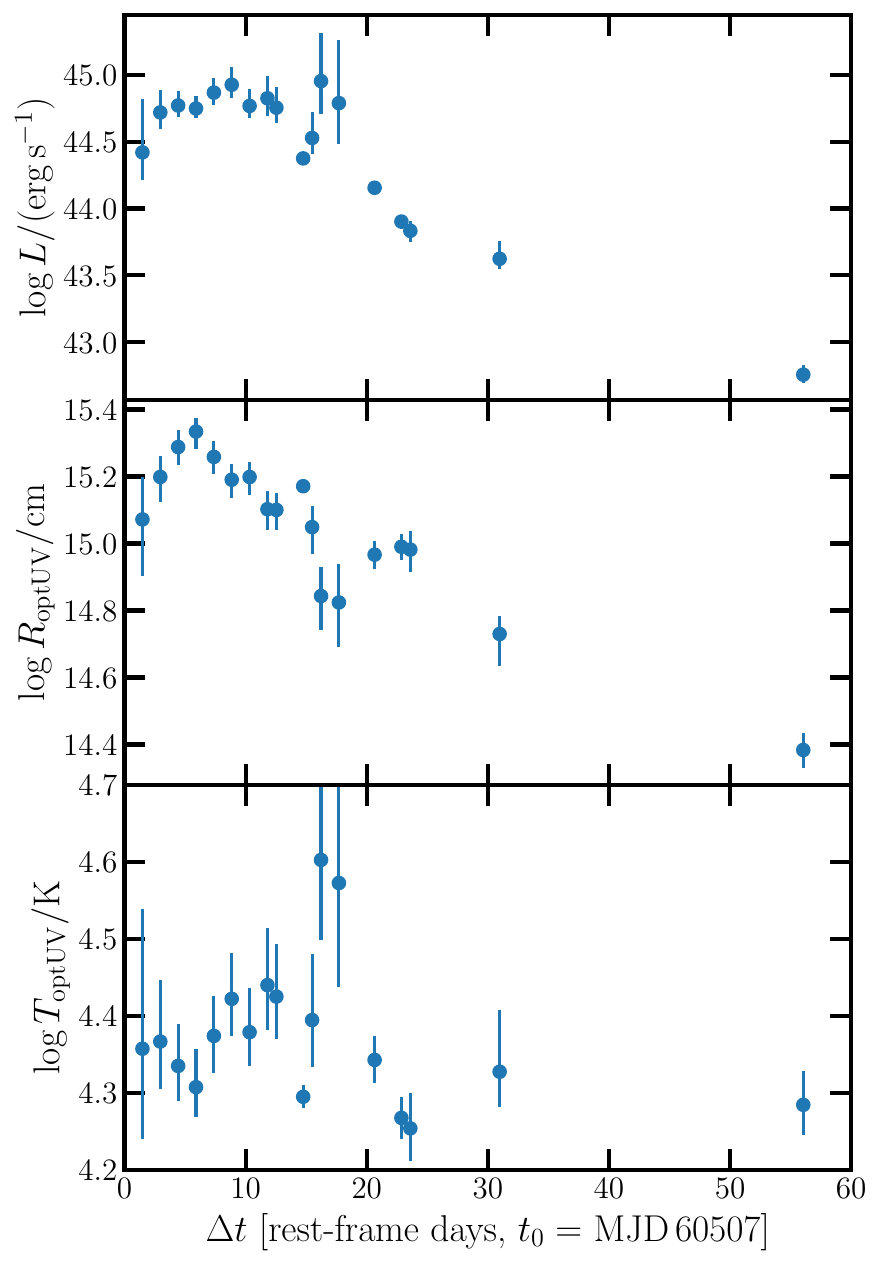}
\caption{The evolution of the best-fit blackbody parameters to the optical/UV photometry of 24puz. The blackbody luminosity, radius, and temperature are shown in the {\it top}, {\it middle}, and {\it bottom} panels, respectively. The luminosity rapidly rises over a few days, then slowly rises/plateaus before rapidly decaying. The radius expands at a velocity of $0.1c$ or a powerlaw ${\sim}t^{0.4}$ and then decays as $t^{-1.3}$ after a ${\sim}$week. The temperature is largely constant and the weighted-mean value is $\log T_{\rm bb}/{\rm K} = 4.32 \pm 0.14$.}\label{fig:optuv_evol}
\end{figure}

All fits are shown in Figure~\ref{fig:uvopt_fits} and best-fit parameters in Table~\ref{tab:OIR_bb}. The luminosity, radius, and temperature evolution are shown in Figure~\ref{fig:optuv_evol}. The luminosity rises over a ${\sim}$day timescale and then approximately plateaus or slowly rises, with a mean value of $\log L_{\rm peak, OptUV} = 10^{44.79 \pm 0.04}\,{\rm erg\,s}^{-1}$. After a ${\sim}12$ days (rest-frame), the luminosity drops as $t^{-3}$, although note that this slope is sensitive to the best-fit luminosity from our {\it HST} observations, which showed a significant red excess for which the appropriate model is uncertain, as we will discuss. The radius is consistent with expanding at $v = (0.082\pm0.02)c$ to a peak of $\log R_{\rm OptUV}/{\rm cm} = 15.33 \pm 0.04$ at 5.9 rest-days post-discovery. 
Alternatively, the radius expansion may be a power-law $\log R_{\rm OptUV}/{\rm cm} \propto t^{0.4}$. The temperature shows slight evolution, but is relatively constant at $\log T_{\rm OptUV}/{\rm K} = 4.32 \pm 0.14$. The error is driven by the standard deviation of the measured temperatures around the mean rather than measured error bars, implying that the temperature does evolve slightly. 

\subsection{Transient spectral features}

\begin{deluxetable}{cccc}
\tablecaption{Intermediate width Balmer line constraints \label{tab:translines}}
\tablehead{Date & MJD & $L_{\rm H \alpha}$ & $L_{\rm H \beta}$ \cr
 }
\startdata
2024-07-29 & 60520.4 & $<4.28$ & $<4.53$ \cr
2024-08-05 & 60527.3 & $<2.26$ & $<3.64$ \cr
2024-09-07 & 60560.3 & $<1.3$ & $<0.12$ \cr
2024-10-07 & 60590.2 & $<2.85$ & $<2.34$
\enddata
\tablecomments{All luminosities are in units $10^{40}\,{\rm erg\,s}^{-1}$. }%
\end{deluxetable}

We constrain narrow (${\lesssim} 3000$\,km\,s$^{-1}$) transient features. We do not consider broad spectral features other than to note that none are apparent by eye in any of our optical spectral (top right panel of Figure~\ref{fig:ztf_lc}). We focus on narrow, transient Balmer H${\alpha}$ and H${\beta}$ lines, as these are sometimes observed at late-times from optical transients, as we will discuss later. These constraints are identical to those from our star-formation constraints on these lines in Section~\ref{sec:SF}, but we allow the line widths to range from $\sigma_v \in [0,0.01c]$. We also fit H${\alpha}$ and H${\beta}$ independently, as transient spectral features need not be photoionized and so the Balmer decrement can vary from the nominal value ${\sim}3$. The resulting luminosity constraints are shown in Table~\ref{tab:translines}. No significant emission is detected.

\subsection{X-ray lightcurve and spectrum} \label{sec:xray}

\begin{figure*}
\centering
\includegraphics[width=0.45\textwidth]{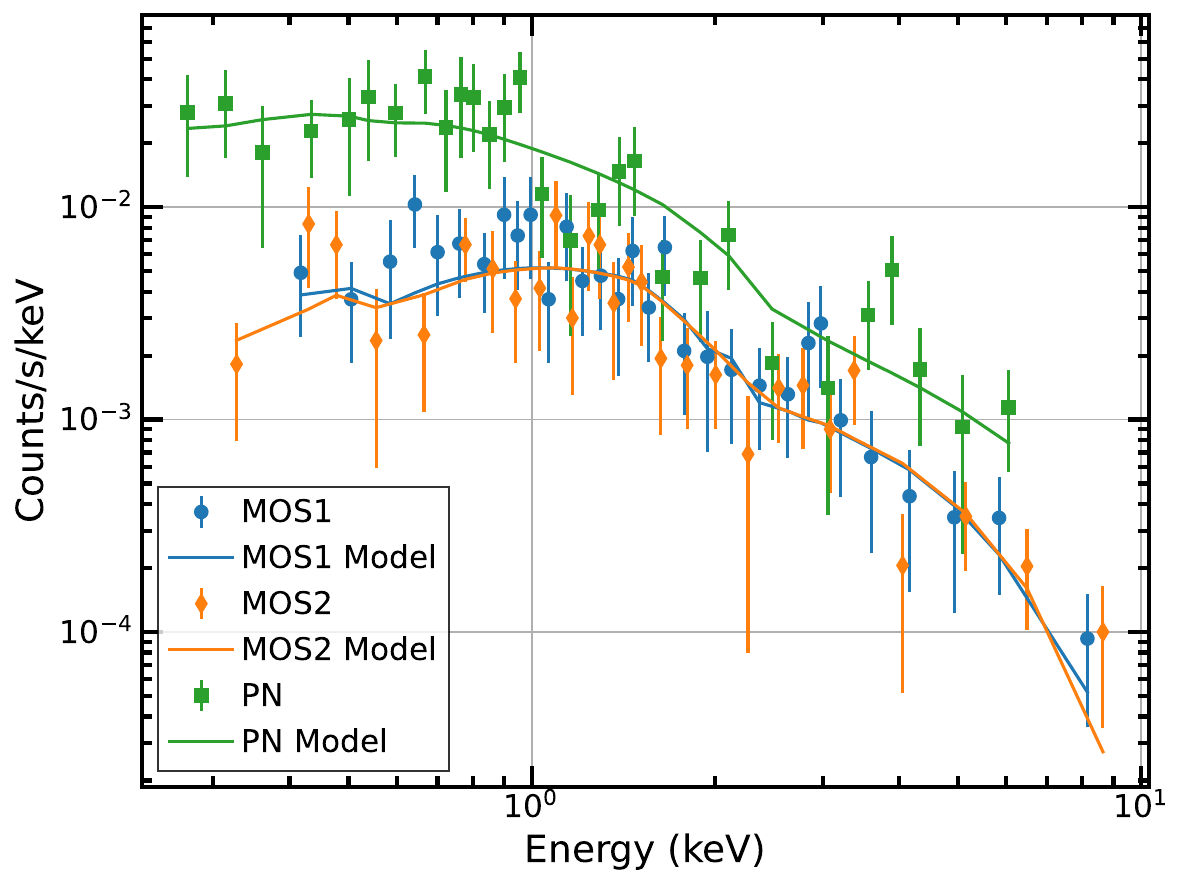}
\includegraphics[width=0.45\textwidth]{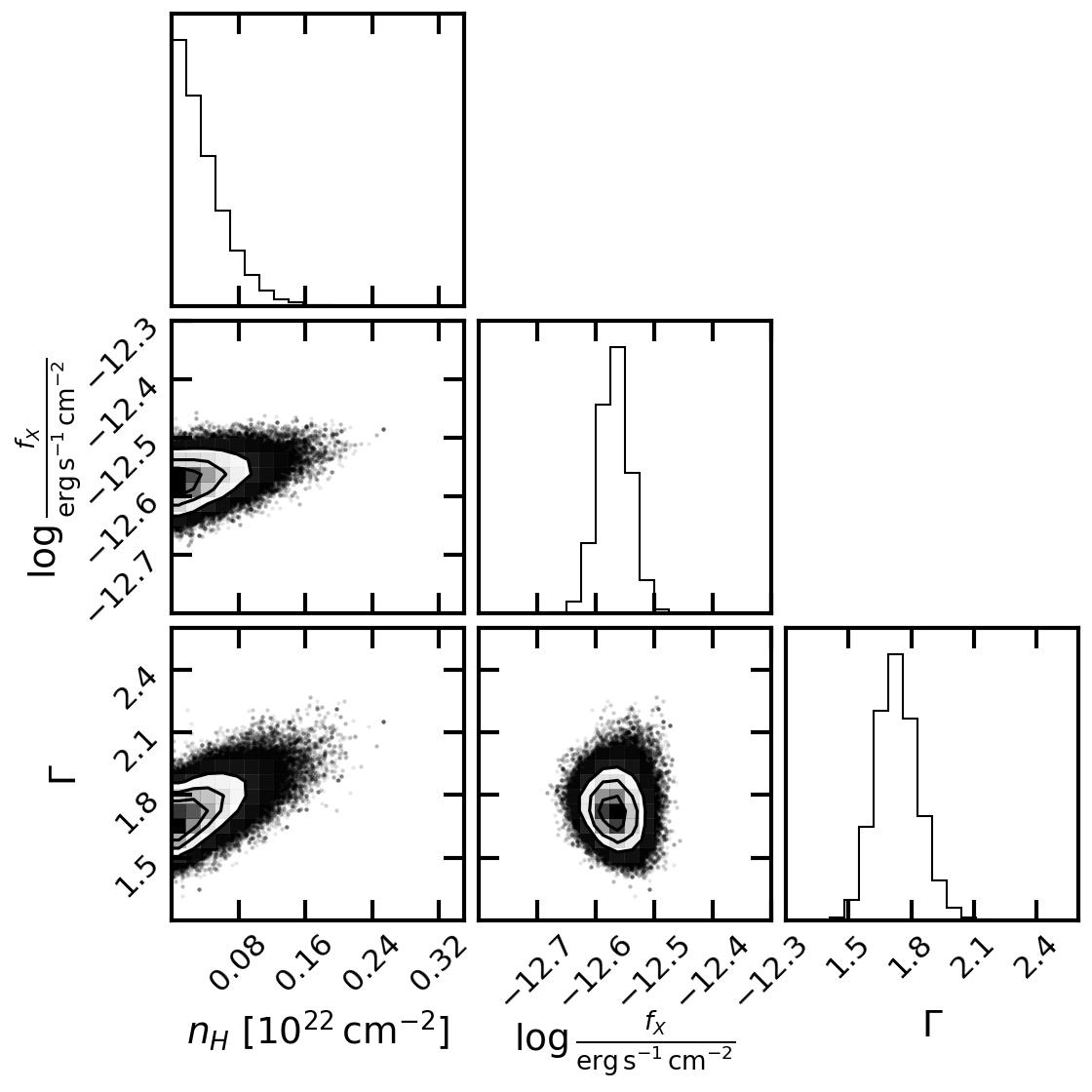}
\caption{{\it Left:} the MJD 60536 (21.4 rest-days) {\it XMM-Newton} spectrum of 24puz (blue scatter) with the  best-fit power-law model overlaid. {\it Right:} corner plot showing the best-fit power-law parameters. The host galaxy absorption is negligible ($n_H \lesssim 3\times10^{21}$\,cm$^{-2}$). The power-law index is $\Gamma = 1.73^{+0.10}_{-0.09}$. 
}\label{fig:xray_spec}
\end{figure*}

We constrained the X-ray evolution of 24puz using our Swift/XRT, XMM-Newton, and {\it NuSTAR} observations. We adopt a Milky-Way Hydrogen density $n_H = 2.88\times10^{20}\,{\rm cm}^{-2}$ and report integrated fluxes in the energy band $0.3-10$\,keV everywhere \citep[][]{BenBekhti2016HI4PI:GASS}. We first constrained the spectral shape using our first XMM-Newton/epic-PN observation (MJD 60536, 21.4 rest-days). We modeled the spectrum using the \texttt{xspec} tool with Wilm abundances \citep[][]{wilm}, Vern cross sections \citep[][]{vern}, and W statistics \citep[][]{cash}. The spectrum was best modeled as a power-law (\texttt{tbabs*zashift*tbabs*cflux*powerlaw}, total W statistic 67.87 for 88 degrees of freedom) rather than a blackbody (\texttt{tbabs*zashift*cflux*bbody}, total W statistic 146.92 for 88 degrees of freedom), so we adopt the power-law model for all epochs. We run a length $300000$ Monte Carlo Markov Chain using \texttt{xspec} and the Metropolis-Hastings algorithm with a temperature of $50$ to fully sample the parameter space. The best-fit power-law and a corner plot showing the best-fit parameter range are shown in the top panels of Figure~\ref{fig:xray_spec}.

The best-fit power-law index is $\Gamma = 1.73^{+0.10}_{-0.09}$. The spectral modeling prefers no intrinsic absorption, with a $3\sigma$ upper limit $n_H \lesssim 2\times10^{21}$\,cm$^{-2}$. 

We also tested a Bremsstrahlung model and note that it is consistent with our spectrum (\texttt{tbabs*apec}, cash statistic 77.63 for 88 degrees-of-freedom), but not preferred over the power-law model. The best-fit temperature is $7.0\pm1.5$ keV. This temperature is low for typical interactions, which produce a forward shock at ${\sim}10^9\,$K, or $100$ keV \citep[][]{Fransson1996Circumstellar1993J}. This temperature may be consistent with emission from the reverse shock, but the reverse shock is expected to be highly absorbed due to rapid (sub-day) cooling in the post-shock region \citep[see Table 1 of][]{Fransson1996Circumstellar1993J}, and so will not dominate the emission. The best-fit normalization corresponds to a volume emission measure $\int n_e n_I {\rm d}V = (3.29 \pm 0.17)\times 10^{66}\,$cm$^{-3}$. Assuming a spherical emitting region with radius ${\sim}10^{15}$\,cm, this corresponds to a high average density ${\sim}10^{10}$\,cm$^{-3}$. This analysis will become relevant when we discuss possible shock origins of the X-ray emission in Section~\ref{sec:xrayem}.

Our second epoch of {\it XMM-Newton} observations ($\Delta t = 37.6$ rest-days) are tentatively softer than the first epoch. We model this epoch as an unabsorbed power-law (\texttt{tbabs*zashift*cflux*powerlaw}, W statistic 28.31 for 26 degrees of freedom), given the lack of evidence for absorption in the first epoch. The results are shown in Appendix Figure~\ref{fig:xray_spec2}, and we find a $1\%$ chance that the photon index is consistent with the first epoch $\Gamma \leq 1.73$. This tentatively suggests that the photon index softened with time. Following the same procedure, the final {\it XMM-Newton} observation was best modeled with $\Gamma = 1.5\pm0.3$, which is consistent with both the first and second epochs.

Given that the evidence for evolution in the photon index is tentative, we construct a lightcurve with better constrained fluxes by simultaneously fiting all epochs together with a constant photon index, following the same procedure as above in \texttt{xspec}. Likewise, we convert the {\it NuSTAR} and Swift/XRT observations into X-ray fluxes, assuming that the underlying spectrum is an unabsorbed power-law with a photon index of 1.7. The resulting lightcurve is shown in the bottom left panel of Figure~\ref{fig:ztf_lc}. We verify that our conclusions do not change if we separately fit the photon indices.

The soft X-ray emission is luminous and highly variable. The peak luminosity detected was in the 21.4 rest-days post-discovery XMM-Newton observation, with $L_X = 10^{44.12 \pm 0.034}$\,erg\,s$^{-1}$. By stacking the Swift/XRT observations at 20.6 and 23.2 rest-days, we obtain a $3\sigma$ upper limit of luminosity $L_X < 10^{43.8}\,{\rm erg\,s}^{-1}$. Our most luminous detection at 21.4 rest-days (i.e., in between the Swift data points) is $6\sigma$ higher than the Swift upper limit. 24puz was variable at a factor of $3.8^{+5.5}_{-1.5}$ assuming normally distributed fluxes ($3\sigma$ limit $>1.3$) on ${\sim}$3 day timescales in the soft X-ray. The \textit{NuSTAR} upper limits exclude any luminous hard emission component. 

\subsection{Radio-mm emission} \label{sec:radio}

\begin{figure*}
\centering
\includegraphics[width=\textwidth]{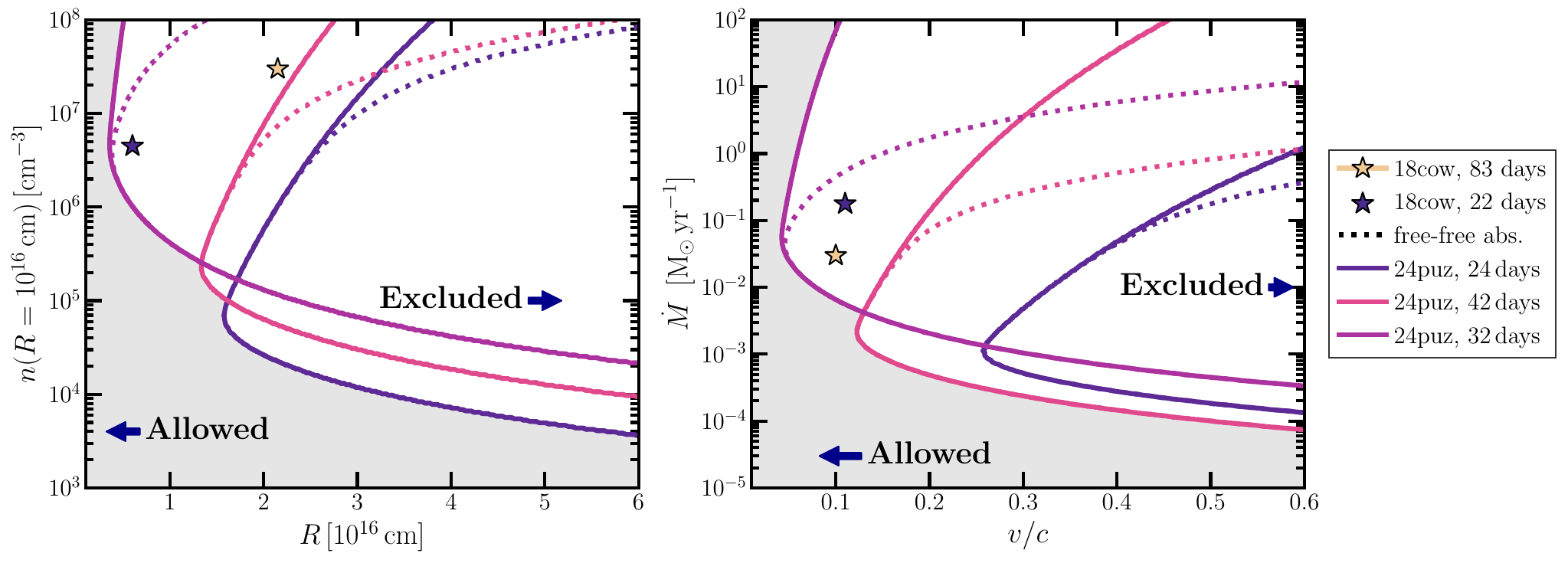}
\caption{Constraints on the physical parameters of any synchrotron-emitting region from our radio/millimeter observations. The {\it left} panel shows the ambient medium density $n$ at outflow radius $R$ for a spherical, non-relativistic outflow. The dashed lines mark the boundary between physical parameters excluded and allowed by a model that includes free-free absorption where the absorbing medium density is assumed to be the same as the emitting region. The grey region is allowed by all models. Each line corresponds to a different observation and the color scales with the time since discovery. The solid lines do not include free-free emission. Regions to the left of these lines are allowed while regions to the right are excluded. The LFBOT AT\,2018cow is shown in stars for comparison \citep[][]{ho_18cow, margutti_18cow}. The {\it right} panel is in the same format as the {\it left} but is shows physical parameters appropriate for a stellar-wind like circum-transient medium. The x-axis shows velocity $v$ and the y-axis the mass-loss rate $\dot{M}$.}\label{fig:radio_limits}
\end{figure*}

24puz was not detected in any of our radio or millimeter observations. Motivated by radio observations of similar transients, as we will discuss in the following section, we constrain the circum-transient medium under the assumption that any radio emission is produced by a non-relativistic, wide-angle outflow colliding with a dense medium. Radio emission in this scenario is generally produced by the synchrotron mechanism, with possible contributions from bremsstrahlung. We follow \cite{Rybicki1986RadiativeAstrophysics} closely in this section.

We consider an electron with Lorentz factor $\gamma$ and pitch angle $\theta$ in a region of uniform density ($n$) and magnetic field ($B$). The synchrotron frequency of this electron is $\nu_s = \frac{eB\gamma^2}{2\pi m_e c}$, where $e$ is the electron charge, $m_e$ is the electron mass, and $c$ is the speed of light. The synchrotron power for this electron is given by 
\begin{equation} \label{eq:sing_elec}
    P_s(\nu \mid B, \gamma, \theta) = \frac{\sqrt{3} e^3 B \sin \theta }{m_e c^2} = F(\nu/\nu_c),
\end{equation}
where $\nu_c = 3/2 \nu_s \sin \theta$, $F(x) = x\int_x^{\infty} K_{5/3}(y)dy$, and $K_{5/3}(y)$ is the modified Bessel function of order 5/3.

For a population of electrons, the synchrotron spectrum is computed by summing Equation~\ref{eq:sing_elec} over all electrons. We adopt the standard assumption of a population of electrons with Lorentz factors drawn from a power law with index $p$: $N(\gamma)d\gamma = N_0 \gamma^{-p} d\gamma$. We adopt the approximation of $F(x)$ used in \cite{Somalwar2023VLASSTDEs}, which allows Equation~\ref{eq:sing_elec} to be analytically integrated for a power-law electron distribution with small errors relative to numerically integrating $K_{5/3}(y)$.

Let the system volume be given by $V$, where for a spherical region of radius $R$ and filling factor $f_V$ we have $V = 4/3 \pi f_V R^3$. The total energy in the system is given by $E$. Then, the total energy stored in the magnetic field is $E_B = \frac{B^2 V}{8\pi}$. We adopt the equipartition assumption, such that the total energy stored in the magnetic field $E_B$ is a fixed fraction of the total energy of the system: $\epsilon_B = E_B / E$. Likewise, the total energy stored in the electrons, $E_e = \frac{N_0 m_e c^2}{(p-2)\epsilon_e}$ is a fixed fraction of the total energy $\epsilon_e = E_e / E$. This assumption allows us to reduce the number of free parameters in the system. We adopt the common assumption that $\epsilon_B = 0.01$ and $\epsilon_E = 0.1$ \citep[][]{margutti_18cow}.

Thus, given a magnetic field, radius, and electron energy distribution index $p$, we can compute the expected synchrotron luminosity at each frequency. The post-shock density, in the strong shock regime, is given by
\begin{equation}
    n_e = \frac{B^2}{6\pi\epsilon_B m_p v^2},
\end{equation}
where we have defined the shock velocity $v$, which we assume is the average velocity $v=R/t$ given a radius $R$ and a time since launch $t$. Whenever we discuss density in the rest of this section, we refer to this post-shock density.

In some cases, we must also include free-free absorption. The free-free optical depth is given by \citep[][Eq{.} 10.13]{Draine:2011}
\begin{multline}\label{eq:freefree}
    \tau_{ff} = \sqrt{\frac{2\pi}{3 m_e^3 k_B T_e}} \frac{4e^6}{3h c \nu^3} \\ 
    \times (1-e^{-\frac{h \nu}{k_B T_e}}) Z_i^2 g_{ff}(\nu,T_e,Z_i) \int n_i n_e dl. 
\end{multline}
We have defined the electron temperature $T_e$, the number of electrons per ion $Z_i$, frequency $\nu$, the ion density $n_i$ and the electron density $n_e$, and the Gaunt factor $g_{ff}(\nu, T_e, Z_i)$. We approximate the Gaunt factor following Section 10.2 of \cite{Draine:2011}. All other variables follow common notation for constants (e.g., electron mass $m_e$). The final integral is over the line-of-sight and is the emission measure ${\rm EM} =  \int n_i n_e dl$. We adopt the common assumptions $Z_i = 1$ and $n_i = n_e$. We adopt a temperature of $T_e= 10^5$\,K, which is higher than the standard photoionization equilibrium temperature due to Compton heating by hard X-ray photons. This is justified in Appendix~\ref{sec:compton}.

We assume the density profile is ${\sim}r^{-2}$, such that the emission measure is $EM = \int_R^\infty n_e n_i dl = n_e^2 R / 3$. As we will discuss in Section~\ref{sec:shockbreakout}, our modeling of the optical/UV emission as a shock prefers a shallower density profile ${\sim}r^{-1}$ in the inner regions of the circum-transient medium ($R\lesssim 10^{15}\,$cm). If the outer density profile is similarly shallow, the emission measure will increase by a factor of ${\sim}3$ (assuming the outer radius is large enough that integrating to infinity causes minimal errors). This does not change our conclusions.

In Figure~\ref{fig:radio_limits}, we show the post-shock density from the 24puz upper limits. We treat each observation independently. In the left panel of this figure, we show, the upper limit on radius (x-axis) for a range of assumed densities. The dashed lines include free-free absorption, while the solid lines only include synchrotron self-absorption. The shaded region is allowed by all observations. The right panel shows the same results but in variables appropriate for stellar winds. We assume a wind-like density profile $\rho = \dot{M}/(4\pi v r^2)$. We convert radius to average velocity for each epoch. The plot is otherwise formatted the same as the left panel.

The observations allow for a non-relativistic outflow colliding with a dense medium. A faster outflow is allowed for low densities ${\lesssim}10^4$\,cm$^{-3}$ or very high densities ${\gtrsim}10^9\,{\rm cm}^{-3}$, although note that we are not consistently treating relativistic effects.

\section{Results} \label{sec:res}

As detailed in the previous section, our observations have shown:
\begin{itemize}
    \item 24puz is most likely associated with G1, a dwarf galaxy ($10^{7.74} \lesssim M_*/M_\odot < 10^{8.25}$) that is located slightly below (${\lesssim}1$\,dex) the star forming main sequence. 24puz is 5\,kpc offset from G1 and shows no evidence for star formation or a massive stellar structure (${\lesssim} 10^9\,M_\odot$) at its location, although space-based follow-up once 24puz has faded is critical for tighter constraints. Both 24puz and G1 are most likely at $z=0.35614\pm0.00009$, based on the detection of ISM or CGM absorption lines, and may be bound to or infalling into a galaxy group. G1 shows tentative evidence for irregularities that could suggest a merger or environmental stripping, but deep imaging is required to confirm that the irregularities are not a background galaxy.
    \item 24puz produced a luminous UVOIR flare. The flare is well-modelled as a single blackbody in observations from $1.5-27$ rest-days post-discovery. The emission peaks at a luminosity $L_{\rm OptUV} = 10^{44.79 \pm 0.04}$\,erg\,s$^{-1}$ after a short, day-timescale rise. The luminosity stays near this value for ${\sim}12$ days and then decreases rapidly, as ${\sim}t^{-3}$. The emitting region radius initially expands rapidly as a power-law or constant velocity ($v=(0.082 \pm 0.02)c$) to a maximum radius of $R_{\rm OptUV} = 10^{15.33 \pm 0.04}$~cm. The temperature of the emitting region is relatively constant at $T_{\rm OptUV} = 10^{4.3 \pm 0.1}$\,K. At $31$ rest-days, a near-infrared excess is detected, which is more significantly detected at $56$ rest-days. A similar excess could have been present on earlier times, but the hot optical/UV emission dominate. 
    \item 24puz is a luminous, highly variable, soft X-ray source. The peak X-ray luminosity observed was $L_X = 10^{44.12\pm0.034}\,{\rm erg\,s}^{-1}$ at 21.4 rest-days. At this time, the photon index was $\Gamma = 1.73^{+0.10}_{-0.09}$ with low intrinsic absorption $n_H \lesssim 3\times10^{21}$\,cm$^{-2}$. There is tentative evidence for a softer photon-index in the 37.6 rest-days spectrum, with a p-value$\sim 1\%$ that this spectrum has the same photon-index as at peak. The X-ray emission is variable on ${\sim}3$ day timescales by a factor of $3.8^{+5.5}_{-1.5}$ ($3\sigma$ limits ${>}1.3$). For the spectrum at peak, a power-law model is preferred over a blackbody or Bremsstrahlung. Bremsstrahlung emission can fit the observations, but the implied temperature ($7\pm1.5$\,keV) is too low for typical emission from a forward shock.
    \item 24puz is non-detected at 15\,GHz and 100\,GHz (observer frame). If an outflow was launched, the circum-transient material must have a density either below ${\sim}10^5$\,cm$^{-3}$ if the average velocity is ${\gtrsim}0.1c$ or a higher density but an average velocity ${\lesssim}0.1c$. 
\end{itemize}

In the rest of this section, we discuss the structure, physics, and energetics of the emitting regions and compare to published classes of transients. 

\subsection{Energetics and emitting region scales} 

The optical/UV photosphere is observed at radii from $(0.25-1.6) \times 10^{15}$\,cm. Numerically integrating the blackbody luminosity over time, we find a total emitted energy from 0-55 rest-frame days post-discovery of $1.4_{-0.2}^{+0.7}\times10^{51}$\,erg. In the case where this is an accretion energy, which we will consider, the accreted mass is $7.7\times10^{-3}\,M_\odot$ for a radiative efficiency of 10\% (note that this efficiency is often lower for super-Eddington accretion, as will become relevant later; \citealp{Takeo2019Super-EddingtonSpectra, mckinney_superedd, jiang_superedd_review, Jiang2014ADisks}). This energy does not include contributions from the red excess observed in later epochs, but  total energy emitting in this red excess is small relative to that emitting in the optical/UV. If we assume that the red excess luminosity traces the optical/UV (as we have the best evolution constraints in this band), then the excess energy emitting at redder wavelengths is ${\sim}10\%$ of that at optical/UV wavelengths. 


The variability of the X-ray emission sets a weak limit on the size of the emitting region ${\lesssim}3\,{\rm light\, days} = 8\times10^{15}\,{\rm cm}$. It is plausible that the X-ray emitting region is much more compact and associated with the central engine that is ionizing the UV-infrared emitting regions, in which case the emitting region size must be smaller than that of the optical emitting region. The total energy detected in the X-rays is ${\gtrsim}4\times10^{49}$\,erg. We computed this value by fitting a power-law in time to the first and last X-ray detections and integrating. This is a lower limit because of the high variability, which is not included in this estimate, and because we integrate over the $0.3-10$\,keV range. From the variability alone, the real energy could be a factor ${\sim}4$ higher. We estimate the energy in the X-rays to be ${\sim}4-16\%$ of the optical/UV energy. 

The lack of transient optical spectra features implies that the optical-infrared emitting region is fully ionized, or that the emitting region density profile rapidly steepens above the photosphere so that the line-forming region is small. Given the luminous UV emission and the sizes of the regions stated above, either is feasible.

\subsubsection{The optical/UV emission} \label{sec:shockbreakout}

\begin{figure*}
\centering
\includegraphics[width=0.49\textwidth]{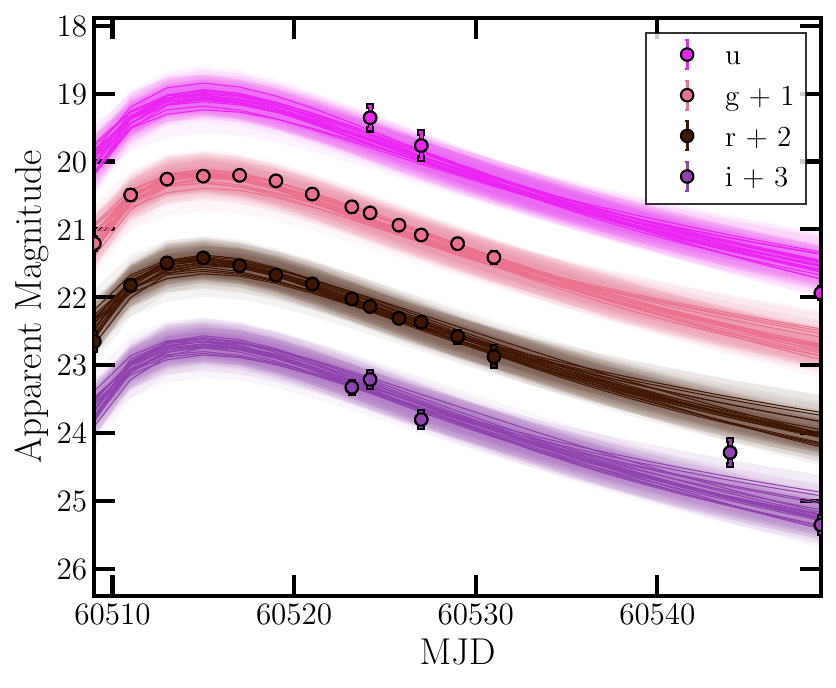}
\includegraphics[width=0.49\textwidth]{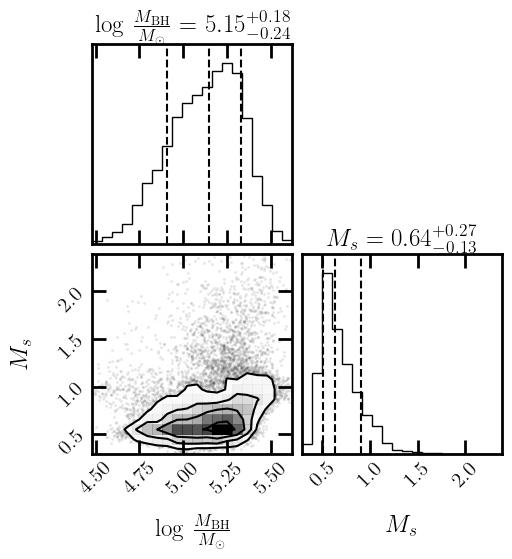}
\caption{ Summary of results from the \texttt{MOSFIT} TDE modeling. Realizations from the MCMC samples are overlaid on select observations in the {\it left} panel. We only include observations in representative bands for clarity, but perform the fit with all data. The constraints on black hole ($M_{\rm BH}$) and disrupted star mass ($M_s$) are shown in the {\it right} panel. }\label{fig:mosfit_sum}
\end{figure*}

We first consider the origin of the optical/UV flare. We rule out a transient powered by $^{56}$Ni decay, following the same logic as \cite{margutti_18cow} used for the LFBOT AT\,2018cow. The $^{56}$Ni mass can be constrained in two ways. First, from the optical/UV peak luminosity and, second, from the rise time. From \cite{Arnett1982TypeCurve}, the rise time $t_{\rm rise}$ can be approximated as the diffusion time $t_{\rm dif}$
\begin{multline}
    t_{\rm rise} \approx t_{\rm dif} \approx \big(\frac{M_{\rm ej} \kappa}{4 \pi v_{\rm ej} c}\big)^{1/2} \\= 10\,{\rm days} \bigg(\frac{M_{\rm ej}}{4.3\,M_\odot}\bigg)^{1/2} \bigg(\frac{v_{\rm ej}}{0.1c}\bigg)^{-1/2} ,
\end{multline}
where $\kappa = 0.1\,{\rm cm}^2\,{\rm g}^{-1}$ is the effective opacity, $v_{\rm ej}$ is the ejecta velocity, and $M_{\rm ej}$ is the ejecta mass. The peak bolometric luminosity is given by \citep[][]{Nadyozhin1994TheDecay}
\begin{multline}
    L_{\rm bol} = \\
    \frac{M_{\rm Ni}}{\tau_{\rm Co}-\tau_{\rm Ni}} \bigg\{ \bigg( Q_{\rm Ni} \big(\frac{\tau_{\rm Co}}{\tau_{\rm Ni}}-1\big) - Q_{\rm Co}\bigg) e^{-\frac{t}{\tau_{\rm Ni}}} + Q_{\rm Co}e^{-\frac{t}{\tau_{\rm Co}}} \bigg\}  \\ 
    = \frac{M_{\rm Ni}}{M_\odot}\bigg(6.45 e^{-\frac{t}{8.8\,{\rm days}}} + 1.45 e^{-\frac{t}{111.3\,{\rm days}}}\bigg)  10^{43}\,{\rm erg\,s}^{-1},
\end{multline}
where $\tau_{\rm Co} = 111.3$\,days is the half-life of $^{56}$Co and $\tau_{\rm Ni} = 8.8$\,days is the half-life of $^{56}$Ni, $Q_{\rm Ni} = 1.75$\,MeV and $Q_{\rm Co} = 3.73$\,MeV are emitted energies per decay, and $t$ is the time since the initial event. $M_{\rm Ni}$ is the Nickel mass. To reproduce the peak optical/UV luminosity of 24puz, we thus require a large $^{56}$Ni mass of ${\sim}30\,M_\odot$. This contradicts constraints from the lightcurve rise time, which would correspond to the diffusion time. The ejecta mass required for a ${\lesssim}10$ day rise with an ejecta velocity of $0.1c$, as implied by the optical/UV photosphere, is ${\lesssim}4\,M_\odot$. The kinetic energy of this ejecta mass is large at $4\times 10^{52}$\,erg. We disfavor a $^{56}$Ni-powered model.

We next consider an accretion-powered flare, where gas surrounding a central source reprocesses high-energy emission to produce the optical/UV flare. Motivated by the similarity of this transient to TDEs that we present in Section~\ref{sec:disc}, we first tested this origin using the \texttt{MOSFIT} code to model the optical/UV emission from 24puz as a TDE-like accretion flare \citep[][]{Guillochon2018MOSFiT:Transients, Mockler2019WeighingEvents}. \texttt{MOSFIT} is based on a grid of hydrodynamical simulations of simulated disruptions of a $1\,M_\odot$ star by a $10^6\,M_\odot$ SMBH, which are then scaled to other stellar and blackhole parameters using analytic relations. The luminosity is assumed to trace the fallback rate measured from these simulations, but with delays from the time for the debris to circularize into an accretion disk, as well as viscosity in that accretion disk. The accretion power is then reprocessed into a blackbody by a reprocessing layer with power-law evolution in the photospheric radius. 

By default, \texttt{MOSFIT} requires sub-Eddington luminosities, but we turned off the Eddington accretion limit as 24puz requires super-Eddington accretion for reasonable BH masses, as we will discuss further. We  otherwise ran with default settings. We used the \texttt{dynesty} sampler and the default priors for the TDE model, except that we set the black hole mass lower limit to $50\,M_\odot$. The results are shown in Figure~\ref{fig:mosfit_sum}. A full corner plot with the results is shown in Appendix Figure~\ref{fig:mosfit_corner}. The best-fit model invokes a black hole with $\log M_{\rm BH}/M_\odot = 5.15^{+0.18}_{-0.24}$ that is fully disrupting a relatively low mass star $M_* = 0.64^{+0.27}_{-0.13}\,M_\odot$. As we discuss in Section~\ref{sec:disc}, such a high BH mass is unlikely for this system, although we cannot exclude it. We disfavor this model.

Regardless of the exact mechanism powering the emission, the peak optical/UV luminosity is likely highly super-Eddington. As we will show in Section~\ref{sec:disc}, we favor a model with accretion onto a BH of mass $M_{\rm BH}/M_\odot \lesssim 10^5$, corresponding to a radiation Eddington ratio $10^2 \lesssim \lambda_{\rm Edd}$. 
Simulations are increasingly suggestion that super-Eddington radiation and accretion are feasible \citep[][]{Takeo2019Super-EddingtonSpectra, mckinney_superedd, jiang_superedd_review, Jiang2014ADisks}. The accretion power may be in the form of mechanical energy carried by optically thick outflows which then convert the kinetic energy into radiation. Observations of ultraluminous X-ray sources (ULXs) find luminosities up to $10^{42}$\,erg\,s$^{-1}$, corresponding to Eddington ratios $10^{2-3}$ \citep[][]{King2023UltraluminousSources}. 

If we are observing emission from gas that is photoionized by an accretion source, then we must consider the lightcurve shape, which begins with a ${\sim}$days rapid rise. The lightcurve then shows a slow rise or a plateau for ${\sim}10$ days before rapidly decaying. The initial turn-on may correspond to accretion beginning. The slow rise/plateau could be explained as a super-Eddington cap \citep[or see][for an accretion-powered model that could produce similar emission. We will discuss this model in more detail shortly.]{Metzger_LFBOT}. If 24puz has reached the limit of super-Eddington radiation, the observed radiation may appear as a plateau until the accretion rate has dropped to sufficiently low values that the radiation begins to trace the accretion rate. Extrapolating from the power-law decay to early times, this implies a remarkable peak accretion rate of ${\gtrsim}10^{46}\,{\rm erg\,s}^{-1}$. Alternatively, the central source may have a plateau in its accretion rate. 

We conclude that an accretion powered model may be able to explain our observations, although with significant uncertainties. The standard modeling code \texttt{MOSFIT} can reproduce our observations, but only with a black-hole mass that is slightly higher than our preferred range, as we will discuss in Section~\ref{sec:disc}.

We next consider a shock breakout. This discussion closely follows \cite{Khatami2024TheMedium}. 
Let us define $M_{\rm ctm} = M_{\rm ctm, \odot} M_\odot$ as the mass of the circum-transient material, $R_{\rm ctm} = R_{\rm ctm, 16} \times 10^{16}\,{\rm cm}$ as the edge of the circum-transient material, $M_{\rm ej}$ as the ejecta mass, and $v_{\rm ej} = \beta_{\rm ej} c$ as the ejecta velocity. Define $\eta = M_{\rm ctm}/M_{\rm ej}$. The shock velocity at $R_{\rm ctm}$ is related to the ejecta velocity as $v_{\rm sh} = \beta_{\rm sh} c = v_{\rm ej} \eta^{-\alpha}$.

\begin{figure*}
\centering
\includegraphics[width=0.75\textwidth]{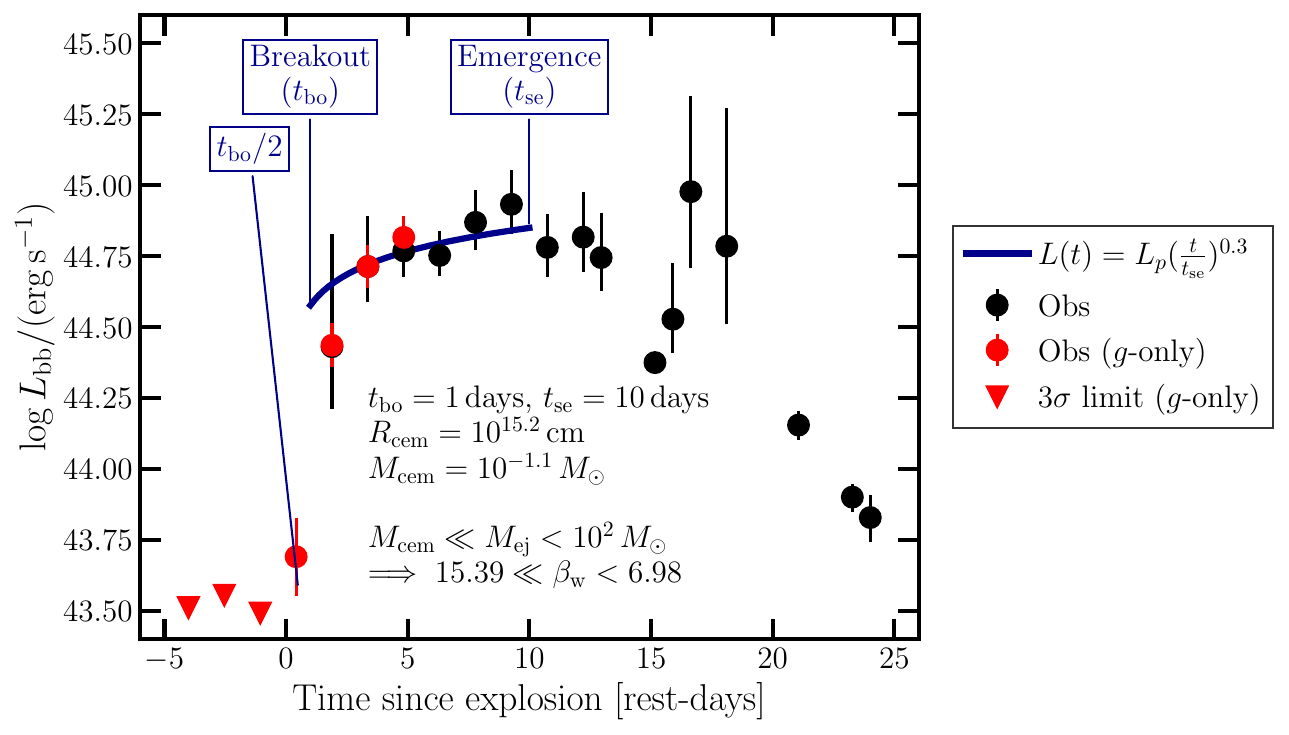}
\caption{ Constraints on a shock breakout model for the optical/UV emission from 24puz, assuming that the ejecta mass ($M_{\rm ej}$) is much larger than the circum-transient medium mass ($M_{\rm ctm}$) and the circum-transient medium has a power-law density profile with slope $s=1$. The optical/UV blackbody luminosity is shown as a function of time in black  in both panels. The red points show the luminosity obtained from early $g$-band-only detections of 24puz, for which we could not fully constrain the luminosity. We assume the temperature is the average of the over the first ${\sim}5$ days. We show the case of small $t_{\rm bo}$ for illustration. The shock emergence is assumed to occur on the same day in both panels, corresponding to the time at which the luminosity begins to decrease. The blue solid line shows the expected power-law evolution of the lightcurve within the shock breakout model, from equation~\ref{eq:lbo}. The resulting inferred physical parameters (circum-transient medium radius $R_{\rm ctm}$ and mass $M_{\rm ctm}$) are listed on the bottom, including a range of possible shock velocities inferred from the assumption that $M_{\rm ej} \gg M_{\rm ctm}$ and the ejecta mass is reasonably small $\lesssim 10^2\,M_\odot$. 
}\label{fig:shock_breakout}
\end{figure*}

As discussed in detail in \cite{Khatami2024TheMedium}, there are four regimes that produce separate lightcurve behavior, determined by (1) whether the shock breakout occurs within or at the edge of the circum-transient medium and (2) whether $\eta$ is large (${\gtrsim}1$) or small (${\ll}1$). The optical emission from the LFBOT AT\,2018cow was reproduced by a shock model for the case of edge shock breakout with a small circum-transient medium ($\eta \ll 1$). 24puz rises rapidly in the optical for a few days, and then the rise rate slows for ${\sim}1$ week. The luminosity then rapidly drops. Unlike AT\,2018cow, the behavior of 24puz is expected in a model where the shock breakout occurs inside the circum-transient material, but still with a light circum-transient medium ($\eta \ll 1$). In this case, the initial, rapid rise corresponds to the shock breakout. The slow rise traces the shock kinetic luminosity and, as we will show, implies a shallow circum-transient medium density profile. The rapid drop begins once the shock reaches $R_{\rm ctm}$. At this point, we would expect to detect some cooling of the emission, which is not the case for 24puz, so we would require a secondary component to heat the ejecta at late-times. We will discuss this later in this section.

We assume a slope $s=1$ fiducially. To reproduce the slope of the slow luminosity rise, we require an ejecta profile $\rho_{\rm ej} = r^{-n}$ with $n=12$. A shallower slope $s$ is also allowed, in which case the ejecta profile will also be slightly shallower (e.g., for $s=0$, we have $n=10$). We tested values of $s=0$ and $s=1$ and our general conclusions do not change. We do not perform any quantitative fitting and instead attempt to roughly reproduce the lightcurve using analytic estimates from \cite{Khatami2024TheMedium}; we encourage work quantitatively fitting the lightcurve to light, interior shock interaction models.

The start date is not directly observable, so we constrain it (and thus $t_{\rm bo}$ and $t_{\rm se}$) as follows. This procedure is imperfect, and constraining the physical parameter using a simulated model grid instead would be ideal. First, we use the analytic expression in \cite{Khatami2024TheMedium} to solve for the ejecta velocity $v_{\rm ej} = \beta_{\rm ej}c$, the circum-transient medium mass $M_{\rm ctm}$, the outer radius of the circum-transient medium $R_{\rm ctm}$, and the breakout luminosity $L_{\rm bo}$ as a function of breakout time $t_{\rm bo}$, shock emergence time $t_{\rm se}$, and $\eta$. We find
\begin{gather}
    v_{\rm ej} = \eta^\alpha \bigg( \frac{\kappa L_p}{4 \pi c t_{\rm bo}} \bigg)^{1/3}; \\
    M_{\rm ctm} = L_p t_{\rm se} \bigg( \frac{4 \pi c t_{\rm bo}}{\kappa L_p} \bigg)^{2/3}; \\
    R_{\rm ctm} = \bigg( \frac{\kappa L_p t_{\rm se}^3}{4 \pi c t_{\rm bo}} \bigg)^{1/3}; \\
    L_{\rm bo} = L_p \bigg( \frac{t_{\rm bo}}{t_{\rm se}} \bigg)^{\frac{(5-s)(n-3)}{n-s}-3} \label{eq:lbo}
\end{gather}
Here, we have defined $\alpha = \frac{1}{n-3}$ and the electron scattering absorption coefficient $\kappa = 0.34$\,cm$^{2}$\,g$^{-1}$. We pick combinations of $t_{\rm bo}$ and $t_{\rm se}$ such that (1) $t_{\rm se}$ corresponds to $9.6$ rest-days post-discovery, which is when the luminosity began to decay and (2) the luminosity at time $t_{\rm bo}/2$ is approximately equal to $0.1L_{\rm bo}$. We consider two extreme values of $t_{\rm bo}$ and $t_{\rm se}$, corresponding to early and late start dates to highlight the range of possible parameters. Earlier start times relative to the detection epoch will bring the ratio of $t_{\rm se}$ and $t_{\rm bo}$ close to one to satisfy condition (2), while later start times would always overproduce the $t_{\rm bo}/2$ luminosity.

The corresponding circum-transient medium parameters are $R_{\rm ctm} \approx 10^{15}\,$cm, which roughly matches the maximum measured blackbody radius, and $M_{\rm ctm} \approx 0.1-1\,M_\odot$. We cannot tightly constrain the ejecta mass without a direct observation of the wind velocity. We have assumed $\eta \ll 1$, so we have $M_{\rm ej} \gg 0.1\,M_\odot$. This corresponds to a shock velocity $\beta_{\rm sh} \gg 0.06$ at $R_{\rm ctm}$, implying a fast shock. If we assume the ejecta mass is not huge, which we will define as ${\lesssim}100\,M_\odot$, we can set a lower limit on the shock velocity $\beta_{\rm sh} \gg 0.03$. The limits for the case of small $t_{\rm bo}$ are shown in Figure~\ref{fig:shock_breakout}. Note that these parameters are consistent with the radio upper limits: the implied densities are sufficiently high (${\sim}10^{10}$\,cm$^{-2}$) that free-free absorption will prevent detection of synchrotron emission, or the density may fall as a steeper power-law outside $R_{\rm ctm}$.

The shock breakout model explains the optical/UV lightcurve shape at early-times, and such a model has been used for events with similar lightcurve evolution, although generally on much longer timescales and with lower luminosities \citep[e.g.][]{Karamehmetoglu2017OGLE-2014-SN-131:Progenitor}. We encounter two problems at later times: the lack of cooling and transient spectral features. The lack of cooling may require an additional ionizing source that dominates by ${\sim}20$\,days \citep[][]{Nakar2014SupernovaeEnvelopes, margutti_18cow}, but we may expect a contribution from reprocessing of a central source given the luminous X-rays detected \citep[e.g.][]{Metzger_LFBOT, margutti_18cow, Lu2020Self-intersectionEvents}. The lack of lines could also explained by an additional ionizing source, if it is sufficient to keep the emitting region fully ionized, or if the circum-transient medium has a rapid density drop-off at the photosphere so that the line-emitting region is small. Full radiative transfer simulations of the shock model with similar physical parameters to those of 24puz would determine whether the lack of cooling and lines are prohibitive.



In summary, the early time optical/UV lightcurve could be explained using both accretion-powered and shock breakout models. At late times, the shock breakout model does not naturally explain the lack of cooling and spectral features, but detailed simulations are required to assess the significance of this problem. The late-time emission likely requires additional accretion power, regardless of the origin of the early-time emission.

\subsubsection{The red excess} \label{sec:redex}

In addition to the optical/UV flare, 24puz showed a red excess above a single blackbody fit in the final two epochs of observations ($31-56$ rest-days post-discovery). There are a few models that are typically considered for similar sources: thermally emitting gas/dust, re-processing of high energy emission by gas, gamma-ray burst-like nonthermal emission, and radioactive decay. We consider the agreement with each of these models with our {\it HST} observations, which have the best coverage of the red excess. 

\begin{figure*}
\centering
\includegraphics[width=0.95\textwidth]{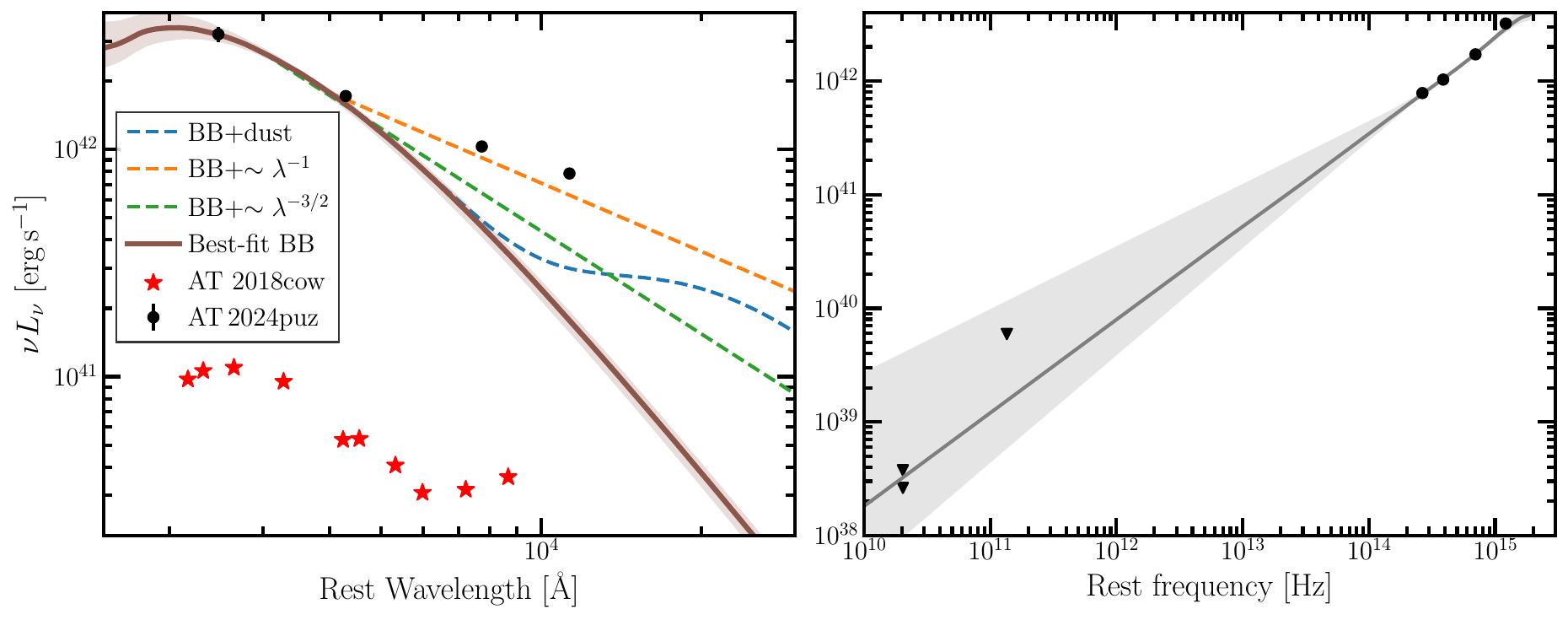}
\caption{Constraints on the origin of the red excess from 24puz. The {\it left} panel shows the {\it HST} spectral energy distribution of 24puz in black and, for comparison, an epoch of observations of the red excess from AT\,2018cow in red. A best-fit hot blackbody is shown in brown. The dashed lines show the hot blackbody with potential red excess emission added on. These are not fits but are meant to guide the eye. In orange, a $\lambda^{-1}$ power-law was added to represent reprocessing in a shallow medium; in green, a $\lambda^{-3/2}$ power-law is summed to represent reprocessing in a wind-like medium. In blue, we add a dust blackbody to show that this model peaks redward of the excess. The {\it right} panel shows our radio upper limits as black triangles. The grey band shows an extrapolation of a power-law fit to the {\it HST} data to radio frequencies. Considering that the {\it HST} data is taken at later-times than the radio, so we would expect it to have faded with time, the power-law extrapolation over-predicts the radio limits.}\label{fig:red_excess}
\end{figure*}

We begin with a thermally emitting region. We first modeled the UVOIR emission as two blackbodies. The temperature of the cooler blackbody was $\log T_r \approx 3600\,$K, which is above the sublimation of all dust grains, but too cool to be consistent with photoionized gas. If the red excess is produced by dust, the temperature inconsistency can be improved by modeling the emission as a hot blackbody and a modified blackbody that accounts for dust absorption \citep[][]{metzger_dust, Tuna2025Time-DependentTransients}. We test both the graphite and silicate models from \cite{Draine2007Era} and report the results for the more conservative graphite model in Table~\ref{tab:OIR_bb}. The temperature is still ${\sim}3500$\,K; i.e., significantly above the sublimation temperature of graphite grains (${\sim}2000$\,K). This discrepancy is highlighted in Figure~\ref{fig:red_excess}, where we show our {\it HST} observations in black and an example spectral energy distribution for a hot blackbody with additional graphite thermal emission in dashed blue. The model shown is not fit to the data, but demonstrates that graphite emission causes a peak redward of the observed red excess (near $2\,{\rm \mu m}$), whereas the observed red excess is present by $9000\,{\rm \AA}$. We thus exclude dust as the main source of the red excess.

We next consider gamma-ray burst-like non-thermal emission. This model is immediately disfavored by the long timescale of the emission, which is much slower evolving than gamma-ray burst-like afterglows. Moreover, a power-law + hot blackbody fit provides a poor model of the {\it HST} data, as seen in the bottom right panel of Figure~\ref{fig:uvopt_fits}. In the right panel of Figure~\ref{fig:red_excess}, the best-fit power-law+blackbody to the {\it HST} data predicts a radio luminosity that is comparable to our $4\sigma$ upper limits, which are described in Section~\ref{sec:radio}. However, our {\it HST} data was obtained ${\sim}40{-}60$ rest-days after the {\it HST} observations. If the emission is caused by an on-axis jet, then the jet emission should be brighter at earlier times: this is ruled out by the luminous radio emission predicted at late-times by the power-law extrapolation from the {\it HST} data. If the emission is caused by an off-axis jet that is now wide-angle and non-relativistic, allowing us to view its emission, it is feasible for the radio emission to be brighter at ${\sim}80$ days than at early times. The shallow observed slope disfavors this scenario. As discussed, we have $\nu F_\nu \sim \nu$, or $F_\nu \sim {\rm const}$. This slope is inconsistent with typical gamma-ray burst synchrotron frameworks. Because of the poor fit, the lack of a radio detection, and the shallow observed slope, we rule out this model.

We next consider high-energy emission that has been reprocessed by gas in the circum-transient medium. We will leave detailed models of this reprocessing to future work, but use simple estimates to assess agreement with our observations. We follow \cite{Chen2024TheTransients} and \cite{Lu2020Self-intersectionEvents}.

In this model, the red excess is caused by a change in the dominant opacity at longer wavelengths, as detailed in \cite{Lu2020Self-intersectionEvents}. At near-infrared wavelengths, free-free absorption dominates. This absorption coefficient is given by
\begin{equation}
    \kappa_{ff} = 0.018 T^{-3/2} \nu^{-2} \rho^2 m_p^{-2}\,[{\rm cm}^{2}\,{\rm g}^{-1}],
\end{equation}
where $T$ is the temperature of the emitting region, $\nu$ is the frequency under consideration, $\rho$ is the density, and $m_p$ is the proton mass. We have assumed that we are in the Rayleigh-Jeans limit and neglected the Gaunt factor.


We assume a power-law density profile
\begin{equation}
    \rho = \rho_0 \bigg(\frac{r}{r_0}\bigg)^{-s}.
\end{equation}

In the near-infrared, free-free absorption dominates the opacity, so the thermalization radius is given by the radius at which the effective optical depth, or the product of the free-free optical depth ($\tau_{ff} \approx \kappa_{ff} \rho r$) and electron-scattering optical depth ($\tau_{es} \approx \kappa_{es} \rho r$), is one. Quantitatively:
\begin{equation}
    \tau_{ff} \tau_{es} \approx \kappa_{ff} \kappa_{es} \rho^2 r_{th, \nu}^2 \approx 1.
\end{equation}
Here, $\kappa_{es} = 0.34$ cm$^2$\,g$^{-1}$ is the electron scattering opacity. The luminosity is then given by
\begin{equation}
    \nu L_\nu = 4\pi r_{th,\nu}^2 \frac{4 \pi \nu B_\nu(T(r_{th,\nu}))}{\kappa_{es} \rho(r_{th,\nu})r_{th,\nu}(s-1)}.
\end{equation}

Combining these four equations, we find
\begin{multline}\label{eq:Lred}
    \nu L_\nu = \\\frac{32 \pi^2 k_B}{\kappa_{es} c^{2}(s-1)} \bigg(\frac{0.018 \kappa_{es}}{m_p}\bigg)^{\frac{1+s}{3s-2}} T^{\frac{7-3s}{4-6s}}(\rho_0r^{s})^{\frac{5}{3s-2}} \bigg(\frac{c}{\lambda}\bigg)^{\frac{7s-8}{3s-2}}.
\end{multline}

For a wind-like density profile ($s=2$), we have $\lambda L_\lambda \sim \lambda^{-3/2}$. As shown by the dashed green line in Figure~\ref{fig:red_excess}, this slope is steeper than the observed red excess, although note that our transition from the hot blackbody to a slope $-3/2$ power-law is ad-hoc and not quantitatively accurate. For a shallow density profile $\rho \sim r^{-1}$, we have $\lambda L_\lambda \sim \lambda^{-1}$, which is closer to the observed slope and is consistent with our optical/UV shock interaction analysis. Many of the approximations quoted above, however, break down for slopes that are much shallower than $s=2$. We will thus adopt an intermediate slope $s=1.5$ for the following analysis to prevent divergences, but urge simulations of the $s=1$ case to test if our conclusions hold.

We can evaluate the required density by matching the red excess luminosity to that predicted by Equation~\ref{eq:Lred}. For a luminosity $\nu L_\nu \sim 10^{42}\,{\rm erg\,s}^{-1}$ at ${\sim}7000\,\AA$ and $s=1.5$, we find that the density is $10^{10.5}\,{\rm cm}^{-3}$ at $10^{15}\,$cm and the total mass must be ${\sim}0.2\,M_\odot (R_{\rm out}/10^{15.1}\,{\rm cm})$. This is consistent with the circum-transient medium mass that was inferred in our shock interaction analysis $M_{\rm ctm}\lesssim 1\,M_\odot$ for $R_{\rm ctm} \approx 10^{15}\,$cm. If this density profile extends farther out, our radio analysis may pose a problem as we find $\rho(R=10^{16}\,{\rm cm}) \approx 10^{9}\,$cm$^{-3}$, which is on the border of our excluded region in Figure~\ref{fig:radio_limits} unless the outflow velocity is ${\lesssim}0.1 c$. Free-free absorption would remove this limit. Alternatively, if the density profile drops rapidly at $10^{15}\,$cm, then the allowed outflow velocity is higher. Given that the implied mass if this profile extends to $10^{16}\,$cm is ${\sim}20\,M_\odot$, which is very large, we believe this latter explanation is likely.

The lack of any significant absorption in the X-ray spectrum poses a problem: the implied column density is ${\gtrsim}10^{25}$cm$^{-2}$ for an outer radius $R_{\rm ctm} \sim 10^{15.1}$\,cm and an inner radius corresponding to the smallest radius measured from our blackbody fitting in Table~\ref{tab:OIR_bb}, $R_{\rm in} \sim 10^{14.4}$\,cm. A smaller $R_{\rm in}$ corresponds to a larger column density. We can alleviate this discrepancy if the emitting region is aspherical. If we are viewing the system along a line-of-sight without significant gas/dust, then the X-rays would appear unabsorbed, as is observed. Alternatively, if the emitting region is fully ionized, it is effectively transparent to X-rays.

There are a number of key caveats to this analysis. First, we adopted common approximations to the photon diffusion time and optical depth that break as the density profile becomes shallower. Moreover, this analysis assumes a spherical, homogeneous medium. If the material is anisotropic or clumpy, then the total mass will be reduced by the volume filling factor \citep[][]{Chen2024TheTransients}. Finally, we have adopted an outer radius based on our optical/UV shock analysis, but accretion-power likely contributes to this emission and this was not included in our modeling. 

\subsubsection{The X-ray emission} \label{sec:xrayem}

We next consider the origin of the X-ray emission. The X-ray spectrum, luminosity evolution, and variability are both very similar to that of the soft component of AT\,2018cow. The X-rays from AT\,2018cow are considered to be associated with a shock or central engine \citep[][]{margutti_18cow}. 

We first consider a central engine model. In this case, the rapid variability is expected given the small scales of an accretion disk. The implied X-rays are highly super-Eddington. Note, however, that with the exception of the luminous {\it XMM-Newton} observation, the typical X-ray luminosity is ${\sim}3\times 10^{43}\,{\rm erg\,s}^{-1}$, which is ${\lesssim}10^3\,$Eddington for the compact object masses we will favor in the following section. As discussed earlier, this Eddington ratio is feasible \citep[][]{King2023UltraluminousSources}. 
The luminous optical/UV emission suggests that most of the X-rays, in a central engine model, are reprocessed. A fraction escape either due to inhomogeneities in the surrounding medium or because they have ionized the ejecta along our line-of-sight. 

We next consider a shock origin of the X-rays. This model faces two problems. First, as discussed in Section~\ref{sec:xray}, our Bremsstrahlung modeling suggests a best-fit temperature of $7\pm1.5$ keV, which is lower than typical X-ray emission from shocks \citep[][]{Fransson1996Circumstellar1993J}. It is also challenging for a shock to produce the ${\sim}$day timescale variability observed from 24puz: this would require significant density variations in the circum-transient medium. Variability on a timescale of days for a shock traveling at $0.1c$ implies spatial variability scales of ${\sim}10^{14}$\,cm. Some circumstellar media do show such variability, but it is not clear whether this is natural for a system like 24puz (e.g., see similar discussion for the LFBOT AT\,2018cow in \citealp[][]{Sandoval2018X-ray2018cow}). We encourage simultaneous modeling of the optical/UV and X-ray emission within a shock framework to assess whether shocks can dominate the observed emission, but we currently disfavor shocks as the dominant source of X-ray emission.

We conclude that a central engine model for 24puz can explain the observed X-rays. A shock model may be feasible, but faces significant challenges (low best-fit temperature and rapid variability).

\section{Discussion} \label{sec:disc}

\subsection{Comparison with known transients}

The persistently blue optical colors and featureless spectra place 24puz in the section of optical transient parameter space occupied by two observational classes: LFBOTs \citep[][]{Metzger_LFBOT, prentice_18cow} and TDEs \citep{hills_tde, lidskii_tde, phinney_tde}. Supernovae are excluded by the rapid evolution and persistently featureless spectra, with a lack of interaction signatures and the high radiated energy (${>}10^{51}$\,erg). The luminous, fast-cooling transients presented by \cite{Nicholl2023ATGalaxies} occupy a similar location in luminosity-timescale space, but show rapid cooling that is excluded for 24puz. They are also not X-ray and radio bright.

TDEs occur when a star is disrupted by a massive black hole (MBH; \citealp[][]{rees_tde, ulmer_tde, evans_tde}). Like LFBOTs, TDEs typically produce blue, constant color optical lightcurves, but they evolve more slowly, rising over few-week timescales and fading over months \citep{vanvelzen_firstztftde, vanvelzen_ztf, arcavi_ptftde, gezari_uvtde, gezari_uvtdes}. Most TDEs produce broad (${\sim}10^4$\,km\,s$^{-1}$) Balmer and Helium spectroscopic features \citep{hammerstein_finalseason, YaoTDESamp, vanvelzen_ztf}, but two, possibly connected subclasses of TDEs produce featureless spectra: featureless TDEs (F-TDEs; \citealp[][]{hammerstein_finalseason}) and jetted TDEs \citep{andreoni_cmc}. F-TDEs have spectra that are featureless for months post-TDE \citep{hammerstein_finalseason,YaoTDESamp}. Jetted TDEs are best known for launching collimated, relativistic jets \citep{bloom_1644, brown_1112, burrows_1644, cenko_2058, pasham_cmc}, and there is some evidence that they also produce featureless spectra and blue, constant color thermal emission (in addition to luminous, nonthermal emission from the jet; \citealp{andreoni_cmc}). By selection, TDEs are traditionally detected in the nuclei of their host galaxies (typically within $1''$; \citealp{vanvelzen_firstztftde, vanvelzen_ztf, hammerstein_finalseason, YaoTDESamp}). 

LFBOTs rise on few-day timescales and fade over ${\sim}1$ week \citep{yao_mrf, chrimes_finch, coppejans_css161010, ho_fbots, ho_koala, ho_2020xnd, perley_xnd, matthews_tsd}. LFBOTs produce hot (${\sim}10^{4-5}\,K$) optical flares without substantial cooling and absolute magnitudes brighter than approximately $-20.5$ mag. They evolve fast (${\lesssim}$week timescale), with no detectable optical spectral lines at early times. LFBOTs never show nebular features, and instead have Hydrogen and Helium features, unlike stellar explosions. They are luminous in the millimeter, suggestive of outflows into dense media (${\sim}10^5\,$cm$^{-3}$). LFBOTs have hard X-rays ($\nu F_\nu \sim \nu^{-0.5}$) likely from the central engine that is also responsible for powering the transients, which could be explained by invoking an asymmetric circum-transient medium and a weak, decelerating jet. They often have a Compton hump feature detected in the hard X-ray, which is common among accretors like X-ray binaries and AGN, and is associated with cold, optically-thick gas \citep[][]{Reynolds1999ComptonCandidates}. The X-ray spectra have been detected to soften at late-times \citep[][]{migliori_cow}.  LFBOTs are hosted in a range of environments, although most have been offset within galaxies near the star forming main sequence. LFBOTs have featureless blue optical spectra that, in some cases, develop broad (${\sim}10^4$\,km\,s$^{-1}$) hydrogen/helium emission after ${\sim}10$s of days and intermediate width (${\sim}10^3$\,km\,s$^{-1}$) lines after ${\sim}1$ month. {\it Hubble Space Telescope} observations of one LFBOT, AT 2018cow, show a late-time (years) plateau in the UV and soft X-ray \citep[][]{inkenhaag_cow, migliori_cow, chen_cow_late, Sun2022AExplosion}. The most offset LFBOT detected thus far is at $1\,{\rm kpc}\approx 3.5r_e$ from its host galaxy, where $r_e$ is galaxy half-light radius \citep{chrimes_fhnlate}. 

The physical origin of LFBOTs is unknown. The late-time UV and possible X-ray plateau detected for AT\,2018cow is reminiscent of late-time emission from the compact accretion disks produced during TDEs \citep[][]{chen_cow_late, inkenhaag_cow, migliori_cow, Sun2022AExplosion}, suggesting that LFBOTs may be TDEs. The faintness of the plateau, combined with the off-nuclear location of the transients, requires TDEs by IMBHs. LFBOTs may alternatively be extreme stellar explosions/mergers. Models are typically required to produce an extended, dense, aspherical medium and a highly energetic and compact central engine. \cite{Metzger_LFBOT} proposes the delayed merger of a black hole and a Wolf-Rayet star. \cite{margutti_18cow} consider a failed explosion of a blue supergiant star, resulting in a stellar-mass BH surrounded by the remains of the star, although \cite{migliori_cow} revise this model to prefer a super-Eddington accreting source. Both models require the presence of massive stars (${\gtrsim}20\,M_\odot$). A millisecond magnetar formed after the electron capture supernova of a ${\sim}8{-}10\,M_\odot$ star is consistent with observations, but still invokes a relatively massive star. Attempts to model the LFBOT AT\,2018cow as a magnetar require neutron stars with near maximal masses and may not be able to explain the multiwavelength emission (e.g., fast X-ray variability, late-time UV plateau) \citep[][]{li_lfbot_magnetar}.

24puz has multiwavelength properties that are consistent with both TDEs and LFBOTs. In Figure~\ref{fig:optcomp}, we compare the optical/UV blackbody luminosity evolution of 24puz, TDEs, and LFBOTs. The left panel shows the blackbody luminosity as a function of time for these objects. 24puz has a lightcurve evolution that is generally consistent with LFBOTs, with a rapid rise followed by a power-law decay. 24puz peaks at a later time relative to other LFBOTs and has a slower rise. This is highlighted in the right panel, which shows the time above half peak luminosity on the x-axis and the peak $g$-band luminosity on the y-axis. 24puz is slower than all LFBOTs. Selection effects likely play a role in this trend: LFBOTs are selected to be fast-evolving, with generally faster than ${\sim}1$ week timescales \citep[e.g.][]{ho_fbots}. 24puz may be a slow evolving LFBOT, suggesting that LFBOT searches should be performed with looser timescale cuts.

The lightcurve shape of 24puz is also consistent with TDEs, although these events show more variation in their blackbody luminosity evolution relative to LFBOTs. The location of 24puz in luminosity-timescale space is unprecedented for TDEs, which generally show a positive correlation between luminosity and timescale. F-TDEs break this correlation, but have long timescales (${\gtrsim}30$\,days). There are no obvious selection effects that would prevent events like 24puz from being discovered in TDEs searches, if 24puz-like objects can occur in galactic nuclei, suggesting that such events are rare.

The featureless optical spectra detected from 24puz are consistent with both early time LFBOT observations and with F-TDEs and jetted TDEs. LFBOTs generally produce some spectral features at late times (both broad and narrow interaction lines). If 24puz is an LFBOT, the lack of lines could be consistent with the high luminosity, as discussed above, but this would be unprecedented for this class. F-TDEs remain featureless at all times, consistent with 24puz. 

A red excess like that from 24puz has been detected for the LFBOT AT\,2018cow \citep[][]{perley_18cow}. The origin of this excess is also unknown, although similar models were proposed as we have explored for 24puz. The optical-IR SED of AT\,2018cow is shown as red stars in Figure~\ref{fig:red_excess}. The peak of the AT\,2018cow excess is redder than that of 24puz, and so a thermally-emitting dust origin is still consistent with the observations \citep[][]{metzger_dust}. Reprocessing can also reproduce the red excess, but requires ${\gtrsim}5\,M_\odot$ of material \citep[][]{Chen2024TheTransients}. TDEs have been theorized to produce near-infrared excesses due to reprocessing \citep[][]{Lu2020Self-intersectionEvents}, with one known example that showed a red excess likely due to reprocessing by a disk-like structure \citep[][]{EarlATEvent}. Thermally emitting dust has been detected for multiple TDEs, but this emission peaks in the mid-infrared and on year-long timescales and is inconsistent with the observations of 24puz. 

The radio and millimeter limits from 24puz are consistent with both TDEs and LFBOTs. ${\gtrsim}30\%$ of optically-selected TDEs produce radio emission at ${\gtrsim}3$\,yrs post-disruption. The fraction detected at early times is lower, suggesting that most TDEs do not produce luminous radio emission at this point (${\lesssim}10^{38}\,{\rm erg\,s}^{-1}$). This is consistent with the 24puz limits. It is also important to note that most radio-observed TDEs are not F-TDEs, which are generally at higher redshifts. A more detailed study of the early-time radio properties of these events would be required to compare to 24puz. 

LFBOTs ubiquitously produce radio and millimeter emission. As shown by the stars in Figure~\ref{fig:radio_limits}, the parameter space occupied by the radio/mm luminous LFBOT AT\,2018cow is excluded for 24puz, but if 24puz has a moderately less dense ambient medium relative to AT\,2018cow, the lack of radio/mm is expected. The lack of transient spectral features would also be expected in this case.

\begin{figure}
\centering
\includegraphics[width=0.48\textwidth]{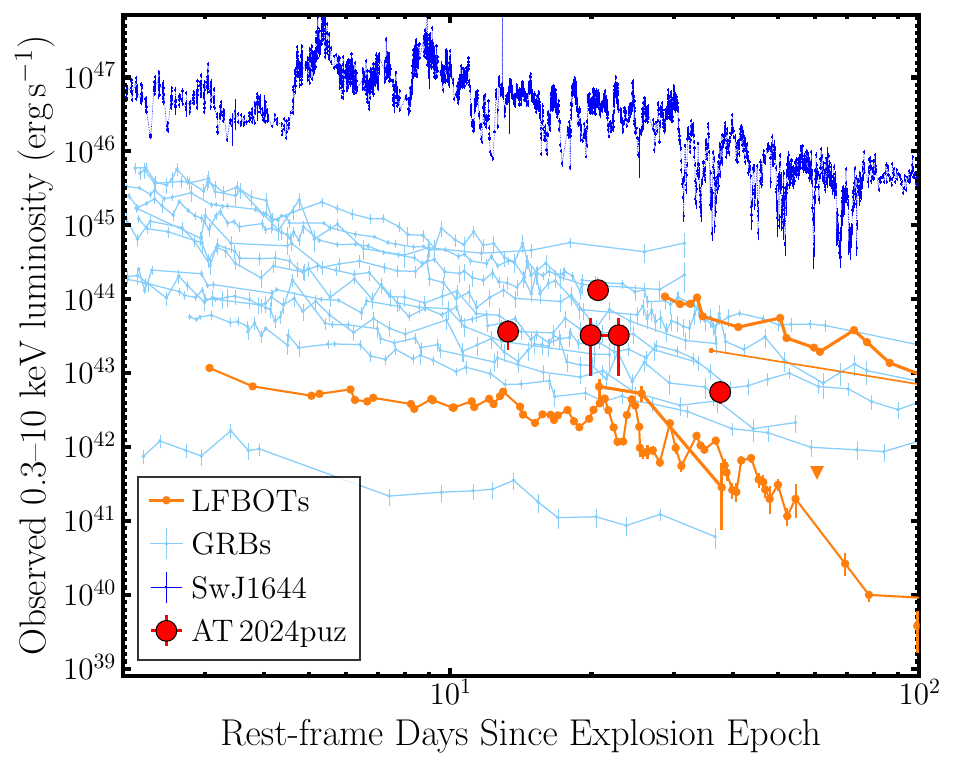}
\caption{X-ray lightcurve comparison, adapted from \cite{yao_mrf}. 24puz is shown in red and LFBOTs in orange. For completeness, a population of gamma-ray bursts (GRBs) is shown in light blue, as these are common comparison points for X-ray transient emission. The jetted TDE SwJ1644 is shown in blue. 24puz is among the most X-ray luminous LFBOTs. It shows a variability and evolution timescale comparable to both LFBOTs and the jetted TDE.}\label{fig:xray_comp}
\end{figure}
The X-ray lightcurve from 24puz is fully consistent with LFBOTs, as shown in Figure~\ref{fig:xray_comp}. The high variability and rapid decline have been observed for multiple LFBOTs. 24puz is among the most X-ray luminous LFBOTs. The hard spectral index $\Gamma = 1.7$ is also consistent with observations of X-rays from LFBOTs. The rapid evolution and hardness of LFBOT X-rays have been used to argue for the presence of a compact central engine from these sources. 

The X-ray properties of TDEs are much more heterogeneous than those of LFBOTs. Most X-ray detected TDEs have soft spectra $\Gamma \gtrsim 3$, although exceptions do exist \citep[][]{guolo_xray, Yao2022TheSystem, Ho2025AAT2024kmq}. There are also X-ray detected TDEs that show rapid variability and fading \citep{guolo_xray}, although this behaviour is not as ubiquitous as for LFBOTs. Jetted TDEs, in particular, show hard, variable X-rays like 24puz \citep[][]{bloom_1644, burrows_1644, andreoni_cmc}. As shown by the blue lightcurve in Figure~\ref{fig:xray_comp}, the X-ray luminosities of these events are much higher than that from 24puz due to beaming effects, but the lightcurve evolution is otherwise similar. X-ray constraints on F-TDEs are weak. 

The host galaxy of 24puz is also fully consistent with LFBOT hosts in terms of stellar mass, star formation rate, and offset from the transient. LFBOT host galaxy are shown relative to that of 24puz and the star forming main sequences in Figure~\ref{fig:Ms_SFR}. While the LFBOT hosts show a range of stellar masses, they tend to lie near the star forming main sequence and, there are three events at comparable host galaxy masses to that of 24puz. As shown in the right panel of Figure~\ref{fig:Ms_SFR}, 24puz would be among the most offset LFBOTs relative to its host, but the LFBOT AT\,2023fhn is at a similarly large offset \citep[][]{chrimes_fhnlate, chrimes_finch}. 24puz and AT\,2023fhn both have large offsets given expectations based on their stellar mass, if they are associated with star formation, as is seen when comparing to the core-collapse supernovae shown in Figure~\ref{fig:Ms_SFR}.

TDE host galaxies, in contrast, are inconsistent with the 24puz host. TDE hosts are shown as blue scatter in Figure~\ref{fig:Ms_SFR}. These hosts are generally more massive than that of 24puz and lie below the star forming main sequence \citep[][]{YaoTDESamp, hammerstein_green}. Selection effects likely play a role here. TDEs are required to be nuclear, whereas massive black holes in dwarf galaxies are often wandering due to a poorly defined gravitational potential. If TDEs from low mass galaxies tend to be off-nuclear, like 24puz, they will not be included in current selections. F-TDE host galaxies are the most discrepant from that of 24puz, as they tend to be the reddest and most massive ($M_* \gtrsim 10^{10}\,M_\odot$) of the TDE hosts \citep[][]{hammerstein_finalseason}

We conclude that 24puz closely resembles LFBOTs, except that it is more luminous in the X-ray through optical, slower evolving, and has no spectral features at late times. 24puz is similar to TDEs in many ways, but that it is fast and luminous in the optical/UV, off-nuclear from a low-mass galaxy, and has harder, faster-evolving X-ray emission than most TDEs. 

\subsection{The rate of 24puz-like transients}

Our search that produced 24puz was incomplete, and a sample size of one leads to large statistical uncertainties in a rate estimate. However, we can set an upper limit on the rate using the ZTF bright transient survey (BTS; \citealp[][]{perley_bts, Fremling2020TheCatalogs}), which has produced complete samples of transients above $m<18.5\,$mag. No sources have been reported with blue, constant colors, extragalactic redshifts, and a timescale of ${\sim}10$ days. We thus use the lack of 24puz-like objects reported in the BTS sample to constrain the rate of such events. Given the peak magnitude of 24puz ($m_g=19.2$\,mag), the magnitude limit of BTS, and the completeness reported by \cite{perley_bts}, we find a $3\sigma$ upper limit on the rate of 24puz-like objects of $<4.6 f_c^{-1}\,{\rm Gpc}^{-3}\,{\rm yr}^{-1}$. Here $f_c$ is the completeness fraction. 

The optical TDE rate is $310^{+60}_{-100}$\,Gpc$^{-3}$\,yr$^{-1}$ \citep[][]{YaoTDESamp}, which is higher than the 24puz rate, but is consistent if 24puz-like objects are a subset of TDEs. The jetted TDE rate is $2^{+4}_{-1} (f_b/1\%)^{-1}$\,Gpc$^{-3}\,$yr$^{-1}$ for a beaming fraction $f_b=1\%$ \citep[][]{andreoni_cmc}, which is consistent with the rate of 24puz-like objects. The rate of LFBOTs has been estimated as $0.3{-}420$\,yr$^{-1}$\,Gpc$^{-3}$ \citep[][]{ho_fbots}, $<300$\,yr$^{-1}$\,Gpc$^{-3}$ \citep[][extrapolated from Palomar Transient Factory data]{coppejans_css161010}, and $700-1400$\,yr$^{-1}$\,Gpc$^{-3}$ \citep[][extrapolated from Pan-STARRS1 Medium-Deep Survey data]{coppejans_css161010}. These are generally consistent with the rate of 24puz-like objects, although the LFBOT rate from the Pan-STARRS1 Medium-Deep Survey is higher. 

\subsection{Nature of the powering source}

We next consider possible explanations to account for nature of the source powering 24puz. We require a system that can produce ${\gtrsim}10^{51}$\,erg of radiated energy, largely in a ${\sim}2$ week span. In the shock model, this energy could be produced in a shorter, ${\sim}$day time-span (${\sim}10^5$\,s), corresponding to the light-curve rise time. The source is surrounded by a dense ${\sim}10^{10}$\,cm$^{-3}$ medium extending to at least $10^{15}$ cm, with a shallow density profile scaling as ${\sim}r^{-1}$. The source produces hard ($\Gamma = 1.7$) X-ray emission that is variable on short, ${\sim}3$ day timescales. We conclude that 24puz has a compact central engine, as has been argued for LFBOTs and given the concordance with the observed properties.

In the next portion of this discussion, we will speculate from the host galaxy properties of 24puz. With a sample size of one, such a procedure is poorly justified. Our goal is not to definitely exclude models or declare a specific model correct. Even if we favor a model, we not that this will always be riddled with uncertainty as we are extrapolating from a single observed source at this time.

The host galaxy of 24puz has a low star-formation rate and stellar mass. The lack of strong star formation is in contrast to expectations from models invoking massive stars or magnetars for the origin of a source like 24puz. The low mass is unexpected for models that invoke neutron stars, which should be more abundant in galaxies with higher stellar masses. Selection effects may of course play a role as we required a faint host galaxy. Star-forming galaxies are detectable to higher redshifts than quiescent galaxies, due to their higher luminosities. If this selection effect is the reason that we have not detected star formation, then the lack of 24puz-like transients identified thus far in, e.g., the BTS sample or supernovae searches is a puzzle. These events should be orders of magnitude more common in star-forming and massive galaxies. Alternatively, we may be simply fortunate to have detected 24puz in a galaxy with very few massive stars. The only way to test this explanation is to continue searching for such events and discovering a sample. Instead, for the rest of this section, we will assume that 24puz is not produced by a massive star. A similar argument holds if 24puz is associated with a globular cluster or a NS, which should be more common in massive galaxies, so we do not consider these scenarios either. Note that this does not rule out a {\it nuclear} stellar cluster as a putative host. 

Instead, we consider accreting black holes. If 24puz is associated with a $<50\,M_\odot$ stellar mass black hole, such events should be orders of magnitude more abundant in massive galaxies ${\sim}10^{11}\,M_\odot$ \citep[][]{Elbert2018CountingLIGO, Sicilia2022TheDistribution}. 
If 24puz is associated with SMBHs $>10^{6}\,M_\odot$, TDE searches should have discovered analogous events, although they have not been focused on off-center events. Moreover, the proximity of a putative dwarf galaxy host and the empirically derived local scaling relations between central black hole mass and stellar mass of galaxies suggests that the BH mass in 24puz is firmly in the IMBH mass range \citep[][]{ReinesMV2015}. Hence, if 24puz is associated with a black hole, it must be associated with a black hole of mass $M_{\rm BH}$ such that $50 \lesssim M_{\rm BH}/M_\odot \lesssim 10^5\,M_\odot$.

We conclude that 24puz is most likely produced by a moderately massive black hole. From the energy budget point of view, accretion events onto massive BHs can readily produce the amount of energy observed, and accreting BHs are known to be able to launch winds/jets, so our shock models are feasible. However, in the context of this explanation, we must contend with two points. First, the transient is highly offset. Second, the shock breakout analysis of the optical emission suggests a dense (${\sim}10^{10}$\,cm$^{-3}$), extended ($10^{15}$\,cm$\approx 10^4$\,AU) medium surrounding the transient. This is much denser than typical environments around nuclear supermassive black holes (see Fig. 2 of \citealp[][]{Alexander2020RadioEvents}). 
Feasible scenarios to produce 24puz must produce this medium, regardless of BH mass.

The high offset from the host galaxy is feasible if the black hole is recoiling and/or has been ejected. At the lower end of this mass range, a recoil is fairly plausibly expected: BHs formed from stellar deaths are expected to have a natal kick similar to that of a neutron star \citep[][]{Repetto2012InvestigatingKicks}. If the BH is formed at the time that most of the stars in the galaxy formed (${\sim}100$\,Myr ago, based on our host stellar population modeling) and receives a kick of ${\sim}100$\,km\,s$^{-1}$, it will travel $10$\,kpc by the time of our observation. Thus, the projected offset observed of 5\,kpc is feasible. 

At the higher end of this mass range, we are met with a challenge. IMBHs, if formed via direct collapse of a gas cloud in-situ, should not necessarily have an associated kick. They are however expected to wander throughout their host galaxy  \citep[][]{ricarte2021a, ricarte2021b}, particularly if the host is a dwarf galaxy with a poorly defined gravitational potential \citep[][]{Weller2022DynamicsSimulation}. This BH is very offset from any light from the host galaxy, which depending on the simulation suite analyzed may or may not be expected frequently, the predictions from the ROMULUS simulation \citep{ricarte2021a} are different from those from the ASTRID simulation wherein wanderer off-set distributions are shorter \citep[][]{Weller2022DynamicsSimulation}. Assuming the dark matter traces the stellar matter, it is unlikely that the BH is close to even a local minimum in the gravitational potential. This model is currently unsettled. Instead, the IMBH could recoil if it has undergone a merger with another IMBH \citep[][]{Gonzalez2007SupermassiveSpins}, which is feasible if the disturbed morphology of G1 indicates a merger. Deep follow-up imaging can confirm or exclude evidence of a merger, but note that the galaxies of stellar mass $10^8\,M_\odot$ are not expected to undergo many mergers specially at late times, with an expected rate of ${<}0.01\,{\rm Gyr}^{-1}$ \citep[][]{Rodriguez-Gomez2015TheModels}. The ejection of an IMBH that has formed and grown in a dense nuclear star cluster \citep{PN-IMBH2021} that has been ejected as a result of a merger is yet another possibility. In that case, we favor the tidal disruption of a white dwarf \citep{2020SSRv..216...39M} over the case of main-sequence disruption, because the circularization (or disk formation) timescale for main-sequence TDEs may be much longer for IMBHs due to weak apsidal precession \citep{2015ApJ...812L..39D}. Another challenge of the IMBH-TDE picture is the need of a dense circum-transient medium, which is not generally available before the disruption. 

The high circum-transient medium density and energetics of the event can be accommodated by the model from \cite{Metzger_LFBOT}. This model nominally invokes a black hole or neutron star in a binary with a massive, Wolf-Rayet star. The compact object enters into a common envelope phase with the massive star and begins inspiraling until it disrupts the stellar core. The lack of star formation from 24puz disfavors a massive star companion, but an main-sequence or evolved stellar companion should produce a similar signal. If the evolved star is a (sub)giant with a large radius, then the timescale may be too long to produce the very short rise time of 24puz. A tidally disrupted main-sequence star may better match the timescale \citep{2021ApJ...911..104K}. We encourage efforts to simulate the emission from 24puz within this model, in particular, taking into account the requirements of an evolved or main sequence stellar companion and a low binary mass ratio. 

We finally consider the case that our assumption that the source cannot be produced by an NS or low-mass, stellar-mass BH due to the host galaxy stellar mass does not hold up to continued searches for 24puz-like sources. Another possible scenario is that the event comes from the explosion of a low-mass helium star with a main-sequence companion star and that the newly born neutron star happens to tidally disrupt the companion star \citep{2025arXiv250103316T}. The low-mass helium star undergoes extreme mass loss before the SN explosion \citep{2022ApJ...940L..27W}, creating the dense circum-transient medium. Such an explosion may be produced after a delay up to 50 Myrs that is consistent with the stellar population age in G1.

\section{Conclusions} \label{sec:conc}

In this paper, we have presented AT\,2024puz, a luminous, multiwavelength transient associated with a dwarf galaxy. 24puz was discovered as a hostless, blue optical flare by the ZTF survey. Deep imaging and spectroscopy showed that it is associated with a dwarf galaxy of mass ${\sim}10^8\,M_\odot$ at $z=0.35614\pm0.00009$ that lies below the star-formation main sequence. The spectra show no features associated with the transient. 

The optical transient was accompanied by luminous UV, IR, and X-ray emission. The early-time optical/UV emission is consistent with either an accretion flare that produces highly super-Eddington radiation, or with a shock that is breaking out within the circum-transient medium. In the latter case, the circum-transient medium mass is $\lesssim 1\,M_\odot$ and it extends to a radius ${\sim}10^{15}\,$cm with a shallow density profile. The shock breakout model may not explain the lack of late-time cooling and line emission, and so would require an additional ionizing source, such as accretion, at ${\sim}20$\,days. A near-infrared excess is seen at late-times ${\sim}80$\,days that may be consistent with a re-processing layer. The X-ray emission could be produced by a shock or a compact central engine, but the rapid evolution and variability is challenging to accommodate within a shock model. 

Based on the lack of star formation and low stellar mass in the host galaxy, we favor a model invoking an BH with mass ${\gtrsim}50\,M_\odot$ (i.e., high mass stellar mass BH or IMBH), although we mildly prefer a BH formed in a stellar explosion to accommodate the high observed offset from the host galaxy. A similar model has been proposed by \cite{Metzger_LFBOT}, who postulate a stellar mass black hole consuming a massive star. Given the low star formation rate of G1, we prefer a main sequence or evolved companion, and we encourage more detailed consideration of this model. However, models that involve ejected IMBHs also remain plausible, in which case we favor the tidal disruption of a white dwarf to match the observed timescale of 24puz.

Pinning down the trigger of 24puz requires a sample of similar sources. The portion of parameter space occupied by this transient has, to our knowledge, barely been explored, and opens up a new section of the landscape of hot, blue transients. These objects are proving to offer a powerful probe of discovering massive stellar mass BHs, IMBHs, and NSs. Instruments and telescopes such as the upcoming Rubin Observatory, the Einstein Probe, and the Ultraviolet Explorer should be sensitive to thousands of these hot blue transients. However, archival searches in ZTF show great promise for identifying and characterizing such previously unexplored transient classes.

In addition to identifying a sample of 24puz-like events, deep late-time follow-up may be able to pin down the event trigger. In particular, deep, multi-band imaging would enable detailed constraints on the star-formation history both of G1 and at the location of 24puz, strengthening our arguments against massive star and NS progenitors. Late-time radio constraints would exclude the late-time emergence of any collimated jet, as is sometimes observed for TDEs and thus may favor an IMBH progenitor \citep[][]{Cendes2024UbiquitousEvents}. Deep space-based spectroscopy can rule out any late-time interaction features, which would provide an additional handle on the circum-transient medium spectrum. 

We conclude by briefly considering implications for LFBOT searches and models, if 24puz is a member of this population and our analysis of the likely progenitor is correct (accreting, ${\gtrsim}50\,M_\odot$ stellar mass BH in a binary). 24puz adds to the increasing fraction of LFBOTs detected in low mass galaxies near the star-forming main sequence, and is the second LFBOT with a very high offset. Models that implicate massive stars in LFBOT models should be reassessed. 24puz suggests that LFBOTs need not be as fast evolving as previously found, and they can be even more luminous than thought. If this high luminosity is a function of the circum-transient medium parameters, it might suggest that LFBOT luminosity will be correlated with the properties of the companion star from which the BH is accreting. While the radio/millimeter constraints from 24puz offer poor constraints at present, they do suggest that, despite the high optical/UV and X-ray luminosities from this event, the radio/millimeter luminosity was not correspondingly high. LFBOTs may not be ubiquitously associated with luminous emission at these frequencies, as was once thought.

\begin{acknowledgments}
We would like to thank Stella Ocker for the use of her WIRC time. We would like to thank Eliot Quataert and Xiaoshan Huang for useful discussions.

This research is based on observations made with the NASA/ESA Hubble Space Telescope obtained from the Space Telescope Science Institute, which is operated by the Association of Universities for Research in Astronomy, Inc., under NASA contract NAS 5–26555. These observations are associated with program 17854. 

Some/all of the data presented in this paper were obtained from the Mikulski Archive for Space Telescopes (MAST) at the Space Telescope Science Institute. The specific observations analyzed can be accessed via \dataset[DOI]{https://doi.org/10.17909/3kvv-bv86}

Based on observations obtained with the Samuel Oschin Telescope 48-inch and the 60-inch Telescope at the Palomar Observatory as part of the Zwicky Transient Facility project. ZTF is supported by the National Science Foundation under Grant No. AST-2034437 and a collaboration including Caltech, IPAC, the Oskar Klein Center at Stockholm University, the University of Maryland, University of California, Berkeley , the University of Wisconsin at Milwaukee, University of Warwick, Ruhr University Bochum, Cornell University, Northwestern University and Drexel University. Operations are conducted by COO, IPAC, and UW. 

The ZTF forced-photometry service was funded under the Heising-Simons Foundation grant \#12540303 (PI: Graham).

The Gordon and Betty Moore Foundation, through both the Data-Driven Investigator Program and a dedicated grant, provided critical funding for SkyPortal. 

SED Machine is based upon work supported by the National Science Foundation under Grant No. 1106171.

Some of the data presented herein were obtained at the W. M. Keck Observatory, which is operated as a scientific partnership among the California Institute of Technology, the University of California and the National Aeronautics and Space Administration. The Observatory was made possible by the generous financial support of the W. M. Keck Foundation.

The authors wish to recognize and acknowledge the very significant cultural role and reverence that the summit of Maunakea has always had within the indigenous Hawaiian community.  We are most fortunate to have the opportunity to conduct observations from this mountain.

This material is based upon work supported by the National Science Foundation Graduate Research Fellowship under Grant No. DGE‐1745301. MWC acknowledge support from the National Science Foundation with grant numbers PHY-2010970 and OAC-2117997. MN is supported by the European Research Council (ERC) under the European Union’s Horizon 2020 research and innovation programme (grant agreement No.~948381) and by UK Space Agency Grant No.~ST/Y000692/1.
\end{acknowledgments}
\begin{acknowledgments}
P.N. acknowledges support from the Gordon and Betty Moore Foundation and the John Templeton Foundation that fund the Black Hole Initiative (BHI) at Harvard University where she serves 
as an external PI.

M.W.C. acknowledges support from the National Science Foundation with grant numbers PHY-2308862 and PHY-2117997.
\end{acknowledgments}

\vspace{5mm}
\facilities{ATLAS, HST (WFC3, ACS), VLA, Keck:1, Keck:2, Hale, XMM, LDT, HEASARC, Swift, NuSTAR, PS1, NOEMA}

\software{astropy, scipy, emcee, dynesty, MOSFIT, numpy, linetools, photutils, casa, lpipe, spectres, prospector, fsps, python-fsps, sfdmap, swifttools, heasoft, extinction, sedpy, statsmodels, scamp, swarp}
\newpage

\appendix

\section{Details of observations}

In Tables~\ref{tab:sedm}, \ref{tab:LT}, \ref{tab:uvot}, \ref{tab:xrt}, \ref{tab:xmm}, \ref{tab:nustar}, \ref{tab:lris}, and \ref{tab:hst}, we provide the details of our observations, as described in Section~\ref{sec:redex}.

\begin{deluxetable}{cccccc}
\tablecaption{P60/SEDM Observation summary. \label{tab:sedm}}
\tablehead{Date & MJD & Filter &  Mag & Mag Error & Limiting Mag }
\startdata
2024-08-01 & 60523.2 & $u$ & 19.58 & 0.13 & 20.14	 \cr
 &  & $u$ & 19.58 & 0.45 & 18.78	 \cr
 &  & $r$ & 20.066 & 0.0485 &	21.69	 \cr
 &  & $i$ & 20.391 & 0.0769 &	21.52  \cr
 \hline
2024-08-02 & 60524.2 & $u$ & 19.51 & 0.16 & 19.87 \cr
 &  & $u$ & 19.51 & 0.18 & 19.75 \cr
 &  &	$r$ & 20.152 & 0.063 & 21.49 \cr
 & &	$i$ & 20.28 & 0.11 & 21.03 \cr
 \hline
2024-08-14 & 60536.2 &	$g$ & 21.16 & 0.19 & 21.32 \cr
 \hline
 & 60543.2 &	$g$ & 21.52 & 0.20 & 21.61 \cr
 \hline
2024-08-22 & 60544.2 & $g$ & 21.78 & 0.22 & 21.75 \cr
 &  & $i$ & 21.35 & 0.18 & 21.55 
\enddata
\tablecomments{Summary of P60/SEDM Observations.}
\end{deluxetable}

\begin{deluxetable}{ccc}
\centerwidetable
\tablecaption{Lowell Discovery Telescope observations \label{tab:LDT}}
\tablehead{\colhead{Filter} & \colhead{Mag} & \colhead{Mag Error}}
\startdata
$u$ & 22.101 & 0.068 \cr
$g$ & 22.220 & 0.054 \cr
$r$ & 22.417 & 0.062 \cr
$i$ & 22.425 & 0.117 \cr
$z$ & 21.935 & 0.314 \cr
\enddata
\tablecomments{Summary of Lowell Discovery Telescope Observations on 2024-08-27 (MJD 60549).}
\end{deluxetable}

\begin{deluxetable}{cccccc}
\centerwidetable
\tablecaption{Liverpool Telescope observations \label{tab:LT}}
\tablehead{\colhead{Date} & \colhead{MJD} & \colhead{Filter} & \colhead{Mag} & \colhead{Mag Error} & \colhead{Limiting Mag}}
\startdata
2024-08-05 & 60527.97 &  $u$ & 19.85 & 0.15 & 19.27\cr
 &  & $g$ & 20.27 & 0.08 & 20.71 \cr
 &  & $r$ & 20.53 & 0.08 & 21.15 \cr
 &  & $i$ & 20.9 & 0.1 & 21.26 \cr
\hline
2024-08-06 & 60528.96 & $u$ & 19.92 & 0.11 & 19.98 \cr
 &  & $g$ & 20.33 & 0.05 & 21.3\cr
 &  & $r$ & 20.91 & 0.07 & 21.61 \cr
 &  & $i$ & 21.04 & 0.06 & 21.98 \cr
\hline
2024-08-08 & 60530.96 &  $u$ & 20.2 & 0.13 & 20.07 \cr
 &  & $g$ & 20.61 & 0.06 & 21.36 \cr
 &  & $r$ & 21.15 & 0.09 & 21.63 \cr
 &  & $i$ & 21.09 & 0.07 & 21.86\cr
\hline
2024-08-15 & 60537.89 & $u$ & 20.84 & 0.31 & 19.59 \cr
 &  & $g$ & 21.83 & 0.29 & 20.86 \cr
 &  & $r$ & 21.58 & 0.2 & 21.17 \cr
 &  & $i$ & 21.32 & 0.13 & 21.5 \cr
\hline
2024-08-16 & 60538.96 & $u$ & 19.97 & 0.26 & 18.53 \cr
 &  & $g$ & 20.78 & 0.19 & 20.2 \cr
 &  & $r$ & 21.6 & 0.32 & 20.36 \cr
 &  & $i$ & 21.62 & 0.28 & 20.54 \cr
\hline
2024-08-18 & 60540.87 & $g$ & 20.41 & 0.22 & 20.1\cr
 &  &  $r$ & 21.45 & 1.07 & 20.4 \cr
 &  & $i$ & 22.52 & 0.92 & 20.87 \cr
\hline
2024-08-19 & 60541.87 & $u$ & 20.34 & 0.45 & 25.93\cr
 &  & $g$ & 21.1 & 0.51 & 27.14 \cr
 &  & $i$ & 21.33 & 0.54 & 27.84 \cr
\hline
2024-08-20 & 60542.88 & $i$ & 21.43 & 0.22 & 19.16 \cr
\enddata
\tablecomments{Summary of Liverpool Telescope Observations.}
\end{deluxetable}

\begin{deluxetable}{cccccc}
\centerwidetable
\tablecaption{Swift/UVOT observations \label{tab:uvot}}
\tablehead{\colhead{Date} & \colhead{MJD} & \colhead{ObsID} & \colhead{Band} & \colhead{Exposure [sec]} & \colhead{Flux [$\mu$Jy]}}
\startdata
2024-08-03 & 60525.8 & 0016746001 & $u$ & $165.1$ & $39.7 \pm 6.2$ \\
$-$ & $-$ & $-$ & $u$ & $165.2$ & $46.1 \pm 6.4$ \\
$-$ & $-$ & $-$ & $m2$ & $495.8$ & $20.5 \pm 3.7$ \\
$-$ & $-$ & $-$ & $w2$ & $601.5$ & $16.9 \pm 2.2$ \\
2024-08-12 & 60535.0 & 0016746002 & $u$ & $148.4$ & $27.2 \pm 6.0$ \\
$-$ & $-$ & $-$ & $m2$ & $444.7$ & $13.0 \pm 5.0$ \\
$-$ & $-$ & $-$ & $w1$ & $296.0$ & $19.3 \pm 4.1$ \\
$-$ & $-$ & $-$ & $w2$ & $544.8$ & $9.9 \pm 1.9$ \\
2024-08-16 & 60538.5 & 0016746003 & $m2$ & $323.6$ & $5.6 \pm 3.6$ \\
$-$ & $-$ & $-$ & $w2$ & $403.0$ & $3.9 \pm 1.8$ 
\enddata
\tablecomments{Summary of Swift/UVOT Observations.}
\end{deluxetable}

\begin{deluxetable}{cccccc}[h!]
\tablecaption{Swift/XRT Observations \label{tab:xrt}}
\tablehead{\colhead{Date} & \colhead{MJD} & \colhead{ObsID} & \colhead{Exposure [sec]} & \colhead{Counts sec$^{-1}$} & \colhead{Flux [$10^{-13}$ erg\,cm$^{-2}$\,s$^{-1}$]}}
\startdata
2024-08-03 & 60525.8 & 16746001 & 3671.3 & $0.0019^{+0.0011}_{-0.0008}$ & $0.8^{+0.5}_{-0.4}$ \cr
2024-08-12 & 60535.0 & 16746002 & 1476.8 & $0.001^{+0.0019}_{-0.001}$ & ${<}4.1$ \cr
2024-08-16 & 60538.5 & 16746003 & 2783.2 & $0.0023^{+0.0017}_{-0.0012}$ & ${<}3.8$ \cr
2024-08-20 & 60542.6 & 16746004 & 2482.2 & $0.0005^{+0.001}_{-0.0005}$ & ${<}2.2$ \cr
2024-10-30 & 60613.0 & 16746005 & 2061.1 & $0.0009^{+0.0012}_{-0.0007}$ & ${<}2.7$ \cr
2024-11-02 & 60616.2 & 16746006 & 2805.7 & $0.0006^{+0.0008}_{-0.0005}$ & ${<}1.9$ \cr
2024-11-05 & 60619.0 & 16746007 & 2657.9 & $0.0^{+0.0009}_{0.0}$ & ${<}1.6$ \cr
2024-11-08 & 60622.4 & 16746008 & 2710.4 & $0.0034^{+0.0024}_{-0.0017}$ & ${<}5.5$ \cr
\enddata
\tablecomments{Summary of Swift/XRT Observations. Fluxes are computed assuming $\Gamma=1.7$. Upper limits are $3\sigma$.}%
\end{deluxetable}

\begin{deluxetable}{ccccccc}[h!]
\centerwidetable
\tablecaption{\textit{XMM-Newton} Observations \label{tab:xmm}}
\tablehead{
Date & MJD & ObsID & Instrument & Exposure [s] & Count Rate [$10^{-2}$\,ct\,s$^{-1}$] & \colhead{$\log \frac{\rm Flux}{{\rm erg\,cm}^{-2}\,{\rm s}^{-1}}$}
}
\startdata
2024-08-14 & 60536 & 0953011201 & EPIC-PN & 3873 & $4.04\pm0.35$ & $-12.577 \pm 0.024$  \cr
 &  &  & EPIC-MOS1 & 12370 & $1.19 \pm 0.12$ &   \cr
 &  &  & EPIC-MOS2 & 13660 & $0.98 \pm 0.09$ &   \cr\hline
2024-09-05 & 60558 & 0953011301 & EPIC-PN & 12940 & $0.34 \pm 0.12$ & $-13.92 \pm 0.15$  \cr
 &  &  & EPIC-MOS1 & 23700 & $0.033 \pm 0.033$ &   \cr
 &  &  & EPIC-MOS2 & 27790 & $0.017 \pm 0.042$ &   \cr\hline
2024-11-30 & 60644 & 0953012101 & EPIC-PN & 27090 & $0.175 \pm 0.052$ & $-13.91 \pm 0.07$  \cr
 &  &  & EPIC-MOS1 & 37470 & $0.045\pm0.020$ &   \cr
 &  &  & EPIC-MOS2 & 39130 & $0.050\pm0.020$ &   \cr
\enddata
\tablecomments{Summary of \textit{XMM-Newton} Observations. Exposure times are quoted after accounting for good time interval flagging. We compute unabsorbed fluxes from $0.3-10$\,keV by jointly fitting the spectra to a power-law model with a photon-index tied between observations and flux tied between instruments for each observation.}
\end{deluxetable}

\begin{deluxetable}{cccccccc}
\centerwidetable
\tablecaption{\textit{NuSTAR} Observations. \label{tab:nustar}}
\tablehead{
\multirow{2}{*}{Date} & \multirow{2}{*}{MJD} & \multirow{2}{*}{ObsID} & \multirow{2}{*}{FPM} & \multirow{2}{*}{Exposure [s]} & \multicolumn{2}{c}{Counts} & \colhead{Flux 90\% upper limit} \cr
\colhead{} & \colhead{} & \colhead{} & \colhead{} &\colhead{} & \colhead{src} & \colhead{bkg} & \colhead{[$10^{-13}$ erg\,cm$^{-2}$\,s$^{-1}$]} 
}
\startdata
2024-09-06 & 60559.2 & 91001632002 & A & 24879.3 & 205 & 207 & 0.52 \cr
           &         &             & B & 24622.1 & 242 & 251 & 0.57 \cr
2024-09-07 & 60560.7 & 91001632004 & A & 37667.2 & 302 & 294 & 0.70 \cr
           &         &             & B & 37284.5 & 376 & 370 & 0.73 \cr
\enddata
\tablecomments{Summary of \textit{NuSTAR} Observations. Flux upper limits are at the 90\% confidence level, computed assuming an unabsorbed power-law with $\Gamma=1.77$ for the 3--79 keV band.}
\end{deluxetable}

\begin{deluxetable}{cccccc}
\tablecaption{Keck I/LRIS observations \label{tab:lris}}
\tablehead{\colhead{Date} & \colhead{MJD} & \colhead{Exposure [sec]} & \colhead{Airmass} & \colhead{Standard (spec{.})} & \colhead{Slit Width (spec{.})} }
\startdata
2024-07-29 & 60520.4 & 600 & 1.33 & BD+28 & $1\farcs0$ \cr
2024-08-05 & 60527.3 & 1800 & 1.20 & BD+28 & $1\farcs0$  \cr
2024-09-07 & 60560.3 & 4500 & 1.42 & Feige 110 & $1\farcs0$  \cr
2024-10-07 & 60590.2 & 7200 & 1.82 & G191-B2B & $1\farcs0$ \cr
\enddata
\tablecomments{Summary of Keck I/LRIS observations.}%
\end{deluxetable}

\begin{deluxetable}{ccccc}
\tablecaption{{\it HST}/WFC3 Observation summary. \label{tab:hst}}
\tablehead{Date & MJD & Detector & Filter & Exposure [sec] }
\startdata
2024-09-30 & 60583 & UVIS & F606W & 994  \cr
2024-09-30 & 60583 & UVIS & F336W & 1073\cr\hline
2024-09-30 & 60583 & UVIS & F105W & 1059 \cr
2024-09-30 & 60583 & UVIS & F160W & 1059  \cr
\enddata
\end{deluxetable}

\section{Chance association probability}\label{sec:chance_assoc}
In this section, we compute the probability that 24puz randomly lies at its location with respect to nearby galaxies. If 24puz were associated with an unrelated background source, it would have a uniform probability of lying at any location in the field. We assess the probability that 24puz, if unassociated with any detected galaxy, would be located at the observed position with respect to the detected galaxies. Because of the limited field-of-view of our {\it HST} observation, we perform this experiment using observations of the Cosmological Evolution Survey (COSMOS; \citealp[][]{Scoville2007TheOverview}) field as part of the Cosmic Assembly Near-infrared Deep Extragalactic Legacy Survey (CANDELS; \citealp[][]{Grogin2011Candels:Survey, Koekemoer2011Candels:Mosaics}, which were performed with the {\it HST} ACS/WFC instrument and include F606W measurements. While the filter throughputs and pixel scales are slightly different for ACS/WFC F606W and WFC3/UVIS F606W images, we estimate that they cause percent-level changes in the measured fluxes and Kron radii, which are negligible for our purposes. We retrieved \texttt{sextractor} catalogs for the CANDELS field from \cite{Nayyeri2017CANDELSFIELD}. 

We compute the random association probability as follows. First, we correct for the increased depth of the CANDELS image. We randomly assign each COSMOS source a per-pixel rms noise from the observed per-pixel noises in our observations and define detectable sources as those with a total signal-to-noise (i.e., integrated over the entire source rather than per-pixel) larger than three, which is our detection threshold. 

We treat the undetected sources as a random sample of background sources and compute the distances from each to the n$^{\rm th}$ nearest-neighbors, in units of Kron radii, for $n=1-10$. We only consider galaxies within $20''$ given the distance from 24puz to the edge of the observation field. We use these distances to compute two probabilities. First, we consider the nearest galaxy to 24puz, which we will denote ``G1'' hereafter. We compute the probability of finding the nearest-neighbor closer than the distance G1 for a randomly located sources. We find a probability $p_1=3\times10^{-4}$ of the nearest neighbor being closer than G1. Second, we compute the probability that all the second through tenth-nearest neighbors are closer than those observed for 24puz: $p_{2-10} = 1.3\times10^{-3}$. It is thus unlikely both that 24puz lies close to the observed galaxies by chance ($3\sigma$) or that 24puz lies close to G1 by chance ($3.4\sigma$).

\section{Ultraviolet-infrared spectral energy distributions and best-fit parameters}

\begin{figure*}[!ht]
\centering
\includegraphics[width=\textwidth]{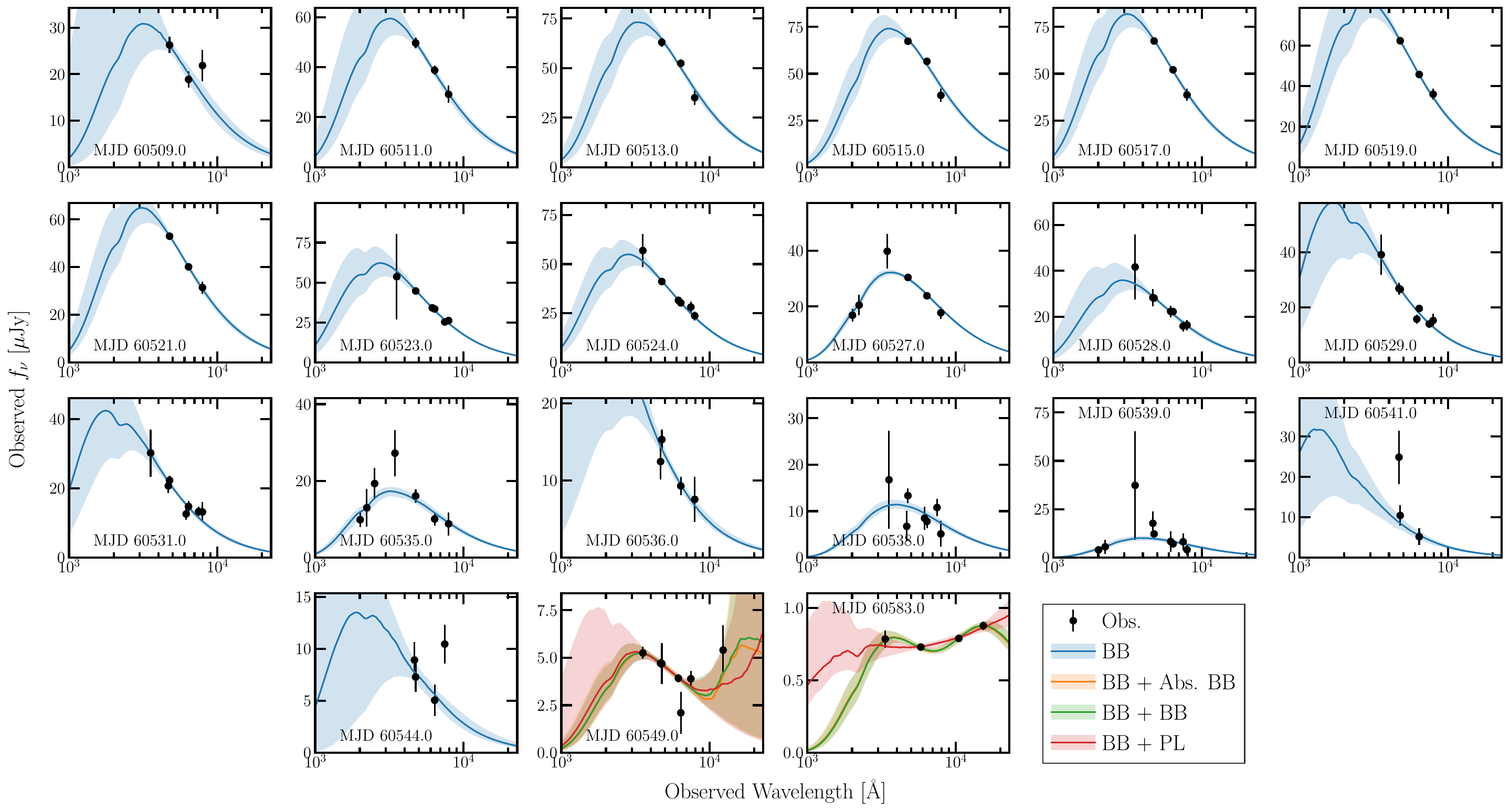}
\caption{Blackbody fits to each epoch of UV/optical imaging of 24puz, as described in Section~\ref{sec:bb_optuv}. The fits are shown as colored bands and the data as black scatter points. }\label{fig:uvopt_fits}
\end{figure*}

In Figure~\ref{fig:uvopt_fits}, we show the multi-epoch, best-fit blackbody models. In Table~\ref{tab:OIR_bb}, we tabulate the best-fit parameters.

\begin{longrotatetable}
\begin{deluxetable*}{ccccccccccc}
\tablecaption{UVOIR blackbody modeling \label{tab:OIR_bb}}
\tablehead{Date & MJD & Model & $\log \frac{T_{bb}}{\rm K}$ & $\log \frac{R_{bb}}{\rm cm}$ & $\log \frac{L_{bb}}{{\rm erg\,s}^{-1}}$  & $M_g$ [mag] & $\log \frac{T_{\rm IR}}{\rm K}$ & $\log \frac{R_{\rm IR}}{\rm cm}$ & $\Gamma_{\rm IR}$ & $\log \frac{L_{\rm IR}}{{\rm erg\,s}^{-1}}$ }
\startdata
2024-07-18 & 60509 & BB & $4.37^{+0.18}_{-0.13}$ & $15.06^{+0.14}_{-0.17}$ & $44.44^{+0.4}_{-0.24}$ & $-20.56^{+0.101}_{-0.069}$ & $-$ & $-$ & $-$ & $-$ \cr
2024-07-20 & 60511 & BB & $4.37^{+0.081}_{-0.064}$ & $15.197^{+0.066}_{-0.074}$ & $44.73^{+0.17}_{-0.13}$ & $-21.249^{+0.052}_{-0.052}$ & $-$ & $-$ & $-$ & $-$ \cr
2024-07-22 & 60513 & BB & $4.336^{+0.055}_{-0.046}$ & $15.288^{+0.049}_{-0.054}$ & $44.774^{+0.114}_{-0.091}$ & $-21.55^{+0.032}_{-0.033}$ & $-$ & $-$ & $-$ & $-$ \cr
2024-07-24 & 60515 & BB & $4.309^{+0.04}_{-0.039}$ & $15.333^{+0.042}_{-0.042}$ & $44.751^{+0.08}_{-0.067}$ & $-21.646^{+0.035}_{-0.033}$ & $-$ & $-$ & $-$ & $-$ \cr
2024-07-26 & 60517 & BB & $4.377^{+0.056}_{-0.048}$ & $15.257^{+0.046}_{-0.054}$ & $44.871^{+0.118}_{-0.094}$ & $-21.581^{+0.037}_{-0.035}$ & $-$ & $-$ & $-$ & $-$ \cr
2024-07-28 & 60519 & BB & $4.425^{+0.056}_{-0.051}$ & $15.189^{+0.046}_{-0.05}$ & $44.93^{+0.12}_{-0.11}$ & $-21.451^{+0.032}_{-0.027}$ & $-$ & $-$ & $-$ & $-$ \cr
2024-07-30 & 60521 & BB & $4.384^{+0.057}_{-0.048}$ & $15.195^{+0.048}_{-0.053}$ & $44.775^{+0.124}_{-0.095}$ & $-21.295^{+0.032}_{-0.031}$ & $-$ & $-$ & $-$ & $-$ \cr
2024-08-01 & 60523 & BB & $4.436^{+0.073}_{-0.055}$ & $15.106^{+0.05}_{-0.063}$ & $44.81^{+0.17}_{-0.12}$ & $-21.077^{+0.022}_{-0.026}$ & $-$ & $-$ & $-$ & $-$ \cr
2024-08-02 & 60524 & BB & $4.423^{+0.068}_{-0.055}$ & $15.102^{+0.051}_{-0.058}$ & $44.75^{+0.15}_{-0.12}$ & $-21.006^{+0.032}_{-0.037}$ & $-$ & $-$ & $-$ & $-$ \cr
2024-08-05 & 60527 & BB & $4.299^{+0.014}_{-0.015}$ & $15.164^{+0.019}_{-0.016}$ & $44.377^{+0.029}_{-0.027}$ & $-20.751^{+0.028}_{-0.039}$ & $-$ & $-$ & $-$ & $-$ \cr
2024-08-06 & 60528 & BB & $4.397^{+0.086}_{-0.068}$ & $15.046^{+0.069}_{-0.078}$ & $44.54^{+0.19}_{-0.14}$ & $-20.613^{+0.049}_{-0.038}$ & $-$ & $-$ & $-$ & $-$ \cr
2024-08-07 & 60529 & BB & $4.61^{+0.15}_{-0.11}$ & $14.84^{+0.09}_{-0.11}$ & $44.96^{+0.37}_{-0.26}$ & $-20.431^{+0.049}_{-0.038}$ & $-$ & $-$ & $-$ & $-$ \cr
2024-08-09 & 60531 & BB & $4.58^{+0.18}_{-0.12}$ & $14.82^{+0.11}_{-0.13}$ & $44.8^{+0.46}_{-0.29}$ & $-20.205^{+0.055}_{-0.061}$ & $-$ & $-$ & $-$ & $-$ \cr
2024-08-13 & 60535 & BB & $4.345^{+0.03}_{-0.031}$ & $14.966^{+0.041}_{-0.046}$ & $44.157^{+0.052}_{-0.055}$ & $-19.966^{+0.093}_{-0.084}$ & $-$ & $-$ & $-$ & $-$ \cr
2024-08-14 & 60536 & BB & $4.66^{+0.23}_{-0.21}$ & $14.66^{+0.17}_{-0.15}$ & $44.83^{+0.6}_{-0.49}$ & $-19.74^{+0.07}_{-0.11}$ & $-$ & $-$ & $-$ & $-$ \cr
2024-08-16 & 60538 & BB & $4.268^{+0.028}_{-0.027}$ & $14.987^{+0.04}_{-0.039}$ & $43.903^{+0.047}_{-0.049}$ & $-19.708^{+0.11}_{-0.073}$ & $-$ & $-$ & $-$ & $-$ \cr
2024-08-17 & 60539 & BB & $4.253^{+0.05}_{-0.045}$ & $14.982^{+0.059}_{-0.075}$ & $43.83^{+0.083}_{-0.088}$ & $-19.62^{+0.15}_{-0.13}$ & $-$ & $-$ & $-$ & $-$ \cr
2024-08-19 & 60541 & BB & $4.78^{+0.17}_{-0.25}$ & $14.47^{+0.16}_{-0.11}$ & $44.91^{+0.47}_{-0.64}$ & $-19.19^{+0.25}_{-0.14}$ & $-$ & $-$ & $-$ & $-$ \cr
2024-08-22 & 60544 & BB & $4.55^{+0.29}_{-0.25}$ & $14.63^{+0.24}_{-0.21}$ & $44.3^{+0.74}_{-0.55}$ & $-19.15^{+0.14}_{-0.15}$ & $-$ & $-$ & $-$ & $-$ \cr
\hline
2024-08-27 & 60549 & BB+BB & $4.33^{+0.07}_{-0.046}$ & $14.728^{+0.055}_{-0.083}$ & $43.624^{+0.118}_{-0.074}$ & $-18.72^{+0.122}_{-0.063}$ & $3.41^{+0.19}_{-0.26}$ & $17.6^{+1.0}_{-1.4}$ & $42.7^{+0.7}_{-5.8}$  \cr
 & & BB+Dust & $4.33^{+0.07}_{-0.046}$ & $14.728^{+0.055}_{-0.083}$ & $43.624^{+0.118}_{-0.074}$ & $-18.72^{+0.122}_{-0.063}$ & $3.42^{+0.2}_{-0.27}$ & $16.0^{+1.07}_{-0.68}$ & $-$ & $42.8^{+0.9}_{-1.2}$  \cr
 & & BB+PL & $4.33^{+0.07}_{-0.046}$ & $14.728^{+0.055}_{-0.083}$ & $43.624^{+0.118}_{-0.074}$ & $-18.72^{+0.122}_{-0.063}$ & $-$ & $-$ & $2.3^{+1.7}_{-1.4}$ & $44.6^{+4.8}_{-7.6}$  \cr
\hline
2024-09-30 & 60583 & BB+BB & $4.284^{+0.045}_{-0.039}$ & $14.385^{+0.051}_{-0.054}$ & $42.756^{+0.078}_{-0.057}$ & $-16.778^{+0.048}_{-0.064}$ & $3.584^{+0.033}_{-0.032}$ & $17.075^{+0.058}_{-0.055}$ & $41.95^{+0.021}_{-0.034}$  \cr
 & & BB+Dust & $4.284^{+0.045}_{-0.039}$ & $14.385^{+0.051}_{-0.054}$ & $42.756^{+0.078}_{-0.057}$ & $-16.778^{+0.048}_{-0.064}$ & $3.586^{+0.029}_{-0.032}$ & $15.375^{+0.057}_{-0.049}$ & $-$ & $41.948^{+0.027}_{-0.028}$  \cr
 & & BB+PL & $4.284^{+0.045}_{-0.039}$ & $14.385^{+0.051}_{-0.054}$ & $42.756^{+0.078}_{-0.057}$ & $-16.778^{+0.048}_{-0.064}$ & $-$ & $-$ & $0.23^{+0.24}_{-0.17}$ & $42.684^{+0.058}_{-0.056}$  \cr
\enddata%
\end{deluxetable*}
\end{longrotatetable}

\section{X-ray spectral modeling}
\begin{figure*}
\centering
\includegraphics[width=0.45\textwidth]{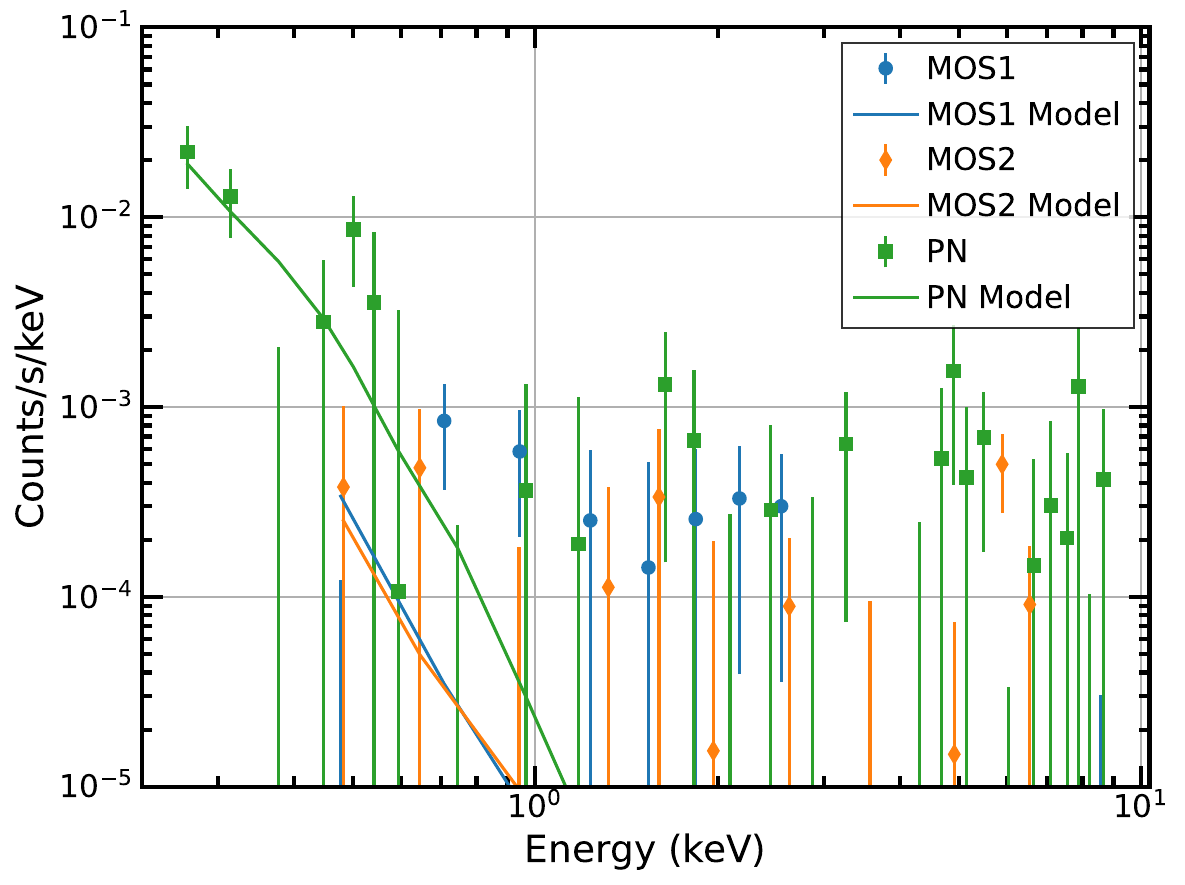}
\includegraphics[width=0.45\textwidth]{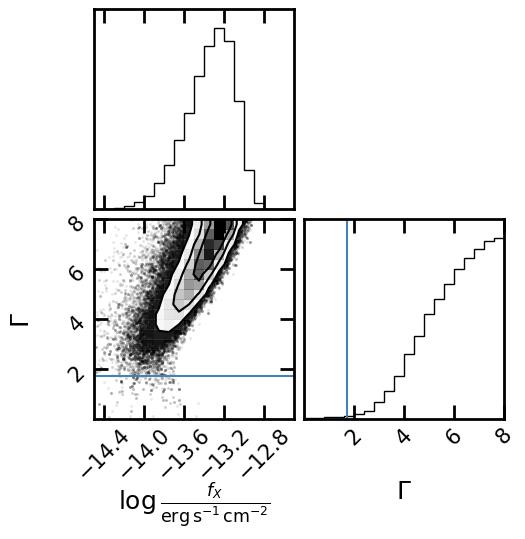}
\caption{Spectral constraints from the MJD 60558 (37.6 rest-days) {\it XMM-Newton} observations, in the same format as Figure~\ref{fig:xray_spec} We assume no absorption in this fit, based on the results from the higher signal-to-noise MJD 60536 observations. The photon index from the first epoch is shown in blue. The photon index has tentatively softened: the probability that the photon-index is consistent with the first epoch is $P(\Gamma \le 1.73) = 1\%$. 
}\label{fig:xray_spec2}
\end{figure*}

The best-fit spectral parameters of our MJD 60558 {\it XMM-Newton} observations are shown in Figure~\ref{fig:xray_spec2}. 

\section{The Compton equilibrium temperature}\label{sec:compton}
Given the strong X-ray source associated with 24puz, the electron temperature may be hotter than the standard photoionization equilibrium temperature ${\sim}10^4\,{\rm K}$ due to Compton heating \citep[][]{ho_18cow}. We compute the Compton equilibrium temperature $T_C$ as follows. The Compton heating rate per unit density is given by \cite{Sazonov2004Quasars:Heating, ho_18cow}
\begin{equation}
    H(\nu) = \frac{\sigma_T}{m_e c^2} \int_0^\infty h \nu f_\nu a\Big(\frac{h\nu}{m_e c^2}\Big) d\nu,
\end{equation}
where $f_\nu$ is the incident flux density, $\nu$ is frequency, and all constants are given with standard symbols. The function $a\Big(\frac{h\nu}{m_e c^2}\Big)$ accounts for Klein-Nishina corrections and, for $k_B T_C \ll m_e c^2$, is given by
\begin{gather}
    a(x) = \frac{3}{8x^4}(x-3)(x+1)\ln(2x+1)\nonumber \\ 
    + \frac{-10x^4+51x^3+93x^2+51+9}{4x^3(2x+1)^3} \nonumber\\
    \approx 1 - \frac{21x}{5}+\mathcal{O}(x^2) .
\end{gather}
For many astropysical systems, including 24puz, the heating is dominated by photons with $h\nu \gtrsim 10\,{\rm keV}$, i.e., hard X-rays.

The cooling rate is given by 
\begin{equation}
    C(\nu) = \frac{4 k_B T \sigma_T}{m_e c^2} \int_0^\infty f_\nu b\Big(\frac{h\nu}{m_e c^2}\Big) d\nu.
\end{equation}
Here, $T$ is the temperature of the irradiated region and $b\Big(\frac{h\nu}{m_e c^2}\Big)$ accounts for Klein-Nishina corrections. The latter is given by
\begin{multline}
    b(x) = \frac{1}{4} \bigg(\frac{3(3x^22-4x-13)}{16x^3}\ln(2x+1) +  \\
    \frac{-216x^6+476x^5+2066x^4}{8x^2(2x+1)^5} \\
    + \frac{2429x^3+1353x^2+363x+39}{8x^2(2x+1)^5}\bigg) \\
    \approx 1 - \frac{47x}{8}+\mathcal{O}(x^2) .
\end{multline}
Once again, this expression is valid for $k_B T_C \ll m_e c^2$. Compton cooling is dominated by photons with $h\nu \lesssim 10\,{\rm keV}$.

The temperature evolution is then given by $(3/2)k_B {\rm d}T/{\rm d}t = H(\nu) - C(\nu)$. The temperature reaches an equilibrium $T_C$ when the heating and cooling rates are equal, or
\begin{equation}
    T_C = \frac{1}{4 k_B} \frac{ \int_0^\infty h \nu f_\nu a\Big(\frac{h\nu}{m_e c^2}\Big) d\nu}{\int_0^\infty f_\nu b\Big(\frac{h\nu}{m_e c^2}\Big) d\nu}.
\end{equation}

Extrapolating from the optical/UV and X-ray observations of 24puz described in Sections~\ref{sec:bb_optuv} and \ref{sec:xrayem} to typical values during our radio observation, we assume that $f_\nu$ is the sum of a blackbody with $T\approx 10^{4.3}\,$K and $\log L_{\rm OptUV}/({\rm erg\,s}^{-1}) \approx 43.5$ and a power-law with $\Gamma=1.77$ and $L_{\rm X}/({\rm erg\,s}^{-1}) \approx 42.2$, where we cut-off the power-law at $0.3$\,keV. These values are most appropriate during our final epoch of radio observations, but we verified that the equilibrium temperature does not change significantly by adopting values appropriate for earlier observations. We find $T_C = 4\times 10^5\,{\rm K}$, and this equilibrium temperature is reached in ${\sim}$days. An increased optical/UV luminosity will decrease the temperature, while an increased X-ray component will increase the temperature. We tested a range of optical/UV and X-ray luminosity and found $T_C \gtrsim 10^5\,$K for most reasonable assumptions. As can be seen from Equation~\ref{eq:freefree} \citep[or Eq. 10.16 of][]{Draine:2011}, in the radio regime, where $h\nu \ll kT_e$, increasing the temperature tends to decrease $\tau_{ff}$, so we conservatively adopt $T_e = 10^5\,$K. 

\section{MOSFIT}

\begin{figure*}[!ht]
\centering
\includegraphics[width=\textwidth]{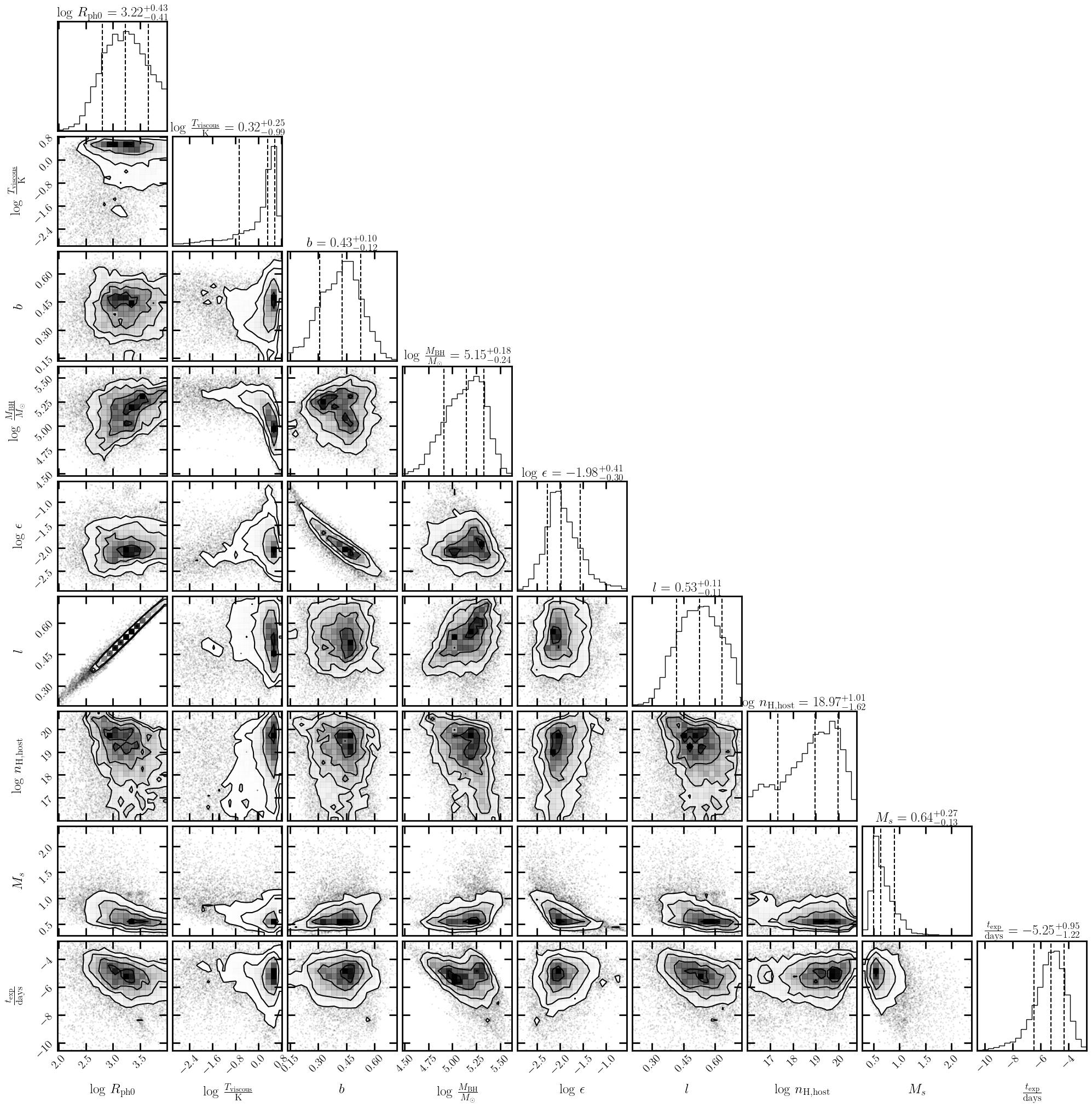}
\caption{Corner plot showing the full \texttt{MOSFIT} parameter set.}\label{fig:mosfit_corner}
\end{figure*}

In Figure~\ref{fig:mosfit_corner}, we show the full set of best-fit parameters found by \texttt{MOSFIT}. The parameter definitions are detailed in \cite{Mockler2019WeighingEvents}.

\bibliography{sample631}{}
\bibliographystyle{aasjournal}

\end{document}